\documentclass[letterpaper, 11pt]{article}

\usepackage{jheppub} 

\usepackage{graphicx}
\usepackage{epstopdf}
\usepackage{amsmath, amssymb}

\usepackage{tikz}
\usepackage{verbatim}
\usepackage{bm}
\usepackage{bbold}
\usepackage{caption}
\usepackage{subcaption}

\usepackage{ytableau}	

\newcommand{\be}{\begin{eqnarray}}
\newcommand{\ee}{\end{eqnarray}}
\newcommand{\bea}{\begin{eqnarray}}
\newcommand{\eea}{\end{eqnarray}}

\newcommand{\nn}{\nonumber}
\newcommand{\bn}{\begin{enumerate}}
\newcommand{\en}{\end{enumerate}}

\def\IZ{\mathbb{Z}}

\def\CB{{\cal B}}
\def\CC{{\cal C}}

\def\CF{{\cal F}}

\def\CL{{\cal L}}
\def\CM{{\cal M}}
\def\CN{{\cal N}}
\def\CO{{\cal O}}

\def\CR{{\cal R}}
\def\CS{{\cal S}}

\def\CU{{\cal U}}

\def\CW{{\cal W}}

\def\a{\alpha}
\def\b{\beta}
\def\g{\gamma}

\def\e{\epsilon}

\def\l{\lambda}

\def\s{\sigma}

\def\G{\Gamma}

\def\half{\frac{1}{2}}

\def\vev#1{\langle #1 \rangle}

\def\Tr{{\rm Tr}}
\def\tr{{\rm Tr}}
\def\det{{\rm det}}

\def\vec#1{\bm{#1}}
\def\fp{\mathfrak{p}}
\def\fq{\mathfrak{q}}
\def\fj{\mathfrak{j}}
\def\fm{\mathfrak{m}}

\title{Infinitely many $\CN=1$ dualities from $m+1-m=1$}

\author{
Prarit Agarwal,
Kenneth Intriligator
and Jaewon Song
}

\affiliation{Department of Physics, University of California, San Diego, La Jolla, CA 92093, USA}

\emailAdd{pagarwal@physics.ucsd.edu}
\emailAdd{keni@physics.ucsd.edu}
\emailAdd{jsong@physics.ucsd.edu}

\preprint{UCSD-PTH-14-10}

\abstract{We discuss two infinite classes of 4d supersymmetric theories, ${T}_N^{(m)}$ and ${\cal U}_N^{(m)}$, labelled by an arbitrary non-negative integer, $m$.  The ${T}_N^{(m)}$ theory arises from the 6d, $A_{N-1}$ type ${\cal N}=(2,0)$ theory reduced on a 3-punctured sphere, with normal bundle given by line bundles of degree $(m+1, -m)$; the $m=0$ case is the ${\cal N}=2$ supersymmetric $T_N$ theory.  The novelty is the negative-degree line bundle.  The ${\cal U}_N^{(m)}$ theories likewise arise from the 6d ${\cal N}=(2,0)$ theory on a 4-punctured sphere, and can be regarded as gluing together two (partially Higgsed) ${T}_N^{(m)}$ theories.   The ${T}_N^{(m)}$ and ${\cal U}_N^{(m)}$ theories can be represented, in various duality frames, as quiver gauge theories, built from $T_N$ components via gauging and nilpotent Higgsing.  We analyze the RG flow of the ${\cal U}_N^{(m)}$ theories, and find that, for all integer $m>0$, they end up at the same IR SCFT as $SU(N)$ SQCD with $2N$ flavors and quartic superpotential.  The ${\cal U}_N^{(m)}$ theories can thus be regarded as an infinite set of UV completions, dual to SQCD with $N_f=2N_c$.  The  ${\cal U}_N^{(m)}$ duals have 
different duality frame quiver representations, with $2m+1$ gauge nodes. }

\begin{document}
\maketitle
\flushbottom

\section{Introduction}

Different 4d $\CN =1$ supersymmetric theories can RG flow to the same IR SCFT \cite{Seiberg:1994pq}.  Such dual descriptions are not merely two similar UV completions of the same IR physics, but rather encode the IR physics quite differently, exchanging strong and weak coupling effects such as Higgsing and mass terms.  The original duality of \cite{Seiberg:1994pq} relates the electric $SU(N_c)$ SQCD theory, with $N_f$ flavors, to a magnetic $SU(N_f-N_c)$ theory, with $N_f$ flavors and added meson singlets and superpotential.  

We will be focussing on $SU(N_c)$ SQCD with $N_f=2N_c$, where the gauge group is self-dual\footnote{Upon adding a quartic $W_{tree}$ on the electric side, the theory is completely self-dual, as the 
meson singlets of the magnetic theory get a mass and can be integrated out.  This theory can be obtained from the self-dual ${\cal N}=2$ SQCD superconformal field theory with $N_f=2N_c$, upon breaking ${\cal N}=2$ to ${\cal N}=1$ by an added mass term for the adjoint chiral superfield; see \cite{Leigh:1995ep,Argyres:1996eh} for discussion of the ${\cal N}=1$ duality from this perspective.}.  In  \cite{Gadde:2013fma}, a new dual of $N_f=2N_c$ SQCD was found, involving two copies of the $T_N$ theory of \cite{Gaiotto:2009we} (see \cite{Tachikawa:2015bga} for a nice, recent review), along with $2N^2+2N$ gauge singlets and a specific superpotential. In  \cite{Agarwal:2014rua}, another new dual of $N_f=2N_c$ SQCD was found, involving a single 
 $T_N$ theory, two quarks/anti-quarks, $N^2 + N$ gauge singlets, and an intricate superpotential.  For $N=2$, the $T_2$ theory reduces to eight free chiral multiplets, the gauging can then be written as a standard Lagrangian, and the duals in this case reduces to ones analyzed in \cite{Intriligator:1995ne, Csaki:1997cu}.

\begin{figure}[h]
	\centering
	\begin{subfigure}[b]{6.0in}
	\centering
	\includegraphics[width=5.2in]{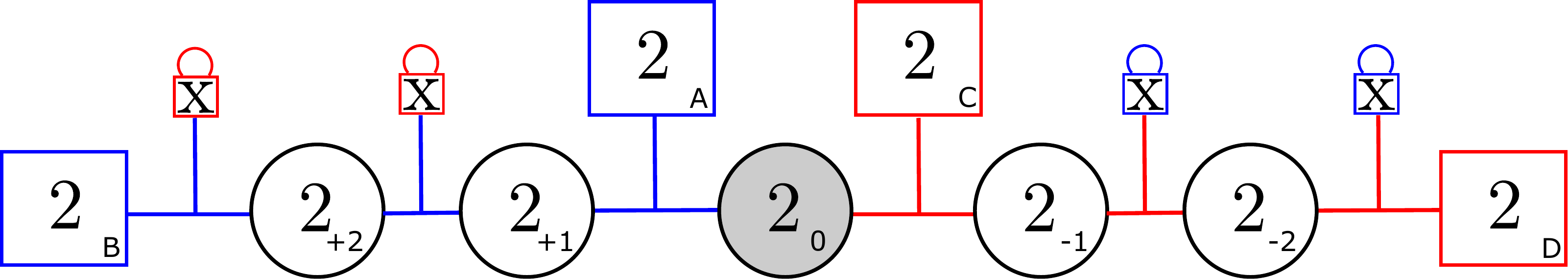}
	\caption{Quiver diagram for $\CU_2^{(2)}$. The edges connecting the nodes denote bifundamental chiral multiplets. A small box with an `x'-mark denotes a singlet chiral multiplet coupled to the bifundamental.}
	\end{subfigure}
	\begin{subfigure}[b]{6.0in}
	\centering
	\includegraphics[width=5.8in]{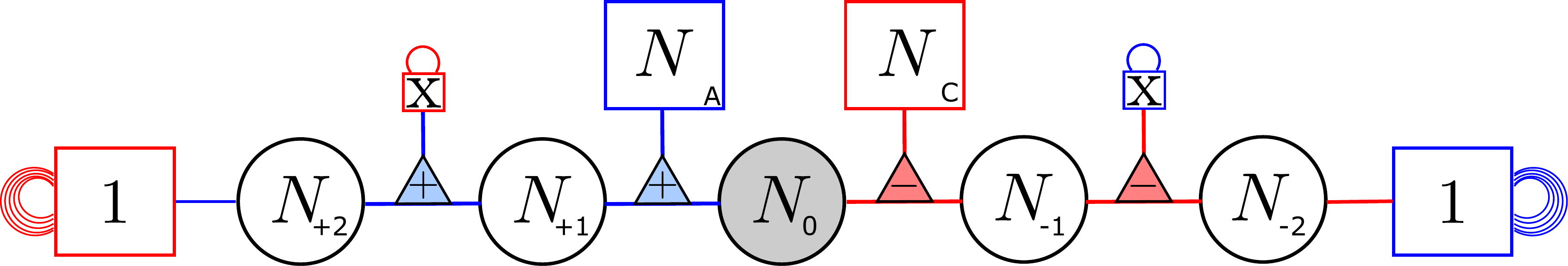}
	\caption{Quiver diagram for $\CU_N^{(2)}$. The triangle refers to the $T_N$ theory. Here a small box with `x'-mark refers to a certain deformation or Higgsing of the theory which breaks one of the $SU(N) \subset SU(N)^3$ global symmetries in $T_N$. There are gauge/flavor singlets as well.}
	\end{subfigure}
 	\caption{Dual descriptions $\CU_N^{(m)}$ of $SU(N)$ SQCD with $2N$ flavors. Here $m=2$, where $m$ refers to the number of white nodes on both sides of the black node in the middle. Black circular nodes denote $\CN=1$ vector multiplets, and white circular nodes denote $\CN=2$ vector multiplets. As usual, square nodes denote global symmetries.}
	\label{fig:dualities}
\end{figure}

In this paper, we argue for the existence of two infinite classes of 4d ${\cal N}=1$ theories, $T_N^{(m)}$ and $\CU_N^{(m)}$, labelled by an arbitrary integer $m\geq 0$. $T_N^{(m)}$ theories are superconformal theories that have several duality frame representations.  We argue that, for all $m$, $\CU_N^{(m)}$ RG flow to the same IR fixed point SCFT as SQCD with $N_f=2N_c \equiv 2N$ fundamentals and quartic superpotential 
\be
W=\lambda ^{i\tilde j; k\tilde \ell}M_{i\tilde j} M_{k\tilde \ell} \ ,\label{WMM}
\ee
where $M_{i\tilde j}=Q_i\tilde Q_{\tilde j}$, and $\lambda ^{i\tilde j; k\tilde \ell}$ are chosen to preserve a $SU(N_c)\times SU(N_c)\times U(1)\times U(1)_B\subset SU(2N_c)_D \times U(1)_B \subset SU(N_f)_L\times SU(N_f)_R\times U(1)_B$; this is a one-complex dimensional conformal manifold of SCFTs. 
The $\CU_N^{(m)}$ 
is a quiver gauge theory consisting of $2m+1$ gauge nodes and components constructed from $T_N$, along with a specific superpotential.  The $m=2$ case is illustrated in the   
 the generalized quiver diagram of figure \ref{fig:dualities}.

The $\CU _N^{(m)}$ can be obtained by gluing (via gauging) two copies of the $T_N^{(m)}$ theories (when $N>2$, we glue partially Higgsed $T_N^{(m)}$).
\begin{figure}[h]
	\centering
	\begin{subfigure}[b]{4.5in}
	\centering
	\includegraphics[height=1in]{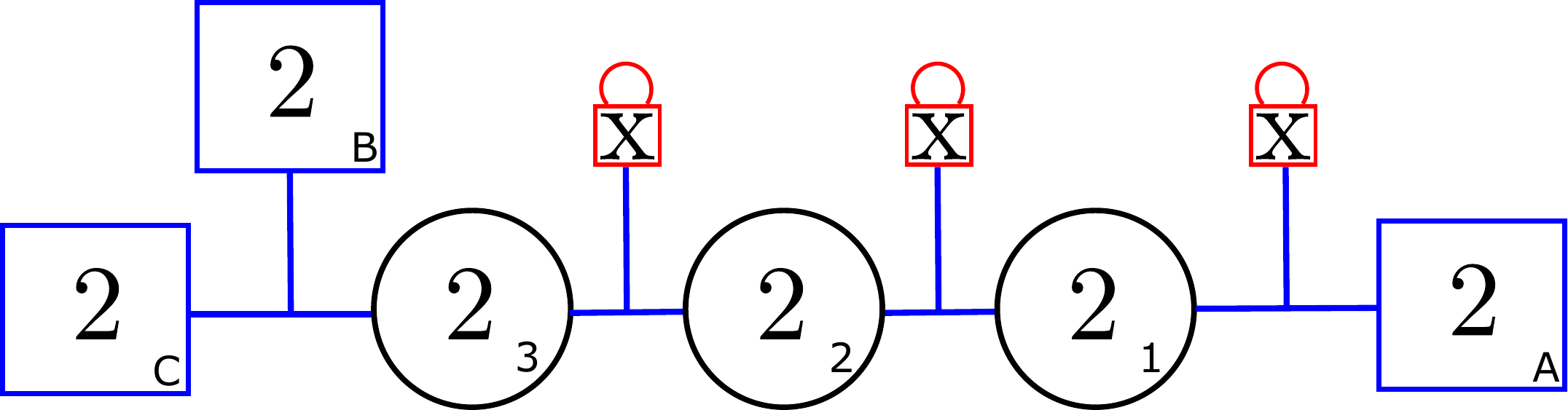}
	\caption{A quiver diagram describing the $T_2^{(3)}$ theory.}
	\end{subfigure}
	
	\begin{subfigure}[b]{4.5in}
	\centering
	\includegraphics[height=1in]{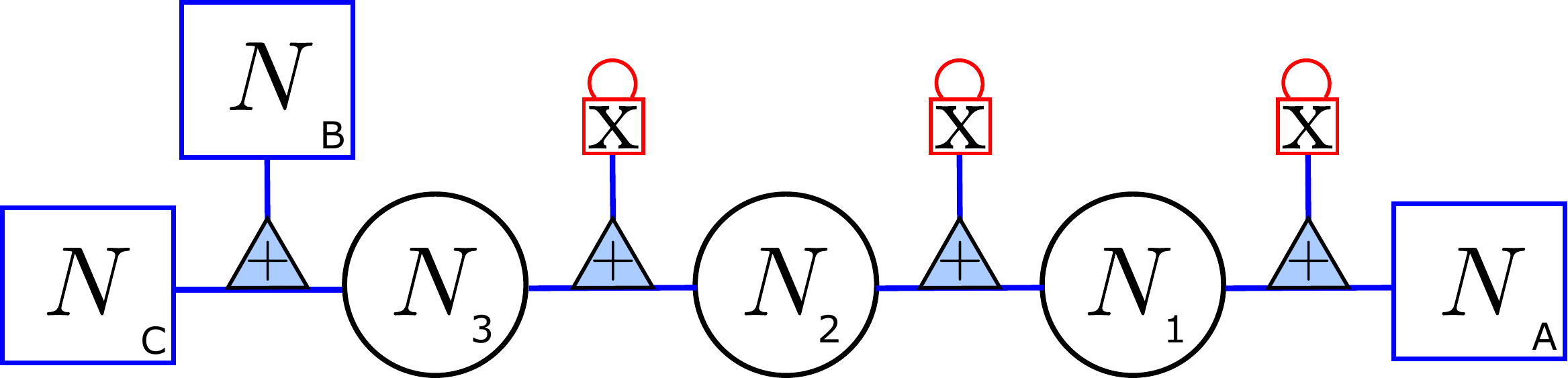}
	\caption{A quiver diagram describing the $T_N^{(3)}$ theory.}
	\end{subfigure}
	\caption{Some examples of the quiver diagram describing the $T_N^{(m)}$ theories. In general, there is a number of dual descriptions for the $T_N^{(m)}$ theory itself.}
	\label{fig:TNm}
\end{figure}
The $T_N^{(m)}$ theories are new $\CN=1$ SCFTs, which like the $\CN =2$ $T_N$ theories only have a Lagrangian description in the $N=2$ case. Nevertheless, for all $N$, results can be obtained via  holomorphy \cite{Seiberg:1994bz, Intriligator:1995au}, much as in  \cite{Gadde:2013fma,Maruyoshi:2013hja} for the $T_N$ case. Also, $a$-maximization \cite{Intriligator:2003jj} enables us to determine exact $R$-charges of the chiral operators and the central charges.  We thus compute the exact $R$ charges, the anomaly coefficients, and the  superconformal index \cite{Romelsberger:2005eg, Kinney:2005ej} of the $T^{(m)}_N$ and the $\CU _N^{(m)}$ theories. 

All of these theories have a natural description as being of class $\CS$: the low-energy limit of the six-dimensional $\CN=(2, 0)$ theory of type $\Gamma = A_{N-1}$, compactified on punctured 
Riemann surfaces $\CC_{g, n}$, generalizing the 4d $\CN=2$ theories of \cite{Gaiotto:2009hg,Gaiotto:2009we}.  For the 4d $\CN =1$ theories, in addition to $\CC_{g, n}$ (called the UV curve)  we need to assign a pair of integers $(p, q)$ 
\begin{equation}\label{pqsum}
\CC _{g, n}^{(p, q)}\equiv \CL (p)\oplus \CL (q)\to \CC _{g, n}, \quad\hbox{with}\quad p+q = -\chi(\CC_{g, n})=2g-2+n,
\end{equation}
where $p\equiv c_1(\CL (p))$ and $q\equiv c_1(\CL (q))$ and the condition is to preserve $\CN =1$ supersymmetry \cite{Benini:2009mz,Bah:2011vv, Bah:2012dg, Xie:2013gma} (as discussed in these references, there are more general possibilities).  From the 6d perspective, various dualities can be understood as arising from different choices of the (generalized) pair-of-pants decompositions of the same Riemann surface \cite{Maruyoshi:2009uk, Benini:2009mz,Bah:2011je,Bah:2012dg, Beem:2012yn, Xie:2013gma, Bah:2013aha, Agarwal:2013uga,Agarwal:2014rua, McGrane:2014pma}. For $\CN =1$ theories, 
 when we decompose $\CC _{g,n}$ into pants, the $(p,q)$ integers are also decomposed into sums over the pants, with each pair of pants also satisfying \eqref{pqsum}, with $g=0$ and $n=3$. 
 
Previous works on class $\CS$ field theories restricted to  $(p, q) \ge 0$, whereas here we consider cases with negative degree.  In particular,  our $T_N^{(m)}$ theory arises from reducing the 6d $A_{N-1}$ $\CN =(2,0)$ theory on the three-punctured sphere $\CC_{0,3}$, with the line bundle degrees 
\begin{equation}\label{tmn}
T_N^{(m)}: \qquad \CL (p)\oplus \CL (q) \to \CC _{g=0, n=3}, \qquad \hbox{with}\qquad (p, q)=(m+1, -m)
\end{equation} 
Some perspectives or expressions that are compatible with negative degree include  gravity duals \cite{Bah:2011vv, Bah:2012dg, Bah:2013qya,Bah:2013wda, Bah:2015fwa}, superconformal indices \cite{Beem:2012yn} and generalilzed Hitchin system associated to the UV curve \cite{Xie:2013gma, Bonelli:2013pva, Xie:2013rsa, Yonekura:2013mya, Xie:2014yya}.   A possible objection to combining positive and negative degree pairs of bundles as in \eqref{tmn} is that they are unstable\footnote{We thank Edward Witten for this remark.} to transitions $m\to m-1$, eventually reducing down to $m=0$.   We find that the $T^{(m)}_N$ theories are stable, but the $\CU _N^{(m)}$ exhibit $m\to m-1$ cascade processes, via renormalization group flows in the associated 4d QFTs.

The 6d $A_{N-1}$, $\CN=(2,0)$ theory on a 4-punctured sphere (with punctures being appropriately decorated) gives 
\begin{equation}\label{sqcd}
\hbox{$SU(N)$ SQCD with $N_f=2N_c$ \ via} \qquad \CL(1)\oplus \CL(1)\to \CC _{g=0, n=4}
\end{equation}
with the $SU(N)^2\times U(1)\times U(1)$-preserving superpotential \eqref{WMM}.   Upon decomposing $\CC _{g=0, n=4}^{p=1, q=1}$ into two pairs-of-pants, one can assign degrees as in \eqref{tmn}, $(m+1, -m)$ to one and $(-m, m+1)$ to the other. This suggests $N_f=2N_c$ SQCD is dual to theories labeled by general $m$, with a RG flow down to $m=0$, leading to an infinite set of duals.  We will flesh out this relation, and provide a number of checks.   Among the checks is a matching of the superconformal index \cite{Dolan:2008qi}, which can be seen easily via the generalized TQFT structure studied in \cite{Beem:2012yn} and in \cite{Gadde:2013fma, Agarwal:2014rua}.

The outline of this paper is as follows. In section \ref{sec:classS}, we review the 4d $\CN=1$ SCFT in class $\CS$ and show how to obtain the theories corresponding to general $(p, q)$ through the nilpotent Higgsing. In section \ref{sec:su2}, we will discuss the construction of $T_2^{(m)}$ theory in detail. For the case of $\Gamma = A_1$, we always get a Lagrangian theory with $SU(2)$ gauge groups. From these building blocks, we show how to obtain the dual theories of $SU(2)$ SQCD.
In section \ref{sec:suN}, we generalize the construction to $T_N^{(m)}$ which involves multiple copies of $T_N$ theory. Using these building blocks, we construct dual theories of $SU(N)$ SQCD.
In section \ref{sec:index}, we compute the superconformal indices of the $T_N^{(m)}$ theory as further checks of our proposed dualities. 

\section{Four-dimensional $\CN=1$ SCFTs and dualities from M5-branes} \label{sec:classS}
In this section, we briefly review the $\CN=1$ class $\CS$ theories, and our particular constructions. 

\subsection{Review of class $\CS$ theories}
For more detail, we refer to the papers \cite{Bah:2011vv,Bah:2012dg,Gadde:2013fma,Xie:2013gma, Agarwal:2013uga,Agarwal:2014rua}. 

\paragraph{Data}
The $\CN=1$ class $\CS$ theories we consider are labelled by:
\begin{enumerate}
\item The choice of a `gauge group' $\Gamma \in ADE$ of the 6d, $\CN =(2,0)$ theory.  
\item The choice of a Riemann surface $\CC_{g, n}$ (UV curve) of genus $g$ and $n$ punctures.
\item The choice of the degree of line bundles $(p, q)$ over $\CC_{g, n}$ satisfying \eqref{pqsum}.
\item We decorate each of the punctures $i=1, \cdots n$ with an $SU(2)$ embedding $\rho_i$ into $\Gamma$ and a $\IZ_2$-valued color $\s_i$.
\end{enumerate}
We will here focus on $\Gamma = A_{N-1}$, though much of the discussion is valid for general $\Gamma$. 
The total space $\CC^{(p, q)}_{g, n}\equiv \CL(p) \oplus \CL(q) \to \CC_{g, n}$ in \eqref{pqsum} is a local Calabi-Yau 3-fold, so M5-branes wrapped on the base $\CC_{g, n}$ preserves 4 supercharges in the 11-dimensional M-theory.  The fourth data labels the punctures that specify the global symmetry of the theory. Here we restrict to the class of punctures that we call the `colored $\CN=2$ punctures', since locally they are of the same type that appear in $\CN=2$ class $\CS$ theories \cite{Gaiotto:2009hg,Gaiotto:2009we}. For $\Gamma = A_{N-1}$, the choice of $\rho_i$ is in one-to-one correspondence with the choice of a partition of $N$, or equivalently a Young diagram of $N$ boxes. The commutant of the $SU(2)$ embedding $\rho_i$ gives the flavor symmetry associated with the $i$-th puncture. 

Such $\CN=1$ class $\CS$ theories admit a $U(1)_+ \times U(1)_- $ global symmetry \cite{Bah:2011vv}, with generators $(J_+,J_-)$, from those Cartans of the $SO(5)$ $R$-symmetry of the $\CN=(2,0)$ theory that can be preserved after a partial topological twist on the UV curve.  Defining 
\begin{equation} \label{eq:RandJ}
    R_0 \equiv  \frac{1}{2} \left(J_+ + J_- \right), \qquad \CF \equiv \frac{1}{2} \left(J_+ - J_- \right) \ 
    \end{equation}
$R_0$ is a $U(1)_R$ symmetry and $\CF$ is a non-R global $U(1)$ symmetry.  The exact superconformal R-symmetry is a linear combination
 \begin{equation}
    R_{\CN=1} = R_0 + \epsilon \CF \ = \frac{1+\e}{2} J_+ + \frac{1-\e}{2} J_-,\label{Re}
    \end{equation}
where $\epsilon$ is fixed by $a$-maximization \cite{Intriligator:2003jj}. For the case $p=q$, this gives $\epsilon =0$.

\paragraph{Pair-of-pants decomposition and duality}

\begin{figure}[h]
\centering
\includegraphics[width=4.5in]{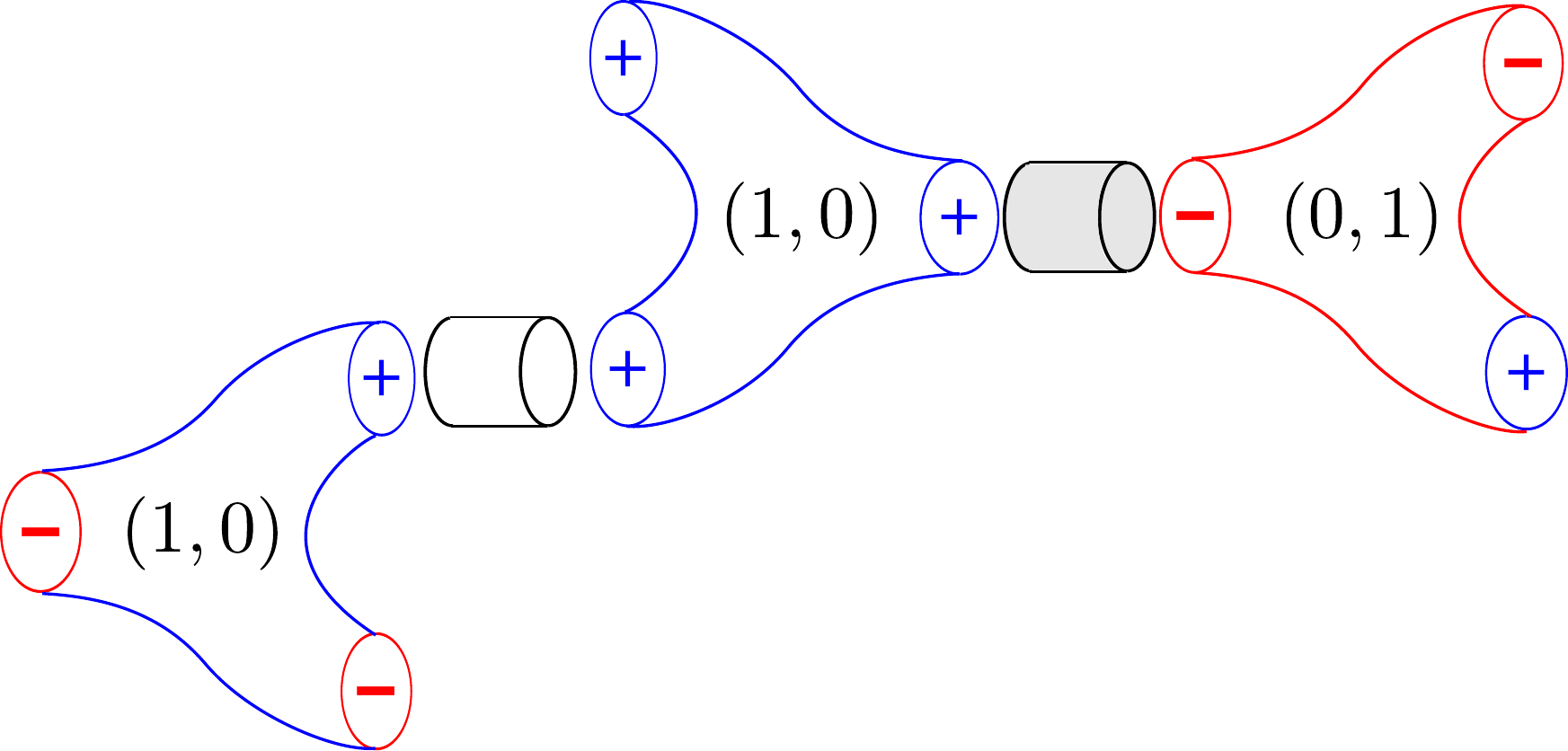}
\caption{An example of colored pair-of-pants decomposition. Here red/blue means $\s=\pm$ respectively. Three red punctures and two blue punctures with $p=2, q=1$. Grey tube denotes $\CN=1$ vector, white tube denotes $\CN=2$ vector multiplet. There are 3 punctures of opposite color. There is an adjoint chiral multiplet attached to each of them.}
\label{fig:pDecompEx}
\end{figure}

\begin{figure}[h]
	\centering
	\includegraphics[width=4.5in]{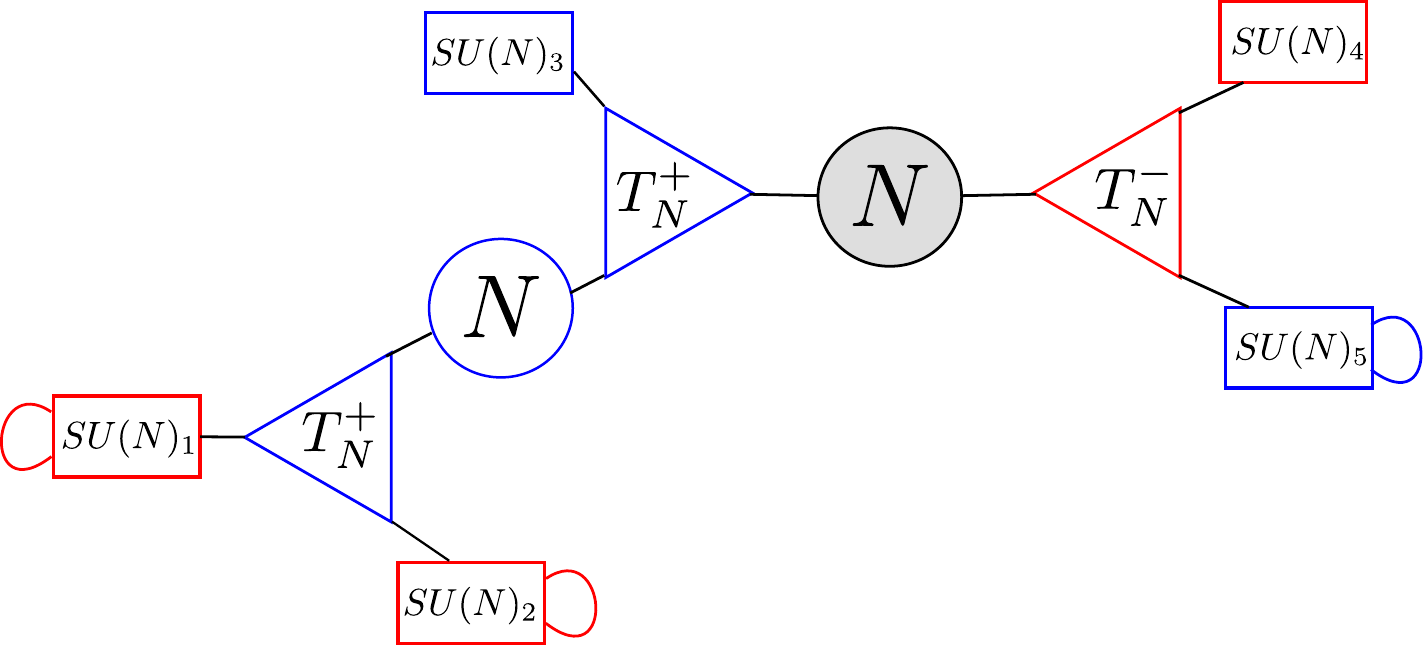}
	\caption{The UV description corresponding to the colored pair-of-pants description of figure \ref{fig:pDecompEx}. Here we assumed all punctures to be maximal. }
	\label{fig:PoPquiver}
\end{figure}

 The pair-of-pants decomposition of (hyperbolic) $\CC_{g,n}$ yields a way to build the theory, and find duals.  One decomposes 
 the total space $\CC^{(p, q)}_{g, n}$, including the normal bundle degrees, with $p+q=1$ for each pant ($g=0$, $n=3$).  If one restricts to $(p, q)$ both non-negative, the two options for each pant are $(1, 0)$ or $(0, 1)$, which are denoted by a coloring  $\sigma=\pm$, with $\CC^{(p, q)}_{g, n}$ then decomposed into $p$ pants of color $\sigma=+$ and $q$ pants with $\sigma=-$. Two pants of same color are glued with an $\CN=2$ vector multiplet, while pants of opposite colors are glued with  an $\CN=1$ vector multiplet. 
  See figure \ref{fig:pDecompEx} for an illustration of the construction. Figure \ref{fig:PoPquiver} gives the theory corresponding to the pair-of-pants decomposition in figure \ref{fig:pDecompEx}.
  Different pair-of-pants decompositions of $\CC_{g,n}$ give IR dual theories.

Each puncture has a $SU(N)$ symmetry, which is unbroken if the puncture is maximal.  In addition to the $\CN =1$ $SU(N)$ current multiplet, there is a $SU(N)$ adjoint-valued chiral superfield multiplet, $\mu$ (often called the ``moment-map'' operator).  The $\CN =1$ current multiplet and $\mu$ combine to form the $\CN =2$ $SU(N)$ current multiplet when $\CN =2$ supersymmetry is preserved.  When the two pants of the same color are glued, the diagonal combination of these $\CN =2$ $SU(N)$ currents is gauged.  When there is an oppositely colored puncture on the pants, we also have extra chiral multiplet $M$ in the adjoint of $SU(N)$, with a superpotential coupling $W= \Tr M \mu$, so $M$ effectively replaces the role of $\mu$ via a Legendre transform. 

Non-maximal punctures are labelled by an $SU(2)$ embedding $\rho$.  We then partially close, or Higgs, the puncture by giving a nilpotent vev $\rho(\s^+)$ to $\mu$ if the color of puncture is the same as the pants, and to $M$ if the puncture has the opposite color. This breaks the global symmetry associated to the puncture from $SU(N)$ to the commutant of the $\rho(SU(2))$ inside $SU(N)$. The building blocks corresponding to a sphere with generic three punctures can be identified from the previous works \cite{Chacaltana:2010ks,Chacaltana:2012zy} for the case of same colored puncture, and \cite{Gadde:2013fma, Agarwal:2014rua} for the oppositely colored puncture.

\subsection{General $(p, q)$ class $\CS$ theories from nilpotent Higgsing}
We aim to find $\CN=1$ class $\CS$ theories corresponding to $\CC_{g, n}^{(p, q)}$ satisfying \eqref{pqsum}, here allowing for negative $p$ or $q$. 
The idea is to start with a theory with positive degrees, $(p', q') \ge 0$, and obtain negative degrees via nilpotent Higgsing of the puncture.  Following the prescription in  \cite{Gadde:2013fma, Agarwal:2014rua}, for the case  $\Gamma = A_{n-1}$, we can identify the  Higgsed theory.
For example, to get the three punctured sphere with degree $(m+1, -m)$, we start with a sphere with $m+3$ punctures, and   line bundles of degree $(m+1, 0)$, with $3$ $+$ punctures and $m$ $-$ punctures. If we Higgs all $m$ of the $-$ punctures, we are left with three $+$ punctures with degrees $(m+1, -m)$.
\begin{figure}[h]
	\centering
	\includegraphics[width=3.3in]{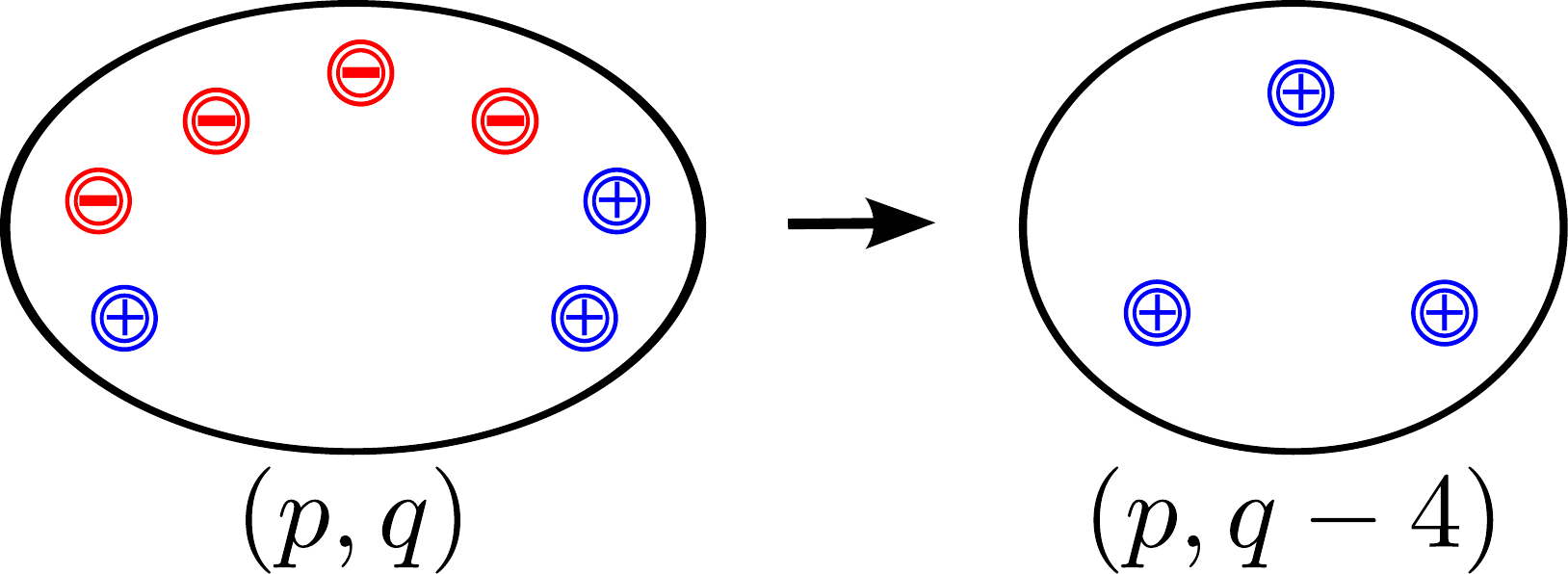}
	\caption{Higgsing the punctures to get the UV curve with lower degrees.}
\end{figure}

This procedure allows us to identify the theory corresponding to non-positive $(p, q)$. In the following, we mainly focus on the three (+ colored) maximal punctured sphere with normal bundle degrees $(m+1, -m)$, which yields the $\CN=1$ theories that we denote by  $T_{N}^{(m)}$, The $m=0$ case reduces to the  $T_N$ theory of \cite{Gaiotto:2009we}.  As we discuss, the $T_{N}^{(m)}$, theory can be constructed from gluing $m+1$ copies of the $T_N$ theory with a number of singlet chiral multiplets and then Higgsing/closing the punctures. The closure of the puncture is implemented via giving a nilpotent vev to associated chiral adjoints $M$.   This can thought of as a nilpotent mass deformation when $\Gamma = A_1$, i.e. for $N=2$. We will discuss this in detail in later sections. 

\section{$SU(2)$ theories} \label{sec:su2}
Let us start with the $SU(2)$ case, coming from the 6d $\Gamma = A_{1}$ theory, and recall that the $T_2$ theory of  \cite{Gaiotto:2009we} reduces to 8 free chiral multiplets.   Likewise, there is a Lagrangian description for every $(p, q)$.  We first consider the $T_2^{(m)}$ theories, and then obtaining duals of $\CN =1$ $SU(2)$ SQCD with $N_f=4$ flavors by gluing two copies of $T_2^{(m)}$.

\subsection{The simplest example: $T_2^{(m=1)}$} \label{subsec:T2m1}
\begin{figure}[t]
	\centering
	\begin{subfigure}[b]{2.9in}
	\centering
	\includegraphics[width=2.3in]{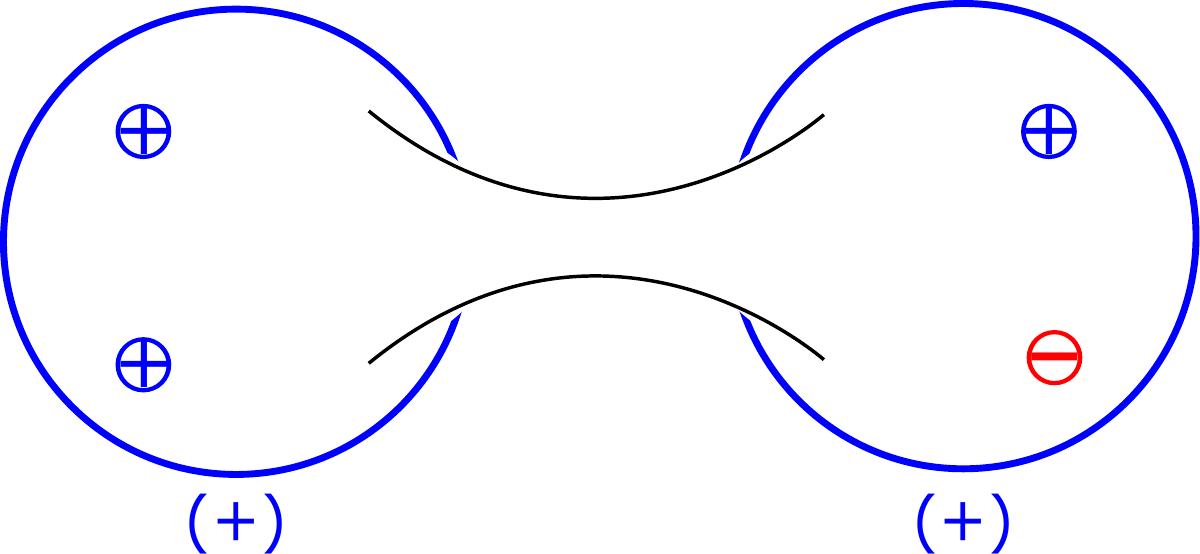}
	\caption{A colored pair-of-pants decomposition for the 4-punctured sphere.}
	\end{subfigure}
	\quad
	\begin{subfigure}[b]{2.9in}
	\centering
	\includegraphics[width=2.7in]{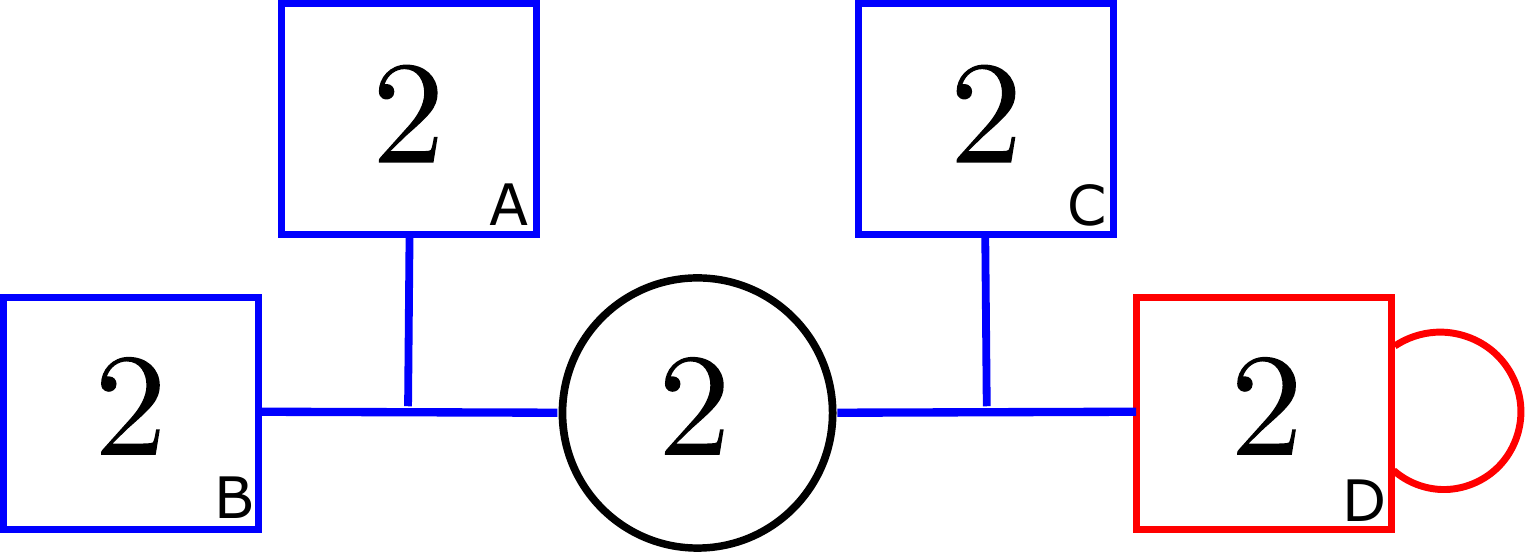}
	\caption{The quiver diagram corresponding to the UV curve and the colored pair-of-pant decomposition on the left.}
	\label{fig:T22m1UnhiggsQuiver}
	\end{subfigure}
	\caption{A colored pair-of-pants decomposition of $\CC_{0, 4}^{(2, 0)}$, with $(n_+, n_-)=(3, 1)$ and its corresponding quiver diagram, see also \cite{Gaiotto:2009we}.  Each node denotes $SU(2)$ global/gauge symmetries.} 	\label{fig:T22m1UnhiggsCurve}
\end{figure}

To obtain the 3-punctured sphere with normal bundle degrees $(m+1, -m)=(2,-1)$, we start with the UV curve $\CC_{0, 4}^{(2, 0)}$ with $(n_+, n_-)=(3, 1)$ where $n_\pm$ denotes the number of $\pm$  punctures. Upon closing the $-$ puncture, we will obtain the UV curve $\CC_{0, 3}^{(2, -1)}$ with all $+$ punctures. Before closing the puncture, the Lagrangian description of the 4d $\CN =1$ theory is given as in figure \ref{fig:T22m1UnhiggsCurve}.
The field content of the theory is given as in the table below:
\be
    	\begin{tabular}{|c|c|c|c|c|c|c|c|c|}
    	\hline
        & $SU(2)_{g}$ & $SU(2)_{A}$ & $SU(2)_{B}$ & $SU(2)_{C} $ & $SU(2)_{D}$ & $R_0$ & $\CF$ & $(J_+, J_-)$ \\
        \hline \hline
        $\phi$ & \textrm{adj} & & & & & $1$ & $-1$ & $(0, 2)$ \\
	$q_{1}$ & $\square$ & $\square$ & $\square$ & &  & $\half$ & $\half$ & $(1, 0)$  \\
    	$q_{2}$ & $\square$ & & & $\square$ & $\square$ & $\half$ & $\half$ & $(1, 0)$ \\
    	$M'$ & & & & & \textrm{adj} & 1 & $-1$ & $(0, 2)$ \\ \hline
	\end{tabular}
	\label{table:A1Unhiggsed2}
\ee
Here $J_\pm$ are combinations of $R_0, \CF$ defined so that $R_0 = \half (J_+ + J_-)$ and $\CF = \half(J_+ - J_-)$. They are the `candidate $R$-charges' which were used in \cite{Agarwal:2014rua}. The exact $R$-charge is given by a linear combination of the two, which is determined by $a$-maximization \cite{Intriligator:2003jj}. In terms of the quiver diagram \ref{fig:T22m1UnhiggsQuiver}, $SU(2)_{A, B}$ refers to the blue flavor nodes on the left, and $SU(2)_C$ refers to the blue flavor node on the right, and $SU(2)_D$ corresponds to the red flavor node on the right. 
The theory has a superpotential $W = \Tr \phi (q_1 q_1 + q_2 q_2) + \Tr M' q_2 q_2$.

\begin{figure}[h]
	\centering
	\includegraphics[width=2.3in]{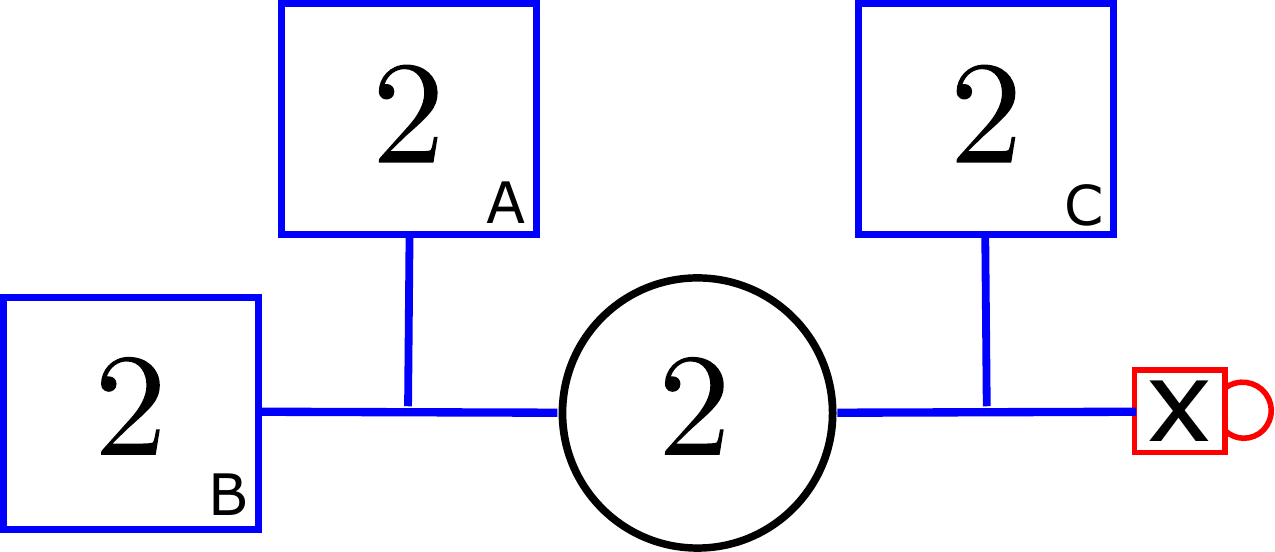}
	\caption{The quiver diagram for the $T^{(1)}_{2}$ theory. The `x'-marked box denotes a closed puncture. It also means there is a singlet coupled to the quarks connected.}
	\label{fig:T22m1quiver}
\end{figure}

We now close the red puncture corresponding to $SU(2)_D$ by giving a nilpotent vev, $M' \sim \s^+$. This triggers a relevant RG flow, giving a mass to some components of the $q_2$ matter multiplet.  Upon integrating them out, we obtain an IR SCFT described by the quiver diagram of figure \ref{fig:T22m1quiver}. It can also be understood as the Fan corresponding to the partition $2 \to 2$ \cite{Agarwal:2014rua}.
The matter content is given as in the table below:\footnote{It was shown in \cite{Gadde:2013fma} that upon Higgsing a puncture labelled by $\rho : SU(2) \to \Gamma$ in the above manner, the $(J_+, J_-)$ charges shift to
$(J_+ ,  J_- - \rho(\sigma^3) )$, where $\rho$ in this case is given by the identity map. This explains the charge assignments of \ref{table:T22m1}.}
\be
    	\begin{tabular}{|c|c|c|c|c|c|c|c|c|}
    	\hline
        & $SU(2)_{g}$ & $SU(2)_{A}$ & $SU(2)_{B} $ & $SU(2)_{C}$ & $R_0$ & $\CF$ & $(J_+, J_-)$ \\
        \hline \hline
        $\phi$ & \textrm{adj} & & & & $1$ & $-1$ & $(0, 2)$ \\
	$q_{1}$ & $\square$ & $\square$ & $\square$ & & $\half$ & $\half$ & (1, 0)  \\
    	$q_{2}$ & $\square$ & & & $\square$ & $0$ & $1$ & (1, -1) \\
    	$M$ & & & &  & 2 & $-2$ & (0, 4) \\ \hline
	\end{tabular}
	\label{table:T22m1}
\ee
The remaining theory has  superpotential
\be
 W = \Tr \phi q_1 q_1 + M \Tr q_2 q_2 \ ,\label{WTtwoone}
\ee
which is generic for the global symmetry with $(J_+, J_-) = (2, 2)$ charges.\footnote{There are no terms of the form $\phi^2 q_2 q_2$, because $(\phi^2)_{\a \b} (q_2)^{\a i} (q_2)^{\b j}$ is identically zero and $\tr (\phi^2) \Tr (q_2 q_2)$ is not in the chiral ring due to the $F$-term for $M$.}  

The charged matter is that of $\CN =2$ $SU(2)$ with $N_f=3$, but the theory is $\CN =1$ supersymmetric because one of the flavors does not couple to the adjoint, instead coupling to the gauge singlet $M$.  This theory has a quantum moduli space of vacua, with several branches.  The $M$ field can have arbitrary expectation value, and $\vev{M}$ gives a mass to the $q_2$ field.  The low-energy theory for $\vev{M}\neq 0$ thus has an accidental $\CN =2$ supersymmetry, given by $\CN =2$ with $N_f=2$ flavors, with global symmetry $SU(2)_A\times SU(2)_B\times SU(2)_R\times U(1)_{\cal R}$.   That theory has \cite{Seiberg:1994aj}
a Coulomb branch, with modulus $u=\Tr \phi ^2$, and two Higgs branches, emanating from the massless monopole and dyon points on the Coulomb branch, at $u\sim \pm \Lambda _L^2\sim \pm M\Lambda$.  Each Higgs branch is a copy of ${\bf C}^2/{\bf Z}_2$, and either $SU(2)_A$ or $SU(2)_B$ is spontaneously broken, depending on which branch.  For $M\to 0$, the two Higgs branches meet at the origin of the Coulomb branch, with additional moduli from $q_2$, subject to the F-term $\Tr q_2q_2=0$. It would be interesting to interpret this moduli space via  geometric construction.

The IR theory at the origin of the moduli space is an $\CN =1$ interacting SCFT.  It has a manifest $SU(2)^3$ flavor symmetry, with three $(J_+, J_-) = (2, 0)$ moment map chiral operators, in the adjoint representations of $SU(2)_{A, B, C}$, given by
\be \label{eq:Momentmap}
 (\mu_A)_i^{~j} = (q_1)_{\a i k} (q_1)^{\a j k} , ~~~ (\mu_B)_i^{~j} = (q_1)_{\a k i} (q_1)^{\a k j} ,  ~~~
 (\mu_C)_i^{~j} = (q_2)_{\a i} \phi^{\a}_{~\b}  (q_2)^{\b j} .
\ee
The operator $\mu_C$ is dressed with the adjoint chiral multiplet $\phi$ to have the correct charges, $(J_+, J_-)=(2, 0)$.  Despite the apparent difference between $\mu _{A,B}$ vs $\mu _C$, the IR SCFT is expected to be $S_3$ permutation symmetric under permutation of the $SU(2)_{A, B, C}$ symmetries. 
Because the theory is $\CN =1$ supersymmetric and not $\CN=2$, these chiral operators are not in the $SU(2)_{A, B, C}$ current multiplets, and they receive anomalous dimension.  The exact superconformal R-charge is as in \eqref{Re}, $R = R_0 + \e \CF$, and then chiral scalar operator dimensions are given by $\Delta ({\cal O})=\frac{3}{2}R({\cal O})$, e.g. $\Delta (\mu _{A, B, C})=\frac{3}{2} (1+\epsilon)$, $\Delta (\Tr \phi ^2)=3(1-\epsilon)$, $\Delta (M)=3 (1-\epsilon)$, with $\epsilon$   determined via a-maximization to be\footnote{ It is outside of the bound $|\epsilon |\leq \frac{1}{3}$ found in \cite{Bah:2013aha}, but here the operator dimensions are above the unitarity bound.}  $\e \simeq 0.52$.  We find that the superconformal index computed from this gauge theory description agrees with the TQFT prediction of \cite{Beem:2012yn}. The index is compatible with the $S_3$ permutation symmetry.  

\subsection{$T_2^{(m=2)}$} \label{subsec:T23m2}
\begin{figure}[h]
	\centering
	\begin{subfigure}[b]{2.9in}
	\centering
	\includegraphics[height=0.8in]{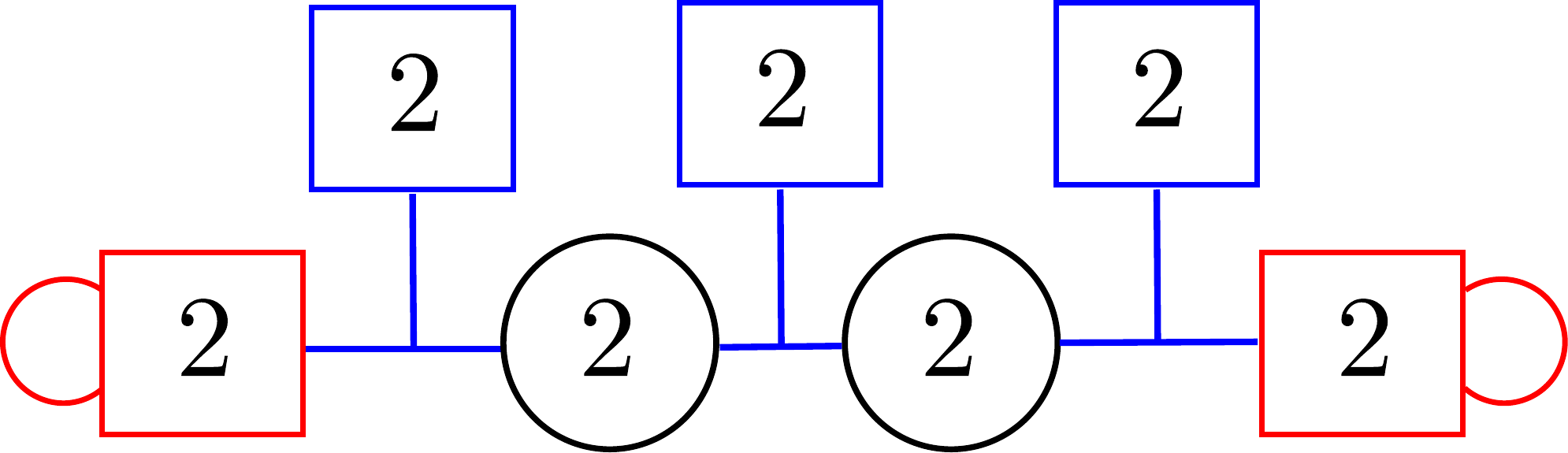}
	\caption{Quiver 1}
	\end{subfigure}
	\begin{subfigure}[b]{2.9in}
	\centering
	\includegraphics[height=1.0in]{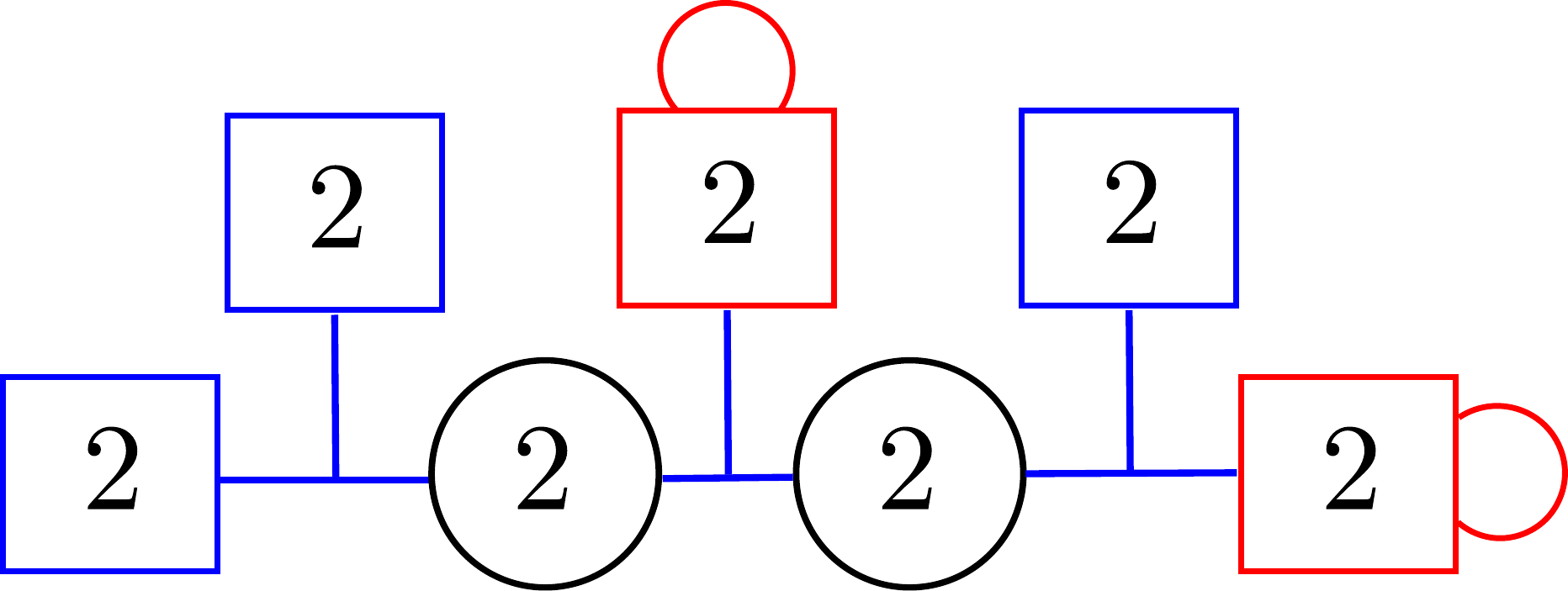}
	\caption{Quiver 2}
	\end{subfigure}
	\begin{subfigure}[b]{2.8in}
	\centering
	\includegraphics[height=1.0in]{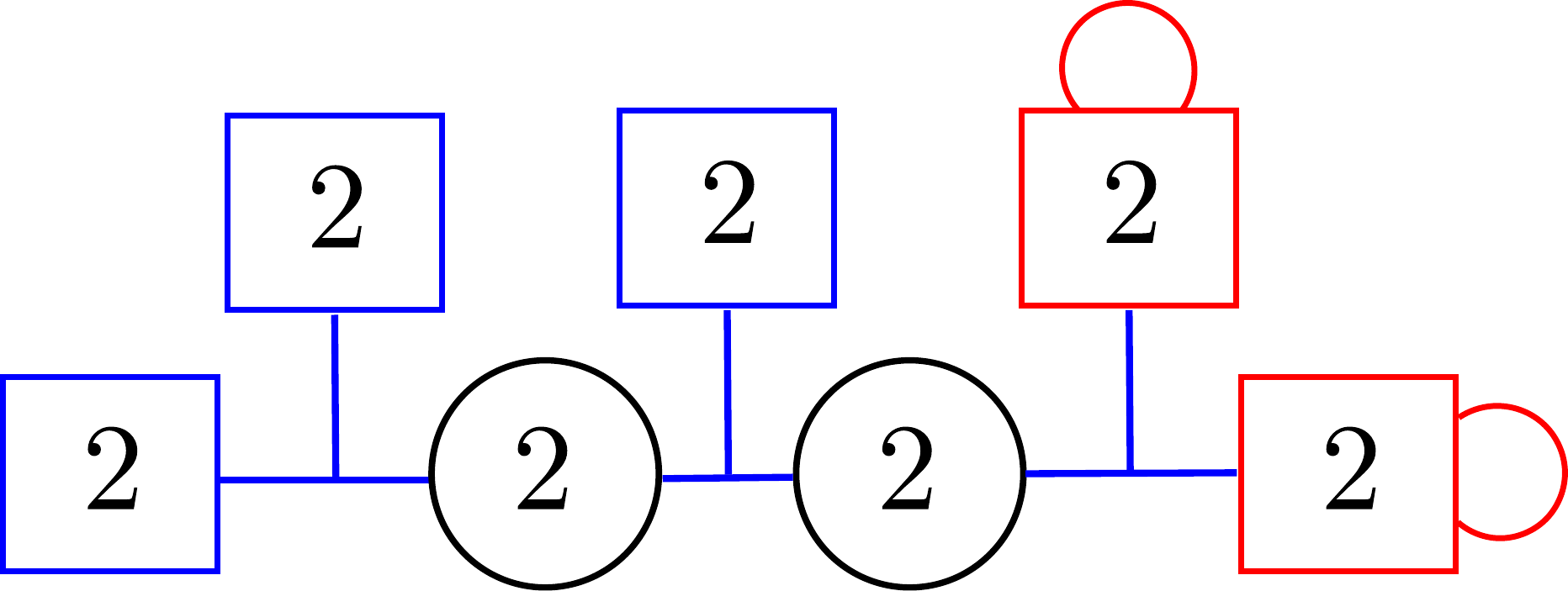}
	\caption{Quiver 3}
	\end{subfigure}
	\caption{Three dual frames corresponding to the UV curve $\CC^{(3, 0)}_{0, 5}$ and $(n_+, n_-)=(3, 2)$ where $n_\pm$ denotes the number of $\pm$ punctures respectively.}
	\label{fig:T23m2UnhiggsQuiver}
\end{figure}

We start from the theory corresponding $\CC_{0, 5}^{(3, 0)}$ with $(n_+, n_-) = (3, 2)$ (unhiggsed theory) and then close the two $-$ punctures to obtain $\CC_{0, 3}^{(3, -2)}$. There are three different ways to do this, starting from the three dual frames of the unhiggsed theory as in the figure \ref{fig:T23m2UnhiggsQuiver}.
The unHiggsed theory has $SU(2) \times SU(2)$ gauge group with bifundamental hypermultiplets and two more fundamentals attached to each of the gauge groups.  The blue parts of the quiver are $\CN =2$ supersymmetric, with chiral adjoints $\phi$ for each gauge group and $\CN =2$ matter couplings.  The red nodes are $\CN =1$ supersymetric, given by two chiral multiplets transforming as adjoints of the flavor groups,  coupled via a superpotential of the form
\be
 W_m = \sum_{a \in \textrm{red nodes}} \Tr  M_a \mu_a \ ,
\ee
where $\mu_a$ is the gauge invariant bilinear of chiral multiplets, in the adjoint of the $SU(2)_a$ global symmetry.
We then close the $-$ punctures by giving nilpotent vevs to the two chiral multiplets $M_a$ attached to the $-$ punctures. This triggers a relevant deformation of the theory which leads to a new SCFT in the IR. Since the three different quivers are dual to each other before Higgsing, they all flow to the same SCFT in the IR.

The nilpotent $M_a$ vev in quivers 1 and 2 gives rise to mass terms for some of the quarks, which we integrate out. 
\begin{figure}[h]
	\centering
	\begin{subfigure}[b]{2.9in}
	\centering
	\includegraphics[width=2.4in]{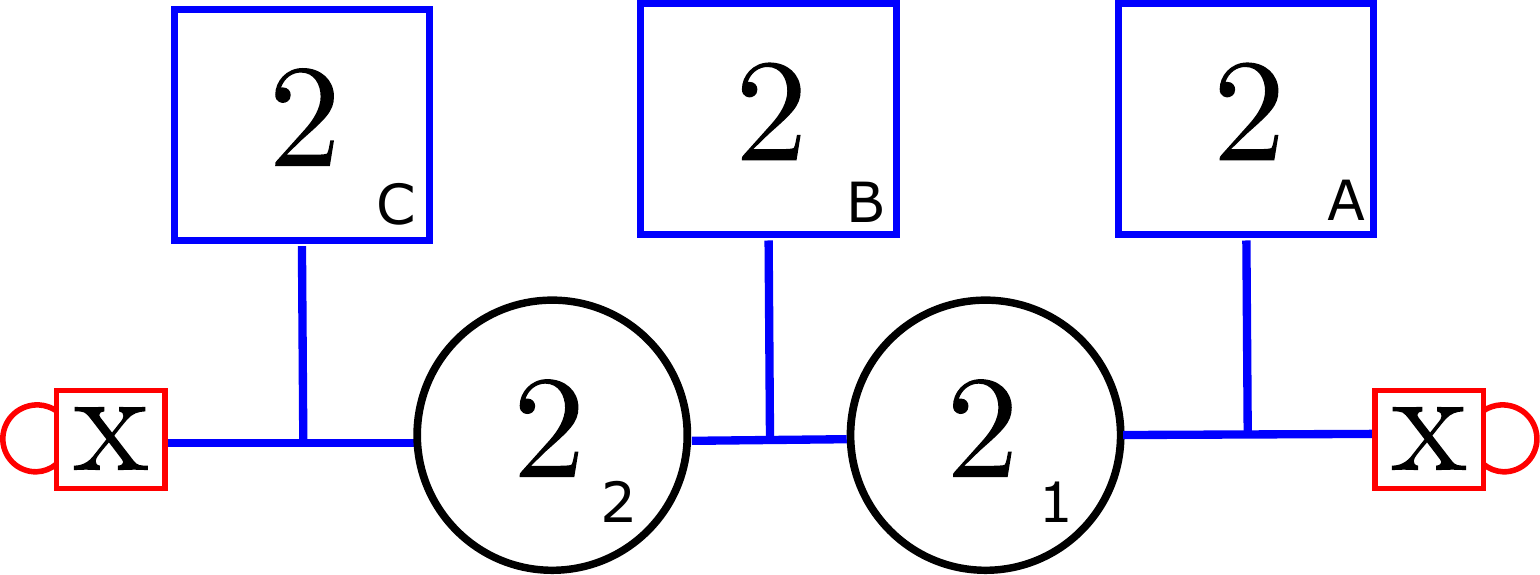}
	\caption{Quiver 1}
	\end{subfigure}
	\begin{subfigure}[b]{2.9in}
	\centering
	\includegraphics[width=2.6in]{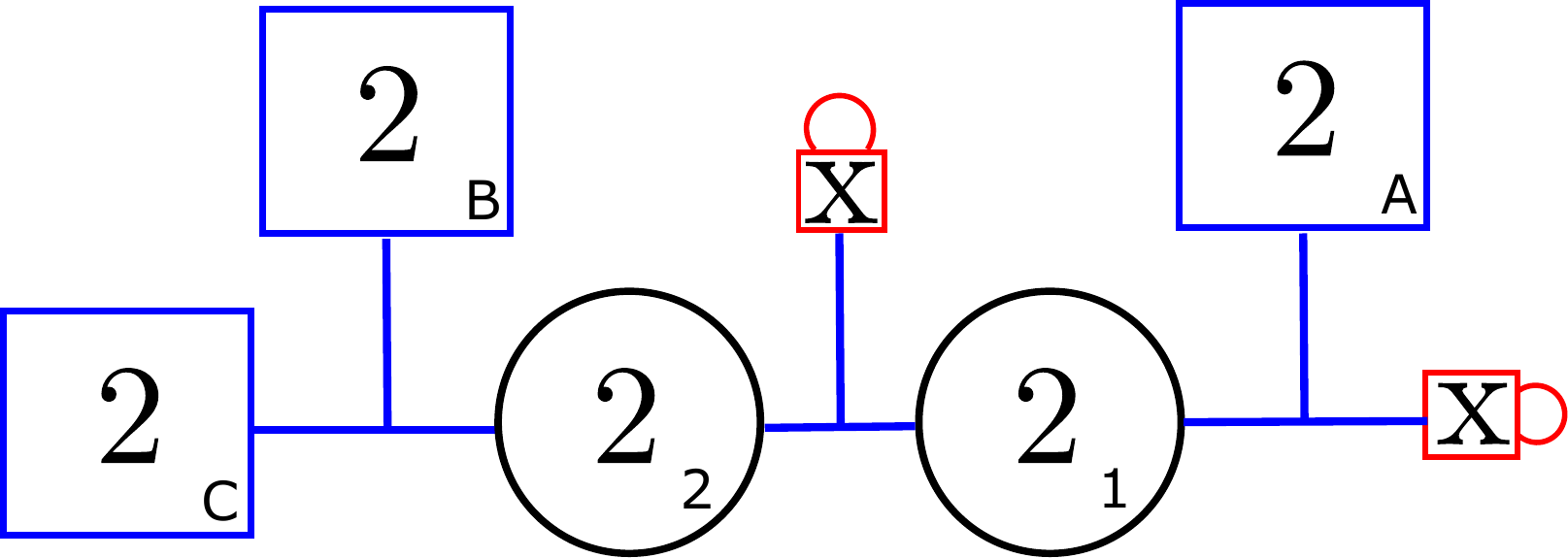}
	\caption{Quiver 2}
	\end{subfigure}
	\begin{subfigure}[b]{2.9in}
	\centering
	\includegraphics[width=2.6in]{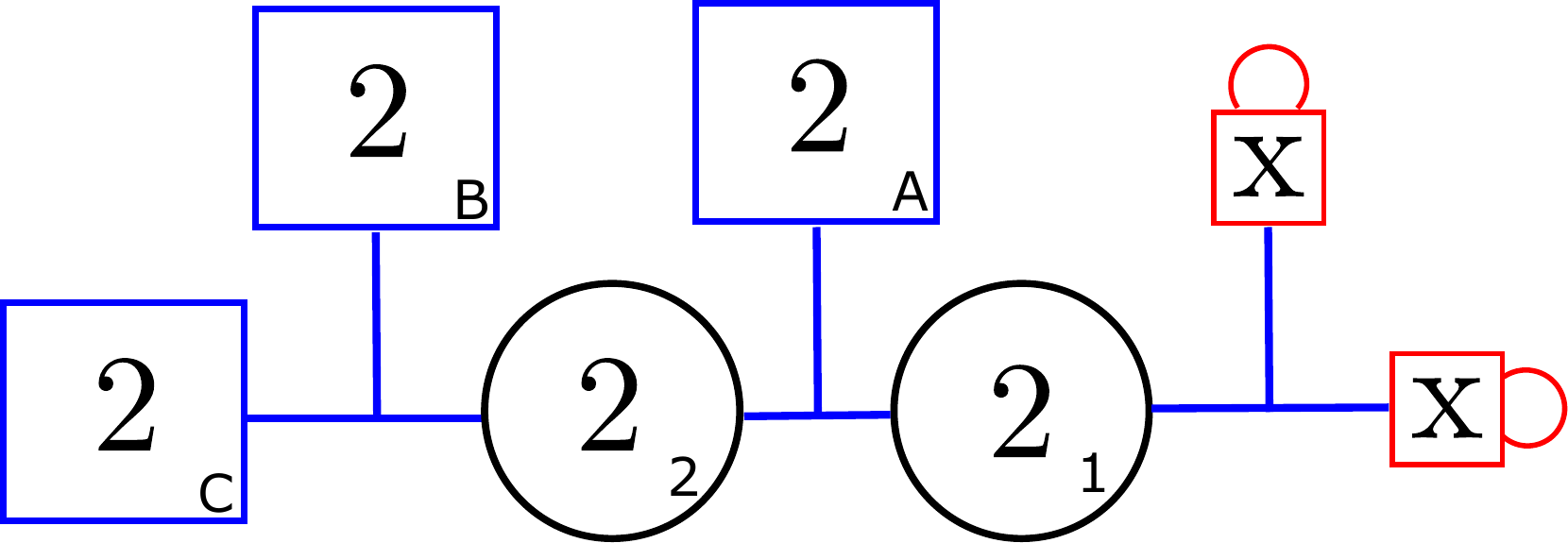}
	\caption{Quiver 3}
	\label{fig:T23m2quiver3}
	\end{subfigure}
	\caption{Three dual frames corresponding to the UV curve $\CC^{(3, -2)}_{0, 3}$ and $(n_+, n_-)=(3, 0)$. }
	\label{fig:T23m2Quiver}
\end{figure}
Figure \ref{fig:T23m2Quiver} describes the quiver after Higgsing. In the figure, an `x'-marked box denotes the remnant of a closed puncture, where a gauge / flavor singlet component of $M_a$ remains, with coupling to the remaining quarks in the theory. Quiver 3 requires a special treatment since the second nilpotent vev does not introduce a mass term.

Consider first quiver 1.  The nilpotent $M_a$ on the right/left-hand side gives the same type of the matter content as in the figure \ref{fig:T22m1quiver},
with matter and charges as in the table below:
\be
    	\begin{tabular}{|c|c|c|c|c|c|c|c|c|}
    	\hline
        & $SU(2)_1$ & $SU(2)_2$ & $SU(2)_A$ & $SU(2)_B $ & $SU(2)_C$ & $R_0$ & $\CF$ & $(J_+, J_-)$ \\
        \hline \hline
        $\phi_1$ & $\textrm{adj}$ & & & & & 1 & -1 & (0, 2)  \\
        $\phi_2$ &  & $\textrm{adj}$ & & & & 1 & -1 & (0, 2)  \\
        $q_1$ & $\square$ & & $\square$ & & & $0$ & $1$ & (1, -1) \\
        $q_2$ & $\square$ & $\square$ & &$\square$ & & $\half $ & $\half $ & (1, 0) \\
        $q_3$ & & $\square$ & &  & $\square$ & $0$ & $1$ & (1, -1) \\
        $M_{1,2}$ & & & & &  & 2 & -2 & (0, 4) \\
        	\hline
	\end{tabular}
\ee
The singlet field attached to the `x'-marked box couples to the neighboring quarks, which gives rise to a cubic superpotential term similar to that in \eqref{WTtwoone}.  In addition, there is a quintic coupling between the quarks and the adjoint chiral multiplets:
\be
W_{\hbox{quiver 1}}=M_1 q_1 q_1+M_3 q_3q_3+\phi _1q_2q_2+\phi _1q_1q_1q_2q_2+\phi _2 q_2q_2q_3q_3
\ee
Quiver 2 can be understood by considering a decoupling limit of the $SU(2)$ gauge group corresponding to the rightmost  gauge node. The left-hand side of the quiver is then the same as the $T_2^{(1)}$ theory. 
We list the matter content and charges of the theory in the table below:
\be
    	\begin{tabular}{|c|c|c|c|c|c|c|c|c|}
    	\hline
        & $SU(2)_1$ & $SU(2)_2$ & $SU(2)_A$ & $SU(2)_B $ & $SU(2)_C$ & $R_0$ & $\CF$ & $(J_+, J_-)$ \\
        \hline \hline
        $\phi_1$ & $\textrm{adj}$ & & & & & 1 & -1 & (0, 2)  \\
        $\phi_2$ &  & $\textrm{adj}$ & & & & 1 & -1 & (0, 2)  \\
        $q_1$ & $\square$ & & $\square$ & & & $0$ & $1$ & (1, -1) \\
        $q_2$ & $\square$ & $\square$ & & & & $0$ & $1$ & (1, -1) \\
        $q_3$ & & $\square$ & & $\square$ & $\square$ & $\half$ & $\half$ & (1, 0) \\
        $M_{1,2}$ & & & & &  & 2 & -2 & (0, 4) \\
        	\hline
	\end{tabular}
	\label{table:T23m2_3}
\ee
The superpotential for the quiver 2 is
\be
 W_{\hbox{quiver 2}} = M_1 q_1 q_1 + M_2 q_2 q_2 + \phi_1 q_2 \phi_2 q_2 + \phi_2 q_3 q_3 \ ,
\ee
where we suppress gauge and flavor indices, which are as determined by the symmetry.  The superpotential is generic given the $(J_+, J_- ) = (2, 2)$ or $R_0 = 2$ and $\CF = 0$ symmetry.

\paragraph{Non-mass deformation}
Let us consider quiver 3. When we close one of the $-$ punctures, we get a similar description as quiver 1 and 2. Now, we need to further close the $-$(red) $SU(2)$ puncture by giving a vev to the chiral flavor adjoint of say $SU(2)_0$. Before closing the last puncture, we have a superpotential term $\Tr M_0 \phi_1 (q_0 q_0)$ where $q_0$ is the quark transforming as a fundamental of $SU(2)_0$, and $\phi_1$ is the chiral adjoint of $SU(2)_0$.  The nilpotent vev $\vev{M_0} = \s^+$ then gives the deformation term $\Tr \s^+ \phi_1 (q_0 q_0)$. Though not a mass term for the quarks, it nevertheless turns out to be a relevant deformation, breaking the $SU(2)_0$ global symmetry. To see that $\Tr \s^+ \phi_1 (q_0 q_0)$ is relevant, note that it has charge $(J_+, J_-)= (2, 0)$ which means the exact $R$-charge (before the deformation) is $R = 1+ \e$, which is relevant, $R<2$, since a-maximization gives $\e \simeq 0.46$.  This gives  $a \simeq 1.55$ before the deformation.

The $SU(2)_0$ breaking $\vev{M_0} = \s^+$ yields a superpotential with terms 
\be \label{eq:T22W}
 W \supset \mu_{\fm=-1} + \sum_{\fm = -1, 0, 1}\mu_\fm M_{-\fm} \ , 
\ee
where $\mu _{\fm =-1, 0, 1}= {\rm Tr} \sigma _m \phi_1 q_0 q_0$ is in the adjoint of $SU(2)_0$.  Much as in \cite{Gadde:2013fma}, the first term in \eqref{eq:T22W} leads to $SU(2)_0$ current non-conservation for the $m=0, \ 1$ components:
\be
 (\bar{D}^2 J)_\fm = \delta_\fm W = \mu_{\fm -1} \ ,
\ee
so, for $m=0, 1$,  $J_m$ and $\mu _{m-1}$ pair up to become long multiplets. The remaining superpotential is
\be
\begin{split}
W =  \phi_1\tilde{q}_0\tilde{q}_0 + M_2 (\phi_1q_0q_0)  + 
     M_1 (q_0\tilde{q}_0) + \phi_1 q_1q_1 + \phi_2 q_1q_1 + \phi_2 q_2 q_2   \ .
\end{split}
\ee
The charges $(J_+, J_-)$ must be shifted to be conserved and unbroken
\be
J_+ \rightarrow J_+, \qquad J_- \rightarrow J_- - 2 \fm \ .
\ee
The matter content after Higgsing is as in Figure \ref{fig:T23m2Quiver}, with charges:
\be
    	\begin{tabular}{|c|c|c|c|c|c|c|c|c|}
    	\hline
        & $SU(2)_1$ & $SU(2)_2$ & $SU(2)_A$ & $SU(2)_B $ & $SU(2)_C$ & $R_0$ & $\CF$ & $(J_+, J_-)$ \\
        \hline \hline
        $\phi_1$ & $\textrm{adj}$ & & & & & 1 & -1 & (0, 2)  \\
        $\phi_2$ &  & $\textrm{adj}$ & & & & 1 & -1 & (0, 2)  \\
        $q_0$ & $\square$ & & & & & $-\frac{1}{2}$ & $\frac{3}{2}$ & (1, -2) \\
        $\tilde{q}_0$ & $\square$ & & & & & $\half$ & $\half$ & (1, 0) \\
        $q_1$ & $\square$ & $\square$ & $\square$ & & & $\half$ & $\half$ & (1, 0) \\
        $q_2$ & & $\square$ & & $\square$ & $\square$ & $\half$ & $\half$ & (1, 0) \\
        $M_1, M_2$ & & & & &  & 2 & -2 & (0, 4) \\
	\hline
	\end{tabular} \nn
	\label{table:T23m2_2}
\ee
We will consider similar type of deformations in section \ref{sec:suN}. 

\paragraph{'t Hooft Anomalies}
The anomaly coefficients of $T_2^{(2)}$, in all three dual frames, are:
\be
\begin{array}{c|cc}
  {\rm Tr} J_+ ,\  {\rm Tr} J_+^3 &&  -2  \\
 {\rm Tr} J_- , J_-^3 &&  -6  \\
  {\rm Tr} J_+^2 J_- && 18 \\
  {\rm Tr} J_+ J_-^2 && -18 \\
\end{array}
\ee
 $a$-maximization yields $\e \simeq 0.534$ and $a \simeq 1.45$ for the $T_2^{(2)}$ theory in all three dual frames.

\subsection{$T_2^{(m)}$} \label{subsec:T2pq}
We can generalize previous subsection to construct a general $T_2^{(m)}$ theory. Start with the UV curve $\CC_{0, m+3}^{(m+1, 0)}$  with $(n_+, n_-) = (3, m)$. By closing all the $-$ punctures, we arrive at the sphere with 3 + punctures and normal bundle degree $(m+1, -m)$. We can consider a number of different dual frames, but let us consider the analog of quiver 2 in figure  \ref{fig:T23m2Quiver}. The resulting theory will be a quiver gauge theory, with $SU(2)^m$ gauge symmetry, bifundamental chiral multiplets for the neighboring nodes, and 2 fundamental chirals at the end nodes. In addition, we have adjoint chiral multiplets for each gauge nodes, and $m$ gauge/flavor singlet chiral multiplets.
\begin{figure}[h]
	\centering
	\includegraphics[width=3.5in]{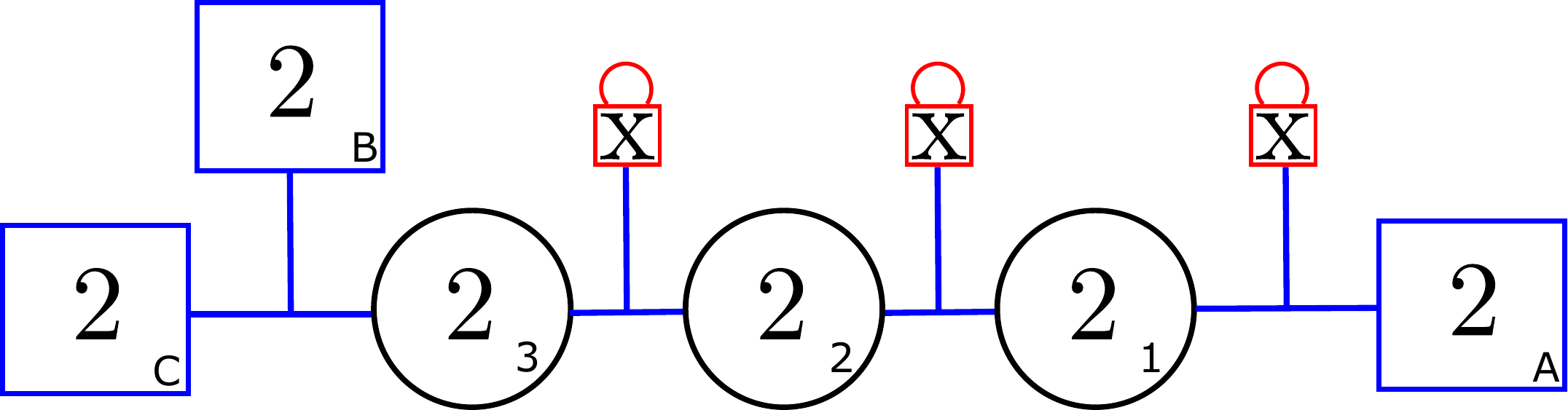}
	\caption{One of the dual frames describing the $T_2^{(3)}$ theory.}
\end{figure}
We summarize the matter contents and their charges in the table \ref{table:T2pqMatter}.
\begin{table}
\centering
\be
    	\begin{tabular}{|c|c|c|c|c|c|c|c|c|}
    	\hline
        & $SU(2)_{i-1}$ & $SU(2)_i$ & $SU(2)_B $ & $SU(2)_C$ & $R_0$ & $\CF$ & $(J_+, J_-)$ \\
        \hline \hline
        $\phi_i$ $(1 \le i \le m)$ & & $\textrm{adj}$ & & & 1 & -1 & (0, 2)  \\
        $q_i$ $(1 \le i \le m)$ & $\square$ & $\square$ & & & $0$ & $1$ & (1, -1) \\
        $q_{m+1}$ $(i=m)$& & $\square$ & $\square$ & $\square$ & $\half$ & $\half$ & (1, 0) \\
        $M_i$ & & & & & 2 & -2 & (0, 4) \\
	\hline
	\end{tabular} \nn
\ee
\caption{The matter content of $T_2^{(m)}$. Here $SU(2)_0$ is the flavor symmetry $SU(2)_A$.}
\label{table:T2pqMatter}
\end{table}
The superpotential is (with indices, and their contractions, suppressed)
\be \label{eq:T2mW}
 W = \sum_{i=1}^m M_i q_i q_i   + \sum_{i=1}^{m-1}\left( \phi_i q_{i+1} \phi_{i+1} q_{i+1} \right) + \phi_m q_{m+1} q_{m+1} \ .
\ee

The 't Hooft anomaly coefficients for this theory are
\be
\begin{array}{c|cc}
 J_+ , J_+^3 && -m  \\
  J_- ,  J_-^3 && m-8  \\
  J_+^2 J_- && 9m \\
  J_+ J_-^2 && -9m  \\
  J_+ SU(2)_{A, B, C}^2 && 0  \\
   J_- SU(2)_{A, B, C}^2 && -2  \\
 \end{array}
\ee
The trial $R$-charge $R = R_0 + \e \CF = \frac{1+\e}{2} J_+ + \frac{1-\e}{2} J_-$ yields the trial $a$-function 
\be
 a(\e) = \frac{3}{32} ( 3 \Tr R^3 - \Tr R ) = \frac{1}{32} \left(3+ 3 (19 m+5) \epsilon -27 \epsilon ^2 + (9-63 m) \epsilon ^3 \right) \ .
\ee
The value of $\e$ is fixed, by maximizing $a(\e)$, to be
\be
 \e (m) = \frac{-3+ \sqrt{133 m^2+16m+4}}{21 m-3}  \lesssim 0.5492 \ .\label{epib}
\ee
As a check, $\e (m=0)= \frac{1}{3}$ which is the value of the free field theory $T_2$. The central charge $a(\epsilon (m))$ grows linearly in $m$, which is not surprising from the quiver gauge theory perspective.

The  $T_2^{(m)}$ theories do not have any exactly marginal deformations: there are $m+(m-1)+1+m = 3m$ couplings from the terms in the superpotential \eqref{eq:T2mW}, and the gauge couplings, and there is no linear relation among their beta functions.   The conformal manifold is an isolated point; this is consistent with geometric construction, since the three punctured sphere has no complex structure modulus. 

\subsection{Infinitely namy $\CN=1$ duals for $SU(2)$ SQCD with 4 flavors} \label{subsec:SU2duals}
$\CN =1$ $SU(2)$ SQCD with 4 flavors can be realized by choosing the UV curve $\CC_{0, 4}^{(1, 1)}$ with $(n_+, n_-)=(2, 2)$. The theory enjoys multiple dualities \cite{Intriligator:1995ne, Csaki:1997cu} which also has a class $\CS$ interpretation \cite{Gadde:2013fma}. Moreover, this theory is known to have 72 dual frames \cite{Spiridonov:2008zr, Dimofte:2012pd}. We now argue that gluing two copies of $T_{2}^{(m)}$ with an $\CN=1$ vector multiplet, for any integer $m \in \IZ_{\ge 0}$, flows to the same SCFT as $SU(2)$ SQCD with 4 flavors.   In the class $\CS$ language, we have chosen two pairs-of-pants labelled by an integer $m$ which gives the same 4-punctured sphere.
\begin{figure}[h]
	\centering
	\includegraphics[width=2.9in]{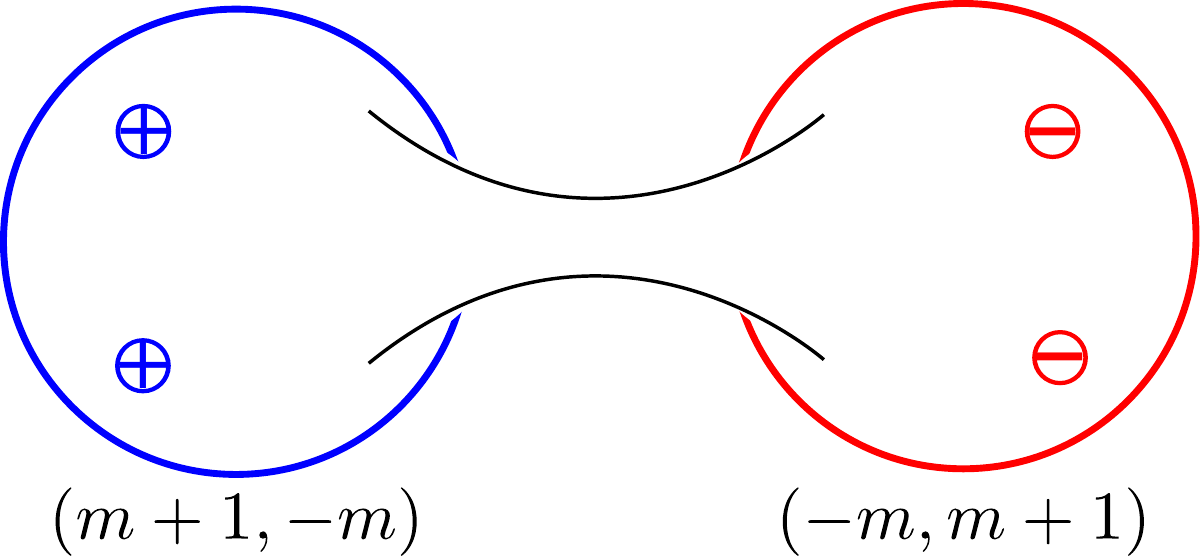}
	\caption{The 4-punctured sphere, with $(p, q)=(1, 1)$, via gluing two pair-of-pants of degrees $(m+1, -m)$ and $(-m, m+1)$. When $m=0$, we get $SU(2)$ SQCD with 4 flavors. The pair-of-pants on the right gives $T_2^{(m)}$, but with reversed $(J_+, J_-)$ charge assignments.}
	\label{fig:SQCDDualCurve}
\end{figure}

For $m=0$, upon gauging an $SU(2)$, each $T_2^{(m=0)}$ factor contributes $N_f=2$ flavors, and the resulting theory is $SU(2)$ with $N_f=4$.  More generally, for all $m$, the effective number of flavors contributed by each $T_2^{(m)}$ theory upon gauging $SU(2)_{X=A,B,C}$ global symmetries is given by the 't Hooft anomaly 
\be
k=-3{\rm Tr}R SU(2)_X^2=3(1-\epsilon)
\ee
e.g. $\epsilon (m=0)=1/3$ gives $k=2$; the gauged $SU(2)$ will be asymptotically free if $2k<3N_c=6$, which is satisfied for all $m$ in  \eqref{epib}.  

\begin{figure}[h]
	\centering
	\begin{subfigure}[b]{6in}
	\centering
	\includegraphics[width=5.2in]{SQCDSU2dual2}
	\caption{The $\CU_2^{(2)}$ quiver, obtained by gauging the $SU(2)$ flavor group on the left-hand side of figure \ref{fig:T23m2quiver3}.}
	\label{fig:SQCDdual1}
	\end{subfigure}
	
	\begin{subfigure}[b]{6in}
	\centering
	\includegraphics[width=5.2in]{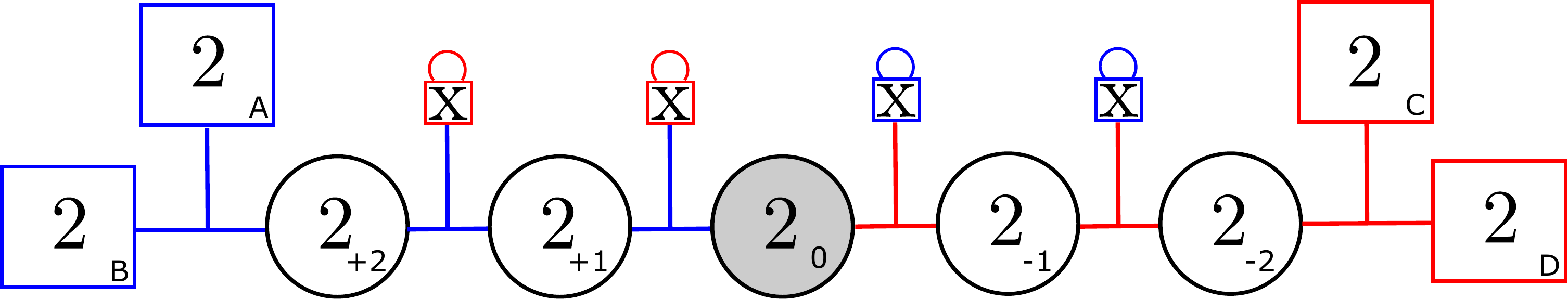}
	\caption{The $\widehat{\CU}_2^{(2)}$ quiver, obtained by gauging the $SU(2)$ flavor group on the right-hand side of figure \ref{fig:T23m2quiver3}. }
	\label{fig:SQCDdual2}	
	\end{subfigure}
	
 	\caption{Two different quivers obtained by gluing two copies of $T_2^{(2)}$. These quiver theories all flow to the same SCFT as $SU(2)$ SQCD with 4 flavors.}
	\label{fig:SQCDDuals}
\end{figure}

There are several, dual descriptions of the resulting theory, corresponding to the dual descriptions of each pair-of-pants discussed in section \ref{subsec:T23m2}.  Let us pick the dual frame referred to there as quiver 2.  As we claimed in section  \ref{subsec:T23m2},  there is a non-manifest $S_3$ permutation symmetry among the $SU(2)_{A,B, C}$ global symmetries.  Correspondingly, there are two dual ways to gauge the 
the $SU(2)$ flavor group; see figure \ref{fig:SQCDDuals}.
Let us pick the dual frame shown in figure \ref{fig:SQCDdual1}. We will label duality frames of this type as $\CU_{2}^{(m)}$. The matter content and their charges are given by two copies of $T_2^{(m)}$ where one copy has flipped $(J_+, J_-)$ charges, as listed in the table. 
\begin{table}
	\centering
	\be
    	\begin{tabular}{|c|c|c|c|c|c|c|c|c|}
    	\hline
        & $SU(2)^{\pm}_{i-1}$ & $SU(2)^{\pm}_{i}$ & $SU(2)_A $ & $SU(2)_B$ & $SU(2)_C $ & $SU(2)_D$ & $(J_+, J_-)$ \\
        \hline \hline
        $\phi^{+}_i$ $(1 \le i \le m)$ & & $\textrm{adj}$ & & & & & (0, 2)  \\
        $q^{+}_1$ $(i=1)$ & $\square$ & $\square$ & $\square$ & & & & (1, 0) \\
        $q^{+}_i$ $(2 \le i \le m)$ & $\square$ & $\square$ & & & & & (1, -1) \\
        $q^{+}_{m+1}$ $(i=m)$ & & $\square$ & & $\square$ & & & (1, -1) \\
        $M^{+}_i$ & & & & & & & (0, 4) \\
	\hline
	$\phi^{-}_i$ $(1 \le i \le m)$ & & $\textrm{adj}$ & & & & & (2, 0)  \\
        $q^{-}_i$ $(i=1)$ & $\square$ & $\square$ & & & $\square$ & & (0, 1) \\
        $q^{-}_i$ $(2 \le i \le m)$ & $\square$ & $\square$ & & & & & (-1, 1) \\
        $q^{-}_{m+1}$ $(i=m)$ & & $\square$ & & & &$\square$  & (-1, 1) \\
        $M^{-}_i$ & & & & & & & (4, 0) \\
	\hline
	\end{tabular} \nn
	\ee
	\label{table:SU2SQCDdualMatter}
	\caption{The $\CU_2^{(m)}$ matter content. $SU(2)_0^\pm$ is the gauge group at the center of the figure \ref{fig:SQCDDuals}.}
\end{table}
In addition to the added gauge multiplet, we have a superpotential term
\be
 W = W_{+} + W_{-} + \l_0 \Tr \mu_+ \mu_-  \ ,
\ee
where $\mu_{\s=\pm} = q_1^\s q_1^\s$ is the operator, with $(J_+, J_-) = (2, 0)$ or $(0, 2)$, associated to the glued punctures and superpotential (with gauge indices contracted and coupling constants $\lambda$) 
\be \label{eq:WUms}
  W_{\s} = \sum_{i=1}^m {\l}_i^\s M^\s_i (q^\s_{i+1} q^\s_{i+1}) + \sum_{i=1}^{m-1} \tilde{\l}_i^\s \left( \phi_i^\s q^\s_{i+1} \phi^\s_{i+1} q^\s_{i+1} \right) + \l'_\s \phi^\s_1 q^\s_{1} q^\s_{1}  \ .   \qquad
\ee

We argue that the $\CU ^{(m)}_2$ theories RG flow to the same IR fixed point as $N_f=4$ $SU(2)$ SQCD, which is the $m=0$ case of $\CU ^{(m)}_2$.  As a first check, we find that the 't Hooft anomaly coefficients of the $\CU ^{(m)}_2$ quiver theory are $m$-independent:\be
\begin{array}{c|cc}
  J_+  , J_+^3, J_-  , J_-^3   &&  -5   \\
  J_+^2 J_-,  J_+ J_-^2  &&  3   \\
  J_+ SU(2)_{A, B}^2,  J_- SU(2)_{C, D}^2  && 0  \\
  J_- SU(2)_{A, B}^2,   J_+ SU(2)_{C, D}^2 && -2  \\
  \end{array}
\ee
The superconformal $U(1)_{R}$ is thus determined by $a$-maximization to be $R = R_0 =\frac{1}{2}(J_++J_-)$.  

\paragraph{Matching of operators} Among the single trace, gauge invariant operators of $\CU ^{(m)}_2$ are 
\be
 \mu_A = q_1^+ q_1^+ , \quad \mu_B = \phi_m^+ q_{m+1}^+ q_{m+1}^+ , \quad \mu_C = q_1^- q_1^- , \quad \mu_D = \phi_m^- q_{m+1}^- q_{m+1}^- \ 
\ee
in the adjoints of $SU(2)_{A,B,C,D}$ respectively, all with superconformal R-charge $R=1$.  These map to meson operators of $N_f=4$ $SU(2)$ SQCD.  The  $N_f=4$ $SU(2)$ SQCD theory has an $SU(8)$ global symmetry (though it is broken by \eqref{WMM} to $SU(2)^4$) with meson / baryon operators in the ${8}\choose{2}$ and the remaining meson/baryon operators are in the $(2,2,2,2)$ of the $SU(2)_A\times SU(2)_B\times SU(2)_C\times SU(2)_D$ subgroup; these operators map to the $R=1$ operators
\be
q_{m+1}^-q_m^-\dots q_2^-q_1^-q_1^+q_2^+\dots q_m^+q_{m+1}^+
\ee

However, there initially appears to be a mismatch in our proposed duality between $\CU ^{(m)}_2$ and $N_f=4$ $SU(2)$ SQCD: each  of the white circle quiver nodes of $\CU ^{(m)}_2$ seems to contribute extra gauge singlet operators, $M_i^\pm$ and $u_i =\tr (\phi^\pm_i)^2$, for $i=1\dots m$.  Classically, these would lead to a mismatch with $N_f=4$ $SU(2)$ SQCD, not only in the spectrum of operators, but also in the moduli space of vacua.    Actually, as we now discuss, the quantum theory does not have the $M_i$ and $u_i$ classical moduli.  They are quantum-lifted in a way similar to what happens in magnetic SQCD, where the classical electric condition $rank(M)\leq N$ arises from non-perturbative 
dynamics in the dual  \cite{Seiberg:1994pq}.  A vev of the would-be moduli would induce a dynamically generated superpotential, which is inconsistent with the $F$-term constraints. 
\begin{figure}[h]
	\centering
	\includegraphics[width=3.5in]{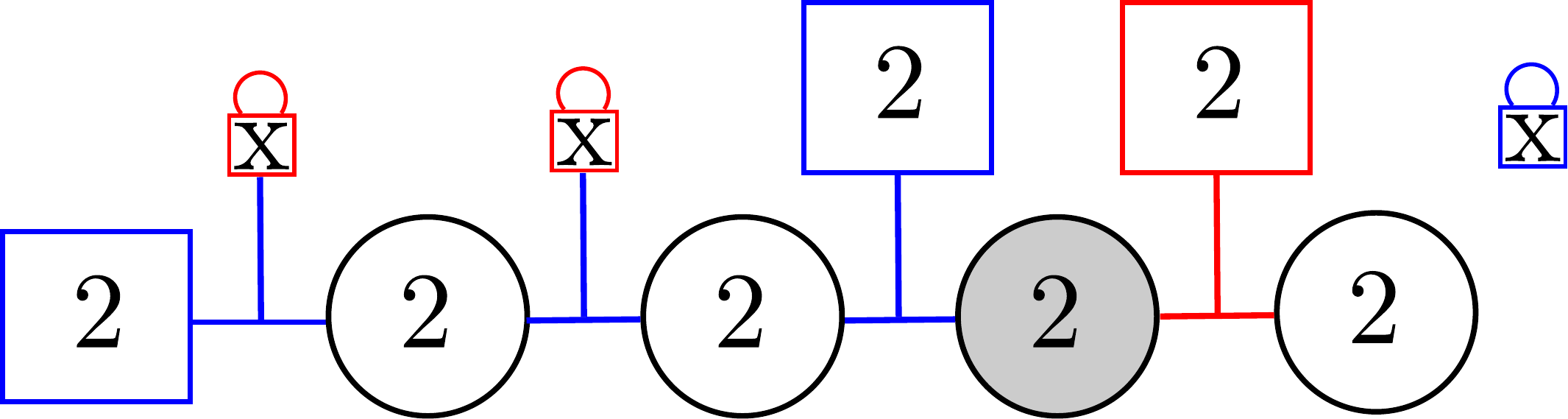}
	\caption{The effective theory after giving a vev to $M_{1}^-$ or $\tr (\phi^-_2)^2$.}
	\label{fig:SQCDSU2dualQ}
\end{figure}

To see this in our setup, suppose first that some $M_{n-1}^-$ has a non-zero vev, which spontaneously breaks $J_+$ and gives a mass to the quarks $q_n^-$ from the first term of \eqref{eq:WUms}. This effectively decouples the side of the $\CU ^{(m)}_2$ quiver in with gauge group $SU(2)_{i\geq n}^-$, as in the figure \ref{fig:SQCDSU2dualQ}. This gives $\Tr J_+ (SU(2)_{n-1}^-)^2 \neq 0$, so the low-energy $SU(2)_{n-1}^-$ instanton factor $(\Lambda ^-_{n-1,L})^{b_L}\sim M^-_{n-1} $ has $J_+$ charge $4$, which allows for superpotential terms 
\be \label{eq:WdynM}
 W_{\textrm{dyn}} \supset \frac{M_{n-1}^-}{q^+_j q^+_j}  \ 
\ee
consistent with the symmetries for all $j$.   This would lead to a $q_j^-$ runaway that is incompatible with $F_{M_i^-}=0$, so the apparent $M_{n-1}^-$ flat direction is actually lifted.   Likewise, if $u_n^-$ gets an expectation value, the associated non-zero $\phi_n^-$ spontaneously breaks $J_+$ and gives a relevant deformation from the second term of \eqref{eq:WUms} (since $q^-_{i} \phi^-_{i-1} q^-_{i}$ has $R$-charge $1$ or $(J_+, J_-) = (0, 2)$). 
In order to preserve $J_+$ symmetry in the IR, the charge of $q^-_{n}$ becomes $(J_+, J_-)=(0, 1)$ and the $SU(2)_{n-1}$ instanton factor gets $J_+$ charged,  $(\Lambda ^-_{n-1,L})^{b_L}\sim u^-_n$ so the theory admits
\be \label{eq:WdynP}
 W_{\textrm{dyn}} \supset  \frac{u^-_n}{q_j^+ q_j^+} \ , 
\ee
which has a runaway for $q_i^\pm$ that is incompatible with $F_{M_j^\pm}$, so the $u_n$ flat direction is lifted.  The superpotentials \eqref{eq:WdynM}, \eqref{eq:WdynP} involves only the quarks on the other $(+)$ side of the quiver, so this quantum effect is present when we couple two $T_N^{(m)}$ theories via $\CN=1$ vector multiplet, but not in the $T_N^{(m)}$ theory itself or when they are coupled via $\CN=2$ vector multiplet.

We give a refined check of operator matching through computing the superconformal index in section \ref{sec:index}. The index of the $\CU_2^{(m)}$ theory agrees with that of the SQCD, which provides a strong check of the duality. Therefore we conjecture that for every choice of $m$, the $\CU_2^{(m)}$ theory flow to the same SCFT as SQCD in the IR.

\paragraph{Exactly marginal deformations}

$\CN =1$ $SU(2)$ SQCD with 4 flavors has a large conformal manifold of exactly marginal deformations
\be
W_{SQCD}=\lambda _{[ij]; [kl]}M^{[ij]}M^{[kl]}, \qquad M^{[ij]}=Q^iQ^j, \qquad i,j=1\dots 8,
\ee
including a one-complex dimensional line of fixed points which preserve $SU(2)^4$ flavor symmetry.  This line of fixed points can also be seen in the $\CU ^{(m)}_2$ theory via the method of \cite{Leigh:1995ep}. The 
 exact NSVZ beta functions for the gauge couplings of $SU(2)_0$ and $SU(2)_i^\pm$ are (with $g_i^\s$ the gauge couplings for $SU(2)_i^\s$)
\be
\begin{aligned}
 \b_{g_0} &\propto -(2 + 2 \g_{q_1^+} + 2 \g_{q_1^-}) \ , &  \\
 \b_{g_1^\s} &\propto -(1+2 \g_{\phi_1^\s} + 2 \g_{q_1^\s} + \g_{q_2^\s}) \ , & \\ 
 \b_{g_i^\s} &\propto  -(2 + 2 \g_{\phi_i^\s} + \g_{q_i^\s} + \g_{q_{i+1}^\s}) \ , & (i=2, \cdots, m)\ .
\end{aligned}
\ee
The exact beta functions for the superpotential couplings are 
\be
\begin{aligned}
 \b_{\l_0} &\propto 1 + \g_{q_1^+} + \g_{q_1^-} \ , 
& \b_{\l^\s_i} &\propto \half \g_{M_i^\s} + \g_{q_i^\s} \ , \\
 \b_{\tilde{\l}_i^\s} &\propto 1 + \half \g_{\phi_i^\s} + \half \g_{\phi_{i+1}^\s} + \g_{q_{i+1}^\s} \ , 
& \b_{\l'_\s} &\propto \half \g_{\phi_1^\s} + \g_{q_{1}^\s} \ ,
\end{aligned}
\ee
where the anomalous dimension $\gamma_{\CO}$ is given by $\Delta(\CO) \equiv \Delta_{\rm classical} (\CO) + \half \gamma_{\CO}$.  Since 
\be \label{eq:betaRel}
 \b_{g_0}  \propto \b_{\l_0} \ , 
\ee
the $\CU ^{(m)}_2$ theory has a one complex dimensional conformal manifold. This can also be seen via the 
the method of \cite{Green:2010da}. There are  $6m+2$ couplings, which break $U(1)^{(6m+2)-1}$ global symmetries (the $-1$ is because we preserve $U(1)_\CF$), so there is a one-complex dimensional conformal manifold that preserves the $SU(2)^4 \times U(1)_\CF\times U(1)_R$ global symmetry.

\begin{figure}[h]
	\centering
	\includegraphics[width=3.5in]{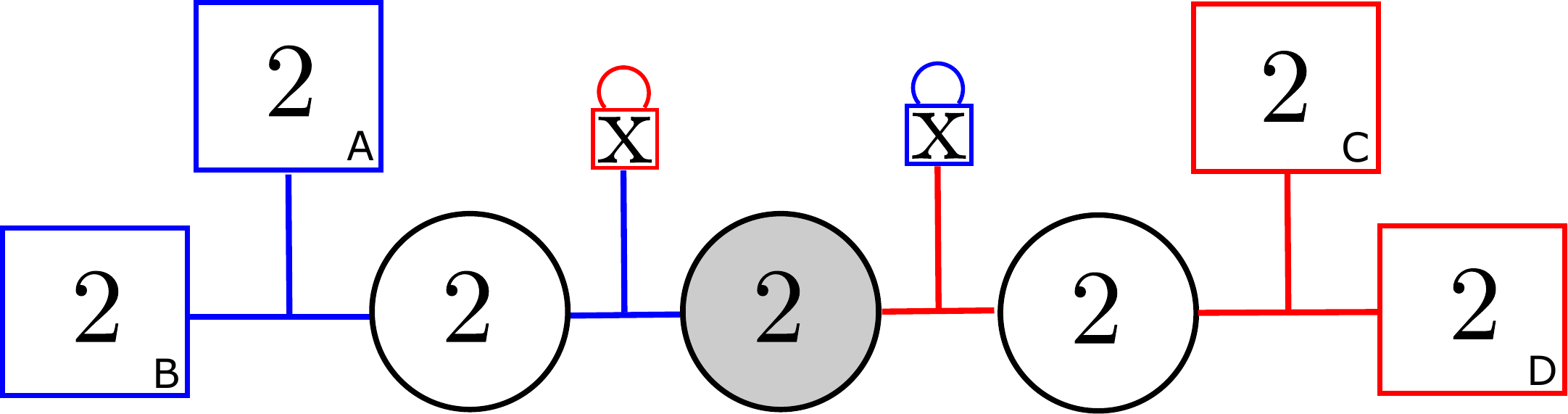}
         \caption{ The low energy description of the theory in figure \ref{fig:SQCDdual1} at scales below $\Lambda_0$}
	\label{fig:gluingt21}
\end{figure}

\paragraph{Cascading RG flow to SQCD}
The duality frame of figure \ref{fig:SQCDdual2} is the  $\widehat\CU ^{(m)}_2$ theory, which we claim is dual to the $\CU ^{(m)}_2$ theory, giving another description of the theory obtained by gluing two copies of $T_2^{(m)}$.
\begin{table}
	\centering
	\be
    	\begin{tabular}{|c|c|c|c|c|c|c|c|c|}
    	\hline
        & $SU(2)^{\pm}_{i-1}$ & $SU(2)^{\pm}_i$ & $SU(2)_A $ & $SU(2)_B$ & $SU(2)_C $ & $SU(2)_D$ & $(J_+, J_-)$ \\
        \hline \hline
        $\phi^{+}_i$ $(1 \le i \le m)$ & & $\textrm{adj}$ & & & & & (0, 2)  \\
        $q^{+}_i$ $(1 \le i \le m)$ & $\square$ & $\square$ & & & & & (1, -1) \\
        $q^{+}_{m+1}$ $(i=m)$& & $\square$ & $\square$ & $\square$ & & & (1, 0) \\
        $M^{+}_i$ & & & & & & & (0, 4) \\
	\hline
	$\phi^{-}_i$ $(1 \le i \le m)$ & & $\textrm{adj}$ & & & & & (2, 0)  \\
        $q^{-}_i$ $(1 \le i \le m)$ & $\square$ & $\square$ & & & & & (-1, 1) \\
        $q^{-}_{m+1}$ $(i=m)$& & $\square$ & & & $\square$ & $\square$ & (0, 1) \\
        $M^{-}_i$ & & & & & & & (4, 0) \\
	\hline
	\end{tabular} \nn
	\ee
	\label{table:SU2SQCDdualMatter}
	\caption{Matter contents of the $\widehat{\CU}_2^{(m)}$ theory; $SU(2)_0^\pm$  is the shaded node in figure \ref{fig:SQCDDuals}.}
\end{table}
The $\widehat{\CU}_m$ theory has superpotential term
\be \label{eq:SU2dual2W}
 W = W_{+} + W_{-} + \Tr \mu_+ \mu_-  \ ,
\ee
where $\mu_{\s =\pm} = \phi_1^\s q_1^\s q_1^\s$ is the operator with $(J_+, J_-) = (2, 0)$ or $(0, 2)$ associated to the punctures that we are gluing and (with implicit gauge index contractions)
\be 
  W_{\s =\pm } = \sum_{i=1}^m M^\s_i (q^\s_i q^\s_i) + \sum_{i=1}^{m-1} \left( \phi_i^\s q^\s_{i+1} \phi^\s_{i+1} q^\s_{i+1} \right) + \phi^\s_m q^\s_{m+1} q^\s_{m+1}  \ .   \qquad
\ee
In this dual frame, the $SU(2)_0$ gauge group has $N_f=N_c$ and no adjoint, so it confines, with a quantum deformed moduli space constraint as in \cite{Seiberg:1994bz}.   At energies below the $SU(2)_0$ dynamical scale, the $SU(2)_0$ node is eliminated, and its adjoining fundamentals are replaced with the $SU(2)_0$ neutral composites 
\be
V^{+} = q_1^+q_1^+, \quad V^{+-} = q_1^+q_1^- \quad \text{and} \quad V^{-} = q_1^-q_1^- \ , 
\ee
where $V^+$ and $V^-$ (the $SU(2)$ analog of baryons) are gauge singlets, while the mesons $V^{+-}$ transform as a bifundamental of $SU(2)_{+1} \times SU(2)_{-1}$, with the constraint   \cite{Seiberg:1994bz}
\be
\det(V^{+-}) -V^+V^- = \Lambda_0^4 \ . 
\ee
The superpotential \eqref{eq:SU2dual2W} becomes (with implicit trace over gauge and flavor indices)
\be
 W =  \phi_1^+ V ^{+-}\phi_1^- V^{+-} +\sum_{\s=\pm}  \left( M^\s_1 V^{\s} + \sum_{i=2}^m M^\s_i q^\s_i q^\s_i + \sum_{i=1}^{m-1}  \phi_i^\s q^\s_{i+1} \phi_{i+1}^\s q^\s_{i+1} +  \phi^\s_m q^\s_{m+1} q^\s_{m+1}  \right) . \qquad
 \label{eq:effsup}
\ee 
We see that $V^\pm$  combine with $M_1^\pm $ to become massive, so they can all be integrated out, setting $V^\pm = M_1^\pm =0$.  The quantum constraint on the moduli space \eqref{eq:SU2dual2W} then implies that $V^{+-}\neq 0$.  The non-zero $V^{\pm}$ bifundamental vev Higgses $SU(2)_{+1} \times SU(2)_{-1}$ to the diagonal $SU(2)$ subgroup. It follows from the superpotential \eqref{eq:effsup} that $\phi_1^\pm $ become massive, and are integrated out.  The resulting low-energy theory is thus similar to the original theory 
(shown in figure \ref{fig:gluingt21}) with $m\to m-1$, i.e. it is $\widehat{\CU}_2^{(m-1)}$.  The above analysis applies to that theory, again reducing $m$, giving a cascading RG flow that eventually ends up at the $m=0$ theory, $\widehat{\CU}_2^{(0)}$, which is simply $SU(2)$ SQCD with $N_f=4$. 

The ${\CU}_2^{(m)}$ dual to $\widehat{\CU}_2^{(m)}$ thus also flows to the same IR SCFT as SQCD. 


\section{$SU(N)$ theories} \label{sec:suN}
We here generalize the discussion in section \ref{sec:su2} to $\CN =1$ $SU(N)$ SQCD with $2N$ flavors. The new element is that we have to replace each bifundamental or trifundamental chiral multiplet, in the links of the quiver, by the $T_N$ theory and its deformations. We first construct the $\CN=1$ $T_N^{(m)}$ theories, which have $SU(N)_A \times SU(N)_B \times SU(N)_C$ flavor symmetry.  We then glue two such theories with $\CN=1$ vector multiplets to construct gauged $T_N^{(m)}$ theories. We argue that this flows to the same theory as obtained from gluing two $T_N$ theories. Then we construct the $\widetilde{T}_N^{(m)}$ theory via partially Higgsing one of the punctures in $T_N^{(m)}$ theory so that we have $SU(N)^2 \times U(1)$ flavor symmetry. We then glue two such theories to obtain
$\CU^{(m)}_N$, and other dual versions, which give new dual descriptions of $SU(N)$ SQCD with $2N$ flavors. 

\subsection{Review of the $T_N$ theory}
Recall that the $T_N$ theory is an $\CN=2$ SCFT with $SU(N)_A \times SU(N)_B \times SU(N)_C$ flavor symmetry. The theory also has $\Delta =2$ ``moment-map'' chiral operators, $\mu_{A, B, C}$, in the adjoint of the $SU(N)_{A, B, C}$ respectively. These operators  satisfy the chiral ring relation \cite{Maruyoshi:2013hja}
\be
 \tr \mu_A^k =  \tr \mu_B^k =  \tr \mu_C^k \ , 
\ee
for $k=2, 3, \cdots N$. 
There are also operators $Q_{ijk}, \tilde{Q}_{ijk}$ which transform as the trifundamental and anti-trifundamental of $SU(N)_A \times SU(N)_B \times SU(N)_C$ with scaling dimension $N-1$. The $T_N$ theory has a Coulomb branch of complex dimension $(N-2)(N-3)/2$, and a Higgs branch, which meet at the origin. 
See \cite{Maruyoshi:2013hja, Tachikawa:2015bga} for more detailed discussion on the chiral ring operators and their relations of the $T_N$ theory. 

Since the $T_N$ theory at the origin is a $\CN =2$ SCFT, it has $U(1)_{R_{\CN=2}} \times SU(2)_R$ symmetry. When we couple this theory to an $\CN=1$ theory, we preserve $(J_+, J_-) = (2 I_3, R_{\CN=2})$, where $I_3$ is the Cartan generator of $SU(2)_R$. As in the previous section, one linear combination of $J_+, J_-$ will become exact $R$-charge, and $\CF = \half(J_+ - J_-)$ will be a charge of the global symmetry of the theory. The $\mu_{A, B, C}$ operators have the charge $(J_+, J_-) = (2, 0)$, and $Q_{ijk}, \tilde{Q}_{ijk}$ have $(J_+, J_-) = (N-1, 0)$.   The 't Hooft anomaly coefficients of the $T_N$ theory are:
\be
\begin{array}{c|cc}
  J_+ ,  J_+^3 &&  0  \\
  J_- , J_-^3 &&  -(N-1)(3N+2)  \\
  J_+^2 J_- && \frac{1}{3}(N-1)(N-2)(4N+3) \\
  J_+ J_-^2 && 0 \\
  J_+ SU(N)_{A, B, C}^2 && 0  \\
  J_- SU(N)_{A, B, C}^2 && -N 
\end{array}
\ee

\subsection{$T_N^{(m)}$ theory}
\begin{figure}[h]
	\centering
	\begin{subfigure}[b]{5in}
	\centering
	\includegraphics[width=4in]{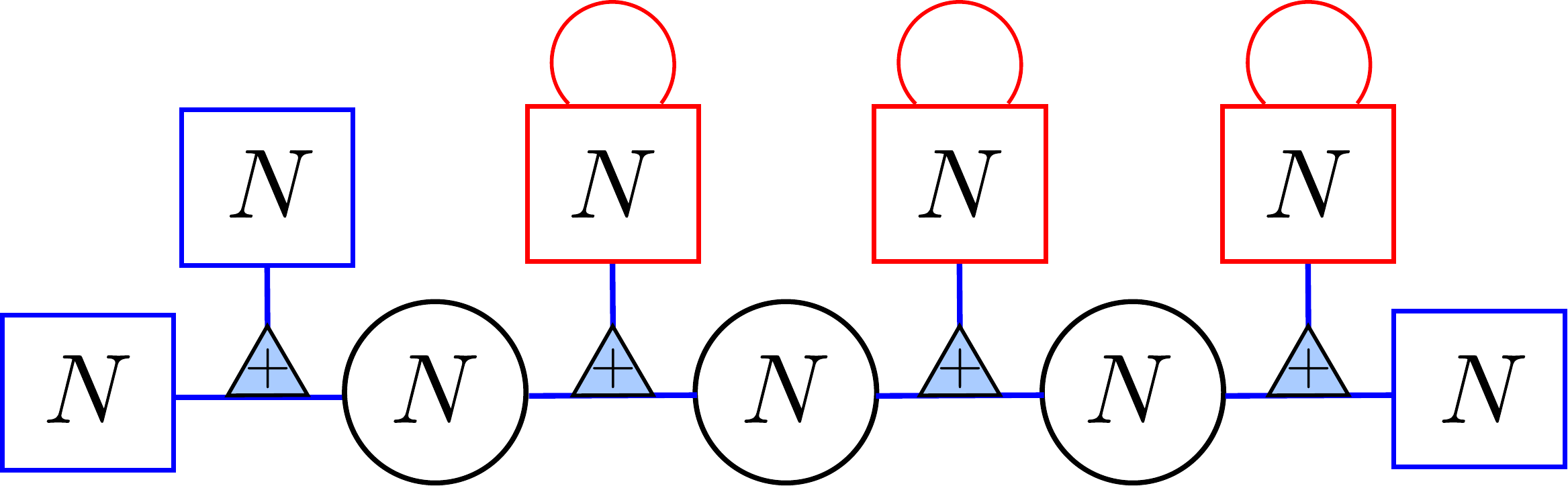}
	\caption{A quiver before Higgsing given by the UV curve $\CC_{3, 3}^{(3, 0)}$ with $(n_+, n_-)=(3, 3)$.}
	\label{fig:TNUnhiggs}
	\end{subfigure}

	\begin{subfigure}[b]{5in}
	\centering
	\includegraphics[width=4in]{TN4m3Quiver}
	\caption{A quiver diagram for the $T_N^{(3)}$ theory, obtained by Higgsing three $-$ punctures above. }
	\end{subfigure}
	\caption{Quiver diagrams for the $T_N^{(3)}$ theory.}
	\label{fig:TNpqQuiver}
\end{figure}

We start with a $m+3$-punctured sphere with 3 $+$ punctures and $m$ $-$ punctures and degrees $(p, q) = (m+1, 0)$. Here we assume all the punctures to be the maximal one carrying $SU(N)$ global symmetry. Let us choose the colored pair-of-pants decomposition so that we get the quiver as described in the figure \ref{fig:TNUnhiggs}.

The theory is composed of $m+1$ copies of $T_N$ theory that are connected via $\CN=2$ vector multiplets and $m$ extra chiral multiplets $M^{(i)}$ $(i=1, \cdots, m)$ transforming under the adjoint of the $SU(N)_i$ global symmetry associated to the $-$ punctures. We denote the moment map operators of the $+$ colored operators by $\mu_{A, B, C}$ and those of $-$ colored operators by $\mu^{(i)}$ $(i=1, \cdots m)$. We use $\phi_i$ for the adjoint chiral multiplets in the $\CN=2$ vector multiplet and $\mu_k$, $\tilde\mu_k$ for the moment map operators for the symmetry group $SU(N)_k$ that are being gauged.
The superpotential is 
\be
 W = \sum_{k=1}^m \Tr \phi_k (\mu_k - \tilde \mu_k ) + \sum_{i=1}^m \Tr \mu^{(i)} M^{(i)} \ .
\ee

Now, we close the punctures by giving a nilpotent vev to $M_i$'s as
\be \label{eq:vev}
 \vev{M^{(i)}} = \rho(\s^+) =
    \begin{pmatrix}
            0 & 1 & \phantom{0} & \phantom{0} &\phantom{0}\\
            \phantom{0}& 0 & 1 & \phantom{0} & \phantom{0} \\
            \phantom{0} & \phantom{0} & \ddots & \ddots & \phantom{0} \\
            \phantom{0} & \phantom{0} & \phantom{0} & 0 & 1 \\
            \phantom{0} & \phantom{0} & \phantom{0} & \phantom{0} &0
    \end{pmatrix} \ ,
\ee
where $\rho$ is the principal embedding of $SU(2)$ into $SU(N)$.
This will induce a relevant deformation to the theory which we name as $T_N^{(m)}$. Here we closely follow the discussion of \cite{Gadde:2013fma}. We can decompose the adjoint representation of $SU(N)$ in terms of sum of the spin-$j$ irreducible representation $V_j$ of $SU(2)$ as $\textrm{adj} = \bigoplus_{j=1}^{N-1} V_j$. Using this, one can write each components of the adjoint of $SU(N)$ in terms of $(\fj, \fm)$ with $\fm=-\fj, -\fj+1, \cdots, \fj-1, \fj$. After giving the vev, the superpotential can be written as
\be
 W = \sum_{k=1}^m \Tr \phi_k (\hat\mu_k - \hat \mu_k' ) + \sum_{i=1}^m \left( \mu^{(i)}_{1, -1} + \sum_{\fj, \fm} \mu^{(i)}_{\fj, \fm} M^{(i)}_{\fj, -\fm} \right) \ .
\ee
This superpotential preserves $(J_+, J_- ) = (2, 2)$ upon the shift
\be \label{eq:TNJshift}
 J_+ \to J_+ \ , \qquad J_- \to J_- - \sum_i 2 \fm^{(i)} \ ,
\ee
where $\fm^{(i)}$ are the weights of the $SU(2)$ representations or the image of $J_3 = \s^3/2$ under $\rho_i$ associated to each puncture $(i)$ being closed.
The vev breaks the original $SU(N)$ global symmetry, with the non-conservation of the current given by
\be
 (\bar{D}^2 J^{(i)} )_{\fj, \fm} = \delta_{\fj, \fm} W = \mu^{(i)}_{\fj, \fm-1} \ .
\ee
The semi-short multiplet $(J^{(i)})_{\fj, \fm}$ and the chiral multiplet $\mu^{(i)}_{\fj, \fm-1}$ combine into a long-multiplet. Therefore all the operators $M^{(i)}_{\fj, -\fm}$ coupled to $\mu^{(i)}_{\fj, \fm}$ decouple, except for $\fm=\fj$. Finally, the remaining superpotential is
\be
W = \sum_{k=1}^m \Tr \phi_k (\hat\mu_k - \hat \mu_k' ) + \sum_{i=1}^m \sum_{\fj=1}^{N-1} \mu^{(i)}_{\fj, \fj} M^{(i)}_{\fj, -\fj} \ .
\ee
We summarize the `matter content' of the theory in the table \ref{table:TNmMatter}. 
\begin{table}	
	\centering
    	\begin{tabular}{|c|c|c|c|c|c|c|c|}
    	\hline
        & $SU(N)_i$ & $SU(N)_A $ & $SU(N)_B$ & $SU(N)_C $ & $(J_+, J_-)$ \\
        \hline \hline
        $\phi_i$ $(1 \le i \le m)$ & $\textrm{adj}$ & & & & (0, 2)  \\
        $\mu_i$ $(1 \le i \le m)$ & adj & & & & (2, 0) \\
        $\tilde{\mu}_i$ $(1 \le i \le m)$ & adj & & & & (2, 0) \\
        $\mu_A$ & & adj & & & (2, 0) \\
        $\mu_B$ & & & adj & & (2, 0) \\
        $\mu_C$ & & & & adj & (2, 0) \\
        $\mu^{(i)}_{\fj, \fj}$ $(1 \le \fj \le N-1)$ & & & & & $(2, -2\fj)$ \\
        $M^{(i)}_{\fj, -\fj}$ $(1 \le \fj \le N-1)$ & & & & & $(0, 2\fj + 2)$ \\
	\hline
	\end{tabular}
	\caption{The `matter content' of the $T_N^{(m)}$ theory. }
	\label{table:TNmMatter}
\end{table}

\paragraph{Anomaly coefficients}
To compute the 't Hooft anomaly coefficients of the $T_N^{(m)}$ theory,  we need to compute effect of the Higgsed $T_N$ block, with the nilpotent vev. Accounting for the above shifts, we find that we simply need to add the contributions from $M_{\fj, -\fj}$ to that of the $T_N$ theory. This gives, 
 for the single puncture Higgsed $T_N$ or equivalently the theory corresponding to the UV curve $\CC_{0, 2}^{(1, -1)}$: 
\be
\begin{array}{c|cc}
  J_+ ,  J_+^3 && 1-N   \\
  J_- , J_-^3 &&  (1-N)(2N+1)  \\
  J_+^2 J_- &&  \frac{1}{3}(N-1)(4N^2-2N-3) \\
  J_+ J_-^2 &&  \frac{1}{3}(1-N)(4N^2 + 4N+3) \\
  J_+ SU(N)_{Z, Z'}^2 && 0  \\
  J_- SU(N)_{Z, Z'}^2 && -N  \\
\end{array}
\ee
Combining this with the known results of the $T_N$ theory and the quiver description depicted in figure \ref{fig:TNpqQuiver} and the charges of the singlets as given in \eqref{table:TNmMatter}, we obtain the anomaly coefficients of the $T_N^{(m)}$ as follows:
\be
\begin{array}{c|cc}
  J_+ ,  J_+^3 &&  m(1-N)  \\
  J_- , J_-^3 &&  (N-1)(m - 3N - 2)  \\
  J_+^2 J_- && \frac{1}{3}(N-1)(4N^2-5N-6 + m(4N^2 + 4N +3)) \\
  J_+ J_-^2 && \frac{1}{3}m(3+N-4N^3) \\
  J_+ SU(N)_{A, B, C}^2 && 0  \\
  J_- SU(N)_{A, B, C}^2 && -N  \\
\end{array}
\ee
Note that the anomalies involving the $SU(N)_{A, B, C}$ are the same as that of $T_N$ theory.
These coefficients can also be obtained from the formula given in the section 5.2 of \cite{Agarwal:2014rua} by extrapolating all the formulas to the negative $p$ or $q$. 

The trial $a$-function is
\be
 a(\e) &=& \frac{3}{64} (N-1) (1 - \e) \left(3 N^2 (\e+1)^2- 3N \left(2 \e^2+ \e +1\right)-2 \left(3 \e^2+3 \e +2\right )\right) \nn \\
 &{ }& + \frac{3}{32} m \e \left(3 N^3 \left(\e^2-1\right)+2N-3 \e^2+1\right) \ ,
\ee
and the value of $\e$ is fixed by $a$-maximization to be
\be
 \e = \frac{- N^2 - N + \sqrt{4 m^2 (N^2+N+1) (3 N^2+N+1)+4 m (3 N^4-5 N^2-5N -2)+(2N^2 - N-2)^2} }{3 (2 m (N^2+N+1)+N^2-2N-2)} . \nn
\ee
For $m=0$, we find $\e = \frac{1}{3}$, which is the expected value for the $\CN=2$ $T_N$ theory. The value of $a$ increases linearly with respect to $m$ and grows cubically with respect to $N$.
We can also determine the $SU(N)$ flavor central charge $k_{SU(N)}$ \cite{Anselmi:1997am,Anselmi:1997ys} to be
\be
 k_{SU(N)} \delta^{ab} = -3 \Tr R T^a T^b = \frac{3}{2}(1 - \e) N \delta^{ab}\ .
\ee
When $\e = \frac{1}{3}$, $k_{SU(N)} = N$ which agrees with the known result of $T_N$ theory. Since $ \frac{1}{3} < \e < \frac{1}{\sqrt{3}}$ for $m>0$, we see the flavor central charge is less than $N$ for $m>0$. In many respect, the $T_N$ theory behaves as $N$ fundamental flavors \cite{Maruyoshi:2013hja} since it contributes the same amount to the beta function of the gauge coupling. For the $T_N^{(m)}$ case, it contributes to the beta function as that of $N_f < N$. 

\subsection{Infinitely many $\CN=1$ duals for gauged $T_N$ theories} \label{subsec:gaugedTn}
\begin{figure}[t]
	\centering 	
	\begin{subfigure}[b]{6in}
	\centering
	\includegraphics[height=0.9in]{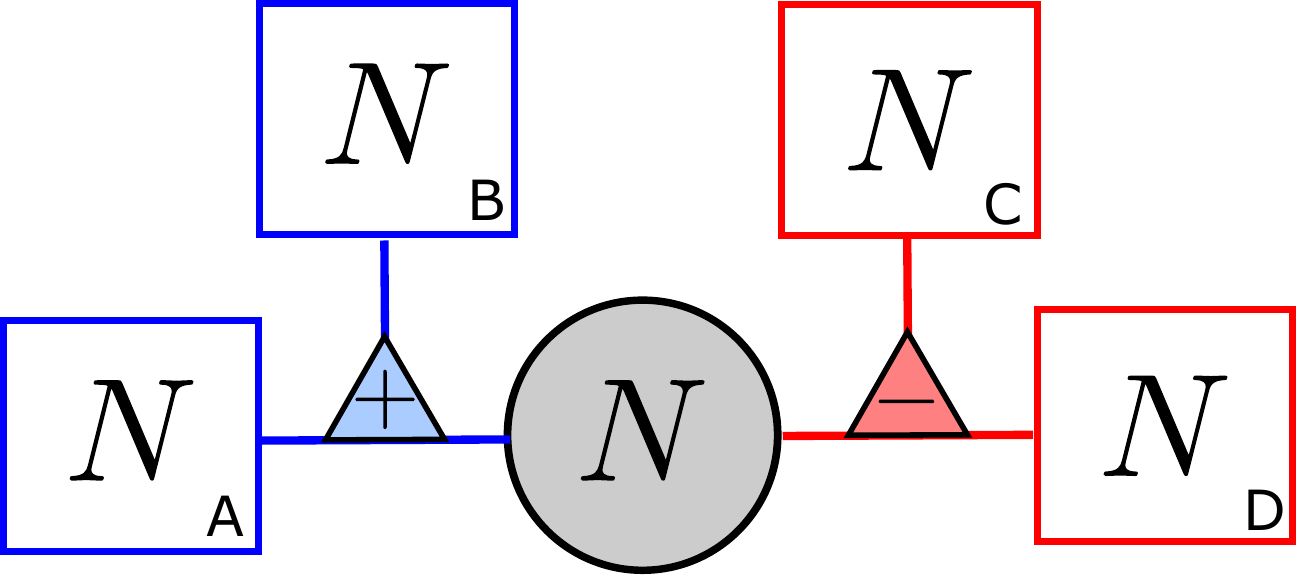}
	\caption{Two $T_N$ theories coupled by gauging the $SU(N)$ flavor symmetry subgroup with an $\CN=1$ vector multiplet.}
	\label{fig:GaugedTn}
	\end{subfigure}
		\begin{subfigure}[b]{6in}
			\centering
			\includegraphics[height=0.9in]{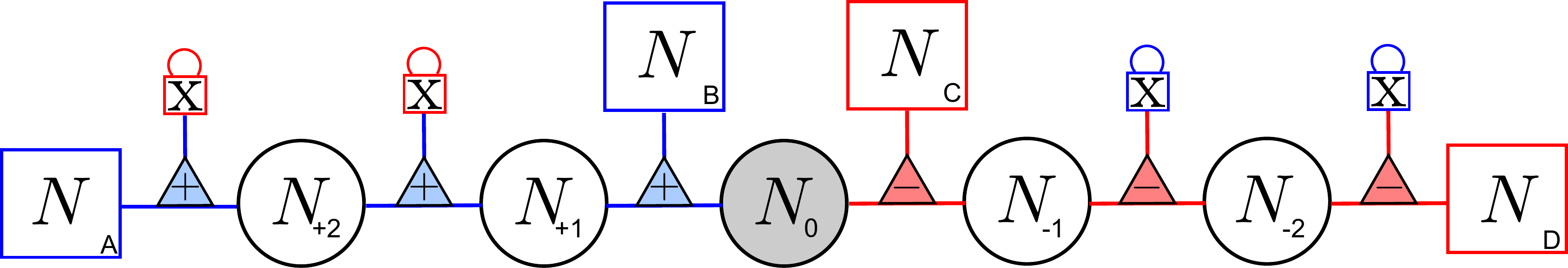}
			\caption{A quiver description obtained by gauging the $SU(N)$ flavor group of two copies of the $T_{N}^{(2)}$ theory.}
			\label{fig:GaugedTnDual1}
		\end{subfigure}
 	\caption{Different quiver descriptions for the 4 maximal-punctured sphere theory with $(p, q)=(1, 1)$. Shaded circular nodes denote the $\CN=1$ vector multiplets and unshaded nodes denote the $\CN=2$ vector multiplets.}
	\label{fig:GaugedTnDuals}
\end{figure}

As a preparation of the SQCD, let us first consider the theory obtained by gluing two copies of $T_N$ theory by gauging one of the $SU(N)$ flavor groups on each of $T_N$. It can be obtained from choosing the UV curve to be the 4-punctured (all maximal, 2 $+$, and 2 $-$ colored) sphere with $(p, q)=(1, 1)$. See the figure \ref{fig:GaugedTn}. This theory and its dualities have been studied in \cite{Benini:2009mz,Gadde:2013fma} which we review here. This theory has $SU(N)_A \times SU(N)_B \times SU(N)_C \times SU(N)_D \times U(1)_\CF \times U(1)_R$ global symmetry with the `matter content' as given in the table \ref{table:gaugedTn}. 
\begin{table}
\centering
    	\begin{tabular}{|c|c|c|c|c|c|c|c|c|}
    	\hline
        & $SU(N)$ & $SU(N)_A $ & $SU(N)_B$ & $SU(N)_C $ & $SU(N)_D$ & $U(1)_R$ & $U(1)_\CF$ & $(J_+, J_-)$ \\
        \hline \hline
        $\mu^+$ & adj & & & & & 1 & 1 & $(2, 0)$ \\
        $\mu^-$ & adj & & & & & 1 & -1 & $(0, 2)$ \\
        $\mu_A$ & & adj & & & & 1 & 1 & $(2, 0)$ \\
        $\mu_B$ & & & adj & & & 1 & 1 & $(2, 0)$ \\
        $\mu_C$ & & & & adj & & 1 & -1 & $(0, 2)$ \\
        $\mu_D$ & & & & & adj & 1 & -1 & $(0, 2)$ \\
        \hline
        \end{tabular} \nn
        \caption{The 'matter content' of the gauged $T_N$ theory. The $SU(N)$ in the first column denotes the gauge group.}
	\label{table:gaugedTn}
\end{table}

For this theory, the superconformal $R$-charge is given by $R_0 = \half (J_+ + J_-)$.
The $\mu_{A, B, C, D}$'s are the operators present in the $T_N$ theory, which are associated to the punctures on the UV curve. The operators $\mu^\pm$ are the operators corresponding to the punctures that we are gluing/gauging. We can write a superpotential term
\be \label{eq:WgaugedTN}
 W = \tr \mu^+ \mu^- \ ,
\ee
which preserves all the global symmetries of the theory.

\begin{table}[h]	
	\centering
    	\begin{tabular}{|c|c|c|c|c|c|c|c|}
    	\hline
        & $SU(N)^{\pm}_i$ & $SU(N)_A $ & $SU(N)_B$ & $SU(N)_C $ & $SU(N)_D$ & $(J_+, J_-)$ \\
        \hline \hline
        $\phi^{+}_i$ $(1 \le i \le m)$ & $\textrm{adj}$ & & & & & (0, 2)  \\
        $\mu^{+}_i$ $(0 \le i \le m)$ & adj & & & & & (2, 0) \\
        $\tilde{\mu}^{+}_i$ $(1 \le i \le m)$ & adj & & & & & (2, 0) \\
        $\mu_A$ & & adj & & & & (2, 0) \\
        $\mu_B$ & & & adj & & & (2, 0) \\
        $\mu^{+, (i)}_{\fj, \fj}$ $(2 \le i \le m+1)$  & & & & & & $(2, -2\fj)$ \\
        $M^{+, (i)}_{\fj, -\fj}$ $(2 \le i \le m+1)$ & & & & & & $(0, 2\fj + 2)$ \\
	\hline
	$\phi^{-}_i$ $(1 \le i \le m)$ & $\textrm{adj}$ & & & & & (2, 0)  \\
        $\mu^{-}_i$ $(0 \le i \le m)$ & adj & & & & & (0, 2) \\
        $\tilde{\mu}^{-}_i$ $(1 \le i \le m)$ & adj & & & & & (0, 2) \\
        $\mu_C$ & & & & adj & & (0, 2) \\
        $\mu_D$ & & & & & adj & (0, 2) \\
        $\mu^{-, (i)}_{\fj, \fj}$ $(2 \le i \le m+1)$ & & & & & & $(-2\fj, 2)$ \\
        $M^{-, (i)}_\fj$ $(2 \le i \le m+1)$ & & & & & & $(2\fj+2, 0)$ \\
	\hline
	\end{tabular}
 	\caption{Matter contents of the quiver obtained by gluing two copies of $T_{N}^{(m)}$. Here $SU(N)_0^\pm$ is identified as the $SU(2)$ gauge group at the center of the figure \ref{fig:GaugedTnDual1}. The operators $\mu^{\pm, (i)}_{\fj, -\fj}$ are the ones in the $i$-th $T_N$ block in the quiver. Here $\fj = 1, 2, \cdots, N-1$.}
	\label{table:gaugedTNdualMatter}
\end{table}
Now let us describe the dual theories of the coupled $T_N$. We couple two copies of $T_{N}^{(m)}$ with an $\CN=1$ vector multiplet to get the theory corresponding to the same 4-punctured (all maximal, 2 $+$ and 2 $-$ colored) sphere with $(p, q)=(1, 1)$. When gluing the two theory with an $\CN=1$ vector, the $(J_+, J_-)$ charge assignment of one of the $T_N^{(m)}$ has to be flipped in order to write the superpotential term \eqref{eq:WgaugedTN}. See figure \ref{fig:GaugedTnDuals}. The `matter content' of the theory is given in the table \ref{table:gaugedTNdualMatter}.

The theory has a superpotential
\be
 W = W_+ + W_- + \tr \mu^{+}_0 \mu^{-}_0 \ ,
\ee
where
\be
 W_\s = \sum_{k=1}^m \Tr \phi^\s_k (\mu^\s_k - \tilde{\mu}^\s_k ) + \sum_{i=2}^{m+1} \sum_{\fj=1}^{N-1} \mu^{\s, (i)}_{\fj, \fj} M^{\s, (i)}_{\fj, -\fj} \ .
\ee
Since the coupled theory for any $m$ comes from the same UV curve, we expect they all flow to the same SCFT in the IR.

Let us compute the anomaly coefficients of the quiver theory. We can use the anomaly coefficients we computed for the $T_N^{(m)}$ and add up with that of $T_N^{(m)}$ with flipped $J_+$ and $J_-$ in addition to the gaugino contributions at the center node. Then we obtain:
\be
\begin{array}{c|cc}
  J_+ ,  J_+^3 , J_- , J_-^3  &&  (2N+1)(1-N)  \\
  J_+^2 J_- , J_+ J_-^2 && \frac{1}{3}(N-1)(4N^2-2N-3) \\
  J_+ SU(N)_{A, B}^2, J_- SU(N)_{C, D}^2 && 0  \\
  J_- SU(N)_{A, B}^2, J_+ SU(N)_{C, D}^2 && -N  \\
\end{array}
\ee
We see that the anomaly coefficients are independent of $m$, therefore it agrees with the gauged $T_N$ which corresponds to the case with $m=0$.

We will match the set of supersymmetric operators by computing the superconformal index in section \ref{sec:index}.

\begin{figure}[h]
	\centering
	\begin{subfigure}[b]{6in}
    \centering
	\includegraphics[height=0.9in]{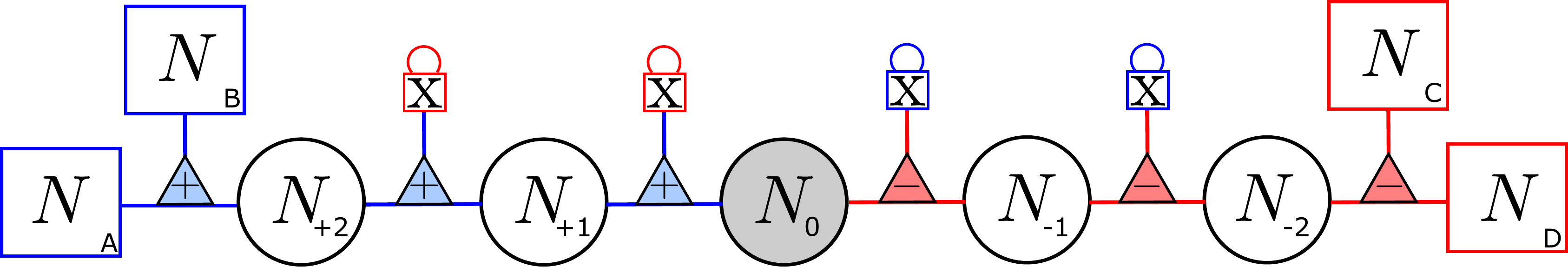}
	\caption{Another quiver description obtained by gluing two copies of the $T_{N}^{(2)}$ theory. This quiver has a cascade of RG flows which reduces it to the quiver of figure \ref{fig:GaugedTn} in the IR.}
	\label{fig:GaugedTnDual}
	\end{subfigure}
	\begin{subfigure}[b]{6in}
		\centering
		\includegraphics[height=0.9in]{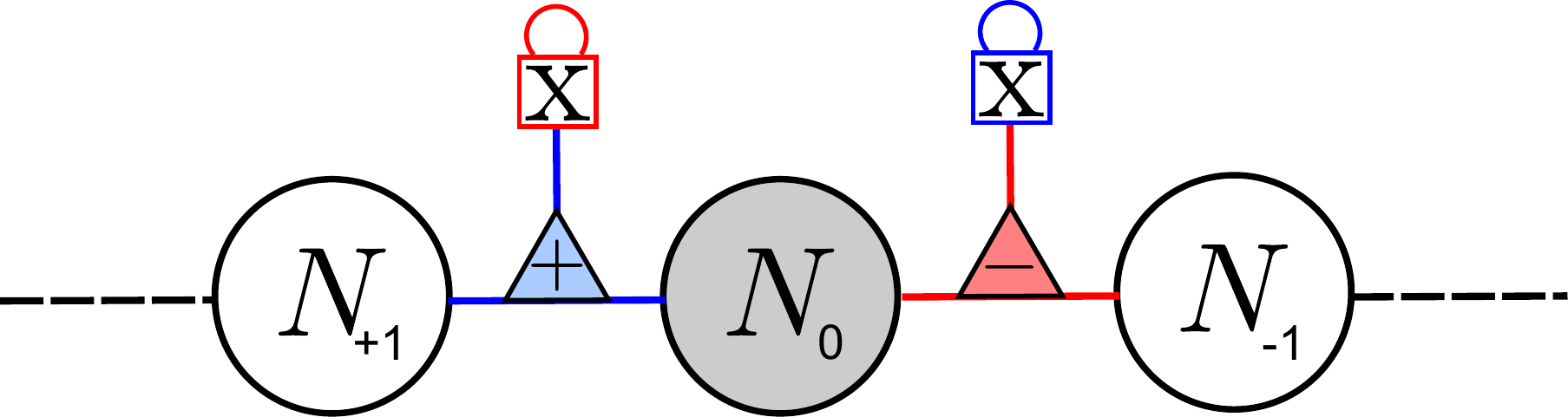}
		\caption{The $\CN=1$ node shown here undergoes confinement, triggering a cascade of RG flows in figure \ref{fig:GaugedTnDual}. The dynamics that lead to this behavior are local to this section of the quiver and do not depend upon the rest of the quiver.}
		\label{fig:confiningTNnode}
	\end{subfigure}	
	\begin{subfigure}[b]{6in}
		\centering
		\includegraphics[height=0.5in]{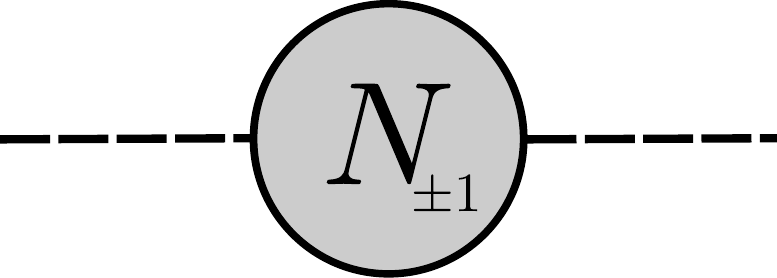}
		\caption{Due to confinement at the $\CN=1$ node in quiver of figure \ref{fig:confiningTNnode}, it reduces to the quiver shown here at low energies.}
		\label{fig:postConfinement}
	\end{subfigure}
	\caption{The quiver in figure \ref{fig:GaugedTnDual} gives an interesting duality frame of the theory obtained by gluing two copies of $T_N^{(2)}$. The sub-quiver shown in figure \ref{fig:confiningTNnode} undergoes confinement at the $\CN=1$ node reducing it to the sub-quiver of figure \ref{fig:postConfinement}. This process triggers a cascade of RG flows in figure \ref{fig:GaugedTnDual} reducing it to the quiver of figure \ref{fig:GaugedTn}.}
	\label{fig:TnmConfinement}		
\end{figure}

\paragraph{Cascading RG flows to the gauged $T_N$ theory}
In section \ref{subsec:SU2duals} we saw that in the dual frame of the form figure \ref{fig:SQCDdual2}, the central gauge node $SU(2)_0$ confines and we get a cascade of RG flows which ultimately reduces the whole system to $SU(2)$ SQCD with 4 flavors. Here, we will argue that a similar mechanism occurs when two $T_N^{(m)}$ blocks are glued to each other to give the duality frame of figure \ref{fig:GaugedTnDual}.  Guided by the $SU(2)$ case, we claim that the $\CN=1$ node in the sub-quiver shown in figure \ref{fig:confiningTNnode} undergoes confinement with a quantum deformed moduli space. At energies below confinement-scale, the spectrum of the quiver will include operators that transform as bifundamentals of the $\pm1$-th nodes of the original quiver. The quantum deformation of the moduli space will imply that these bifundamentals have a non-zero expectation value, breaking the product gauge group $SU(N)_{+1} \times SU(N)_{-1}$ down to the diagonal $SU(N)$. The expectation value will also make the adjoint chiral fields coupled to the $\pm1$-th nodes massive, which will therefore get integrated out. The upshot will be a reduction of $m\to m-1$: at low energies, the quiver shown in figure \ref{fig:confiningTNnode} reduces to that shown in figure \ref{fig:postConfinement}. This process triggers a cascade of RG flows which reduces the quiver of figure \ref{fig:GaugedTnDual} down to that shown in figure \ref{fig:GaugedTn}.

\begin{figure}[h]
	\centering
	\includegraphics[width=6.2in]{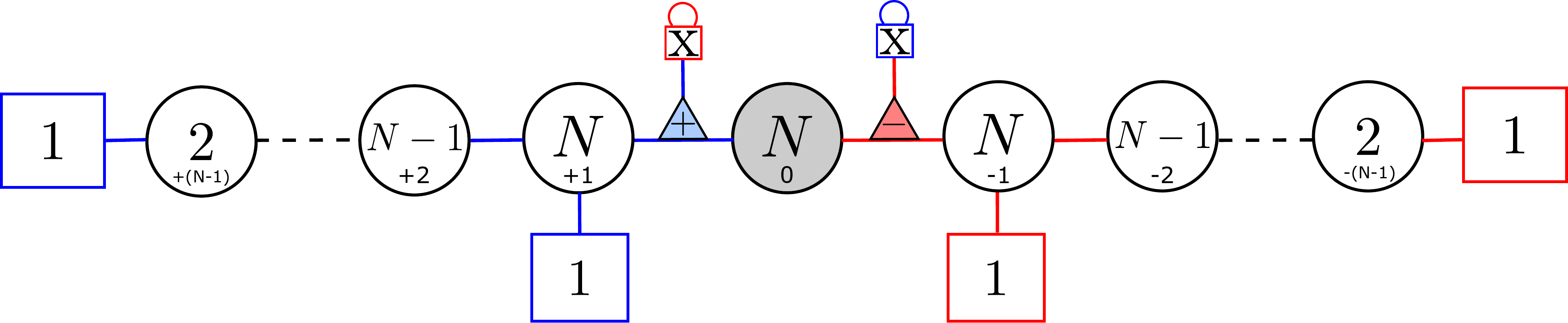}
	\caption{The quiver obtained by gluing two $T_N^{(1)}$ blocks and $\CN=2$ quiver tails corresponding to the minimal puncture. The $T_N^{(1)}$ blocks are glued to each other via an $\CN=1$ vector multiplet along one of their full punctures. The other full puncture of each block is glued, via an $\CN=2$ vector multiplet, to an $\CN=2$ tail corresponding to the minimal puncture.}
	\label{fig:evidenceofconfinement}
\end{figure}

\begin{figure}[h]
		\centering
		\includegraphics[width=4.5in]{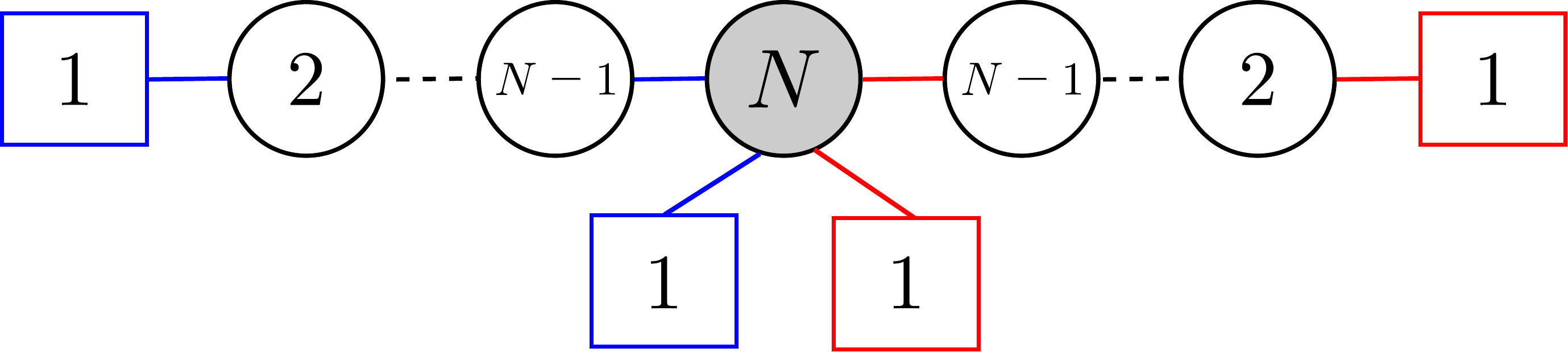}
		\caption{The expected low energy theory if the central $\CN=1$ node in figure \ref{fig:evidenceofconfinement} undergoes confinement.}
		\label{fig:evidenceofconfinementlowenergy}
\end{figure}

As an evidence to support our claim about figure \ref{fig:confiningTNnode}, we consider the theory obtained by gluing two $T_N^{(1)}$ blocks via an $\CN=1$ vector multiplet along one of their full punctures. The other full puncture of each block is glued (via an $\CN=2$ vector) to an $\CN=2$ quiver tail corresponding to the minimal puncture, giving the quiver in figure \ref{fig:evidenceofconfinement}. 
If our claim is correct then the central $\CN=1$ node of this quiver should also exhibit confinement, and the theory will then flow to the quiver of figure \ref{fig:evidenceofconfinementlowenergy}. We now argue that this is indeed the case. 

\begin{figure}[h]
	\centering
	\begin{subfigure}[b]{6in}
		\centering
		\includegraphics[width=5.5in]{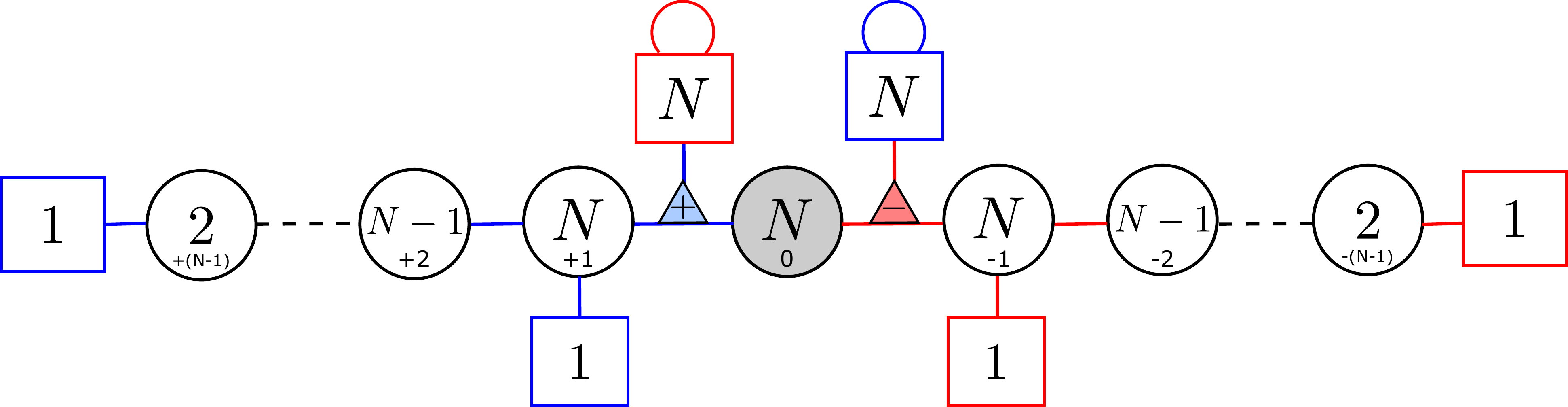}
		\caption{Gauged $T_N$ theory with quiver tails attached. }
		\label{fig:TNwithTail}
	\end{subfigure}
	\begin{subfigure}[b]{6.3in}
		\centering
		\includegraphics[width=6.3in]{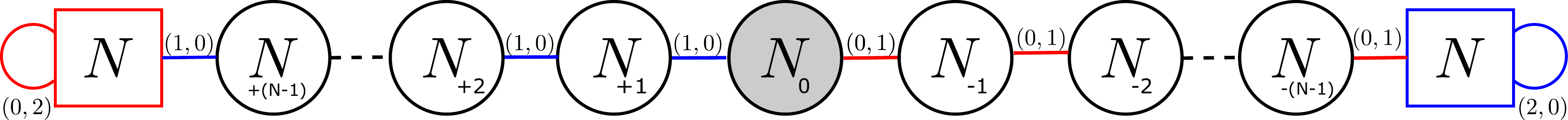}
		\caption{Linear quiver dual of the above quiver.}
		\label{fig:LinearQuiverDual}
	\end{subfigure}
	\caption{Quiver theory of figure \ref{fig:evidenceofconfinement} before closing the punctures. It is given by a gauged $T_N$ theory with quiver tails attached. }
\end{figure}

\begin{figure}[h]
	\centering
	\includegraphics[width=6.3in]{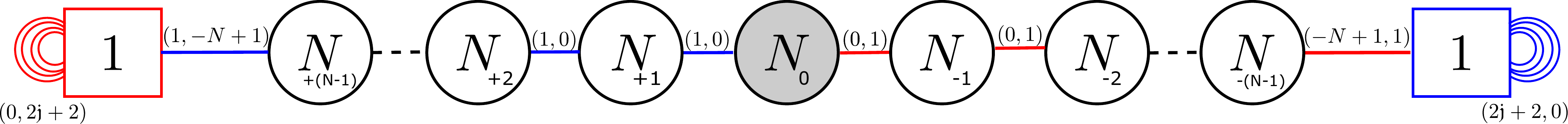}
	\caption{The linear quiver dual to the duality frame of figure \ref{fig:evidenceofconfinement}. We have $N-1$ singlets attached to each ends. Here $\fj = 1, \cdots, N-1$.}
	\label{fig:GMSUNlinearquiver}
\end{figure}

Note that the quiver of figure \ref{fig:evidenceofconfinement} is dual to the linear quiver shown in figure \ref{fig:GMSUNlinearquiver}. When the `x'-marked punctures of the figure \ref{fig:evidenceofconfinement} are not closed, as in figure \ref{fig:TNwithTail}, the theory is dual to the linear quiver of figure \ref{fig:LinearQuiverDual} \cite{Gaiotto:2009gz}. The only difference here is that we added gauge singlets to the punctures.
From here, we close the punctures at each ends by a nilpotent Higgsing to get the linear quiver theory as given in the figure \ref{fig:GMSUNlinearquiver} \cite{Agarwal:2014rua}. 
We have also shown the $(J_+,J_-)$ charges of the various fields in the same figure.
The superpotential terms of this quiver are given by all the single trace gauge singlet local operators with charges $(J_+,J_-) = (2,2)$.

\begin{figure}[h]
	\centering
	\includegraphics[width=6.2in]{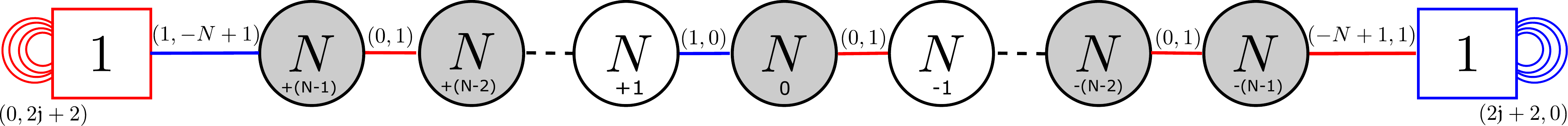}
	\caption{A duality frame of figure \ref{fig:GMSUNlinearquiver} obtained by dualizing, the $0$-th node, then the $\pm1$-st nodes, followed by $\pm2$-nd nodes and so on until we finally dualize the $\pm(N-2)$-th nodes.}
	\label{fig:GMSUNlinear2}
\end{figure}

\begin{figure}[h]
	\centering
	\includegraphics[width=6.0in]{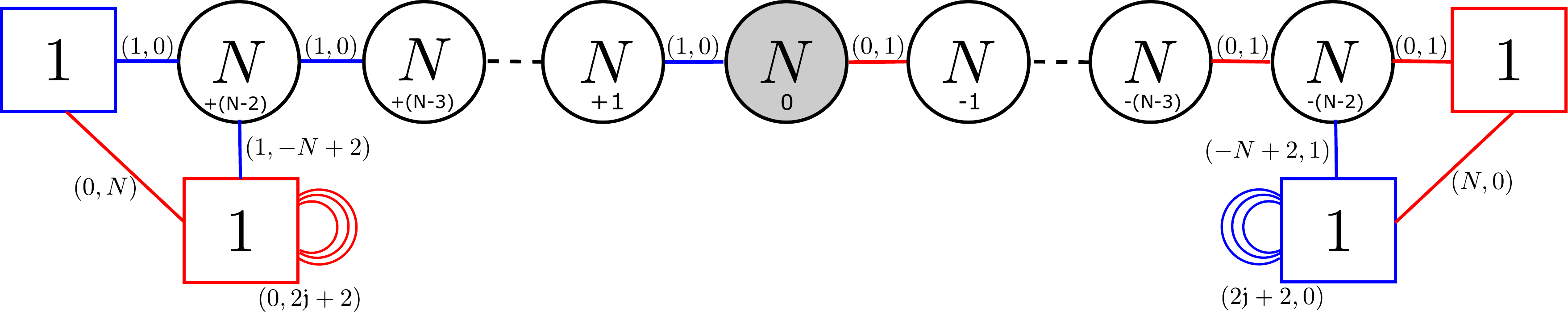}
	\caption{The low energy theory of the quiver in figure \ref{fig:GMSUNlinear2} obtained by noticing that the nodes at its left and the right ends undergo s-confinement. Here $\fj=1, \cdots, N-2$. }
	\label{fig:GMSUNl2lowenergy}
\end{figure}

Let us now dualize the central $\CN=1$ node of figure \ref{fig:GMSUNlinearquiver}, followed by dualizing the $\pm1$-st nodes, then dualize the  $\pm 2$-nd nodes and so on until we finally dualize the $\pm(N-2)$-th nodes of the quiver. This will land us on a linear quiver which has an $\CN=1$ vector multiplet at the $0$-th, $\pm(N-2)$-th and $\pm(N-1)$-th nodes while the rest of the nodes have an $\CN=2$ vector multiplet as shown in figure \ref{fig:GMSUNlinear2}. 
Notice that the $\CN=1$ node at either ends of the quiver in the current duality frame is equivalent to an SQCD with $N_f=N_c+1$ flavors. These nodes will therefore undergo s-confinement. The low energy theory of this quiver will then be given by fields describing the mesonic and baryonic fluctuations of the end nodes. Equivalently, we can Seiberg dualize this node to get the theory of free chiral multiplets. This corresponds to the quiver of figure \ref{fig:GMSUNl2lowenergy}.
Once again the superpotential of this quiver can be written down by considering all the chiral gauge invariant operators which have charges $(J_+, J_-)=(2, 2)$. This will include the low energy superpotential of $N_f=N_c+1$ SQCD that is expected to be there after s-confinement of the edge nodes in figure \ref{fig:GMSUNlinear2}.

\begin{figure}[h]
	\centering
	\includegraphics[width=6.2in]{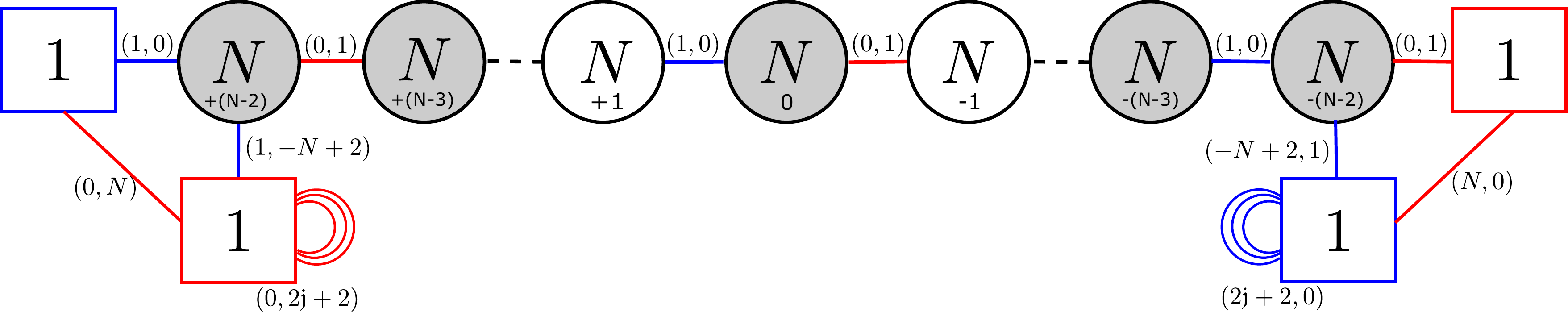}
	\caption{The duality frame of the theory in figure \ref{fig:GMSUNl2lowenergy} obtained by dualizing its $0$-th node, followed by the $\pm1$-th nodes and so on until we dualize the $\pm(N-3)$-th nodes. }
	\label{fig:GMSUNl3}
\end{figure}
In order to proceed we will first have to go through the following series of dualities: dualize the $0$-th node in the quiver of figure \ref{fig:GMSUNl2lowenergy} followed by the $\pm 1$st nodes, then the $\pm 2$nd nodes and so on until we finally dualize $\pm(N-3)$-th nodes. This series of dualities will produce a quiver whose central and last two nodes on either sides are gauged using an $\CN=1$ vector multiplet while the rest of the nodes are gauged using an $\CN=2$ vector multiplet. This quiver is depicted in figure \ref{fig:GMSUNl3}.

\begin{figure}[h]
	\centering
	\includegraphics[width=6.2in]{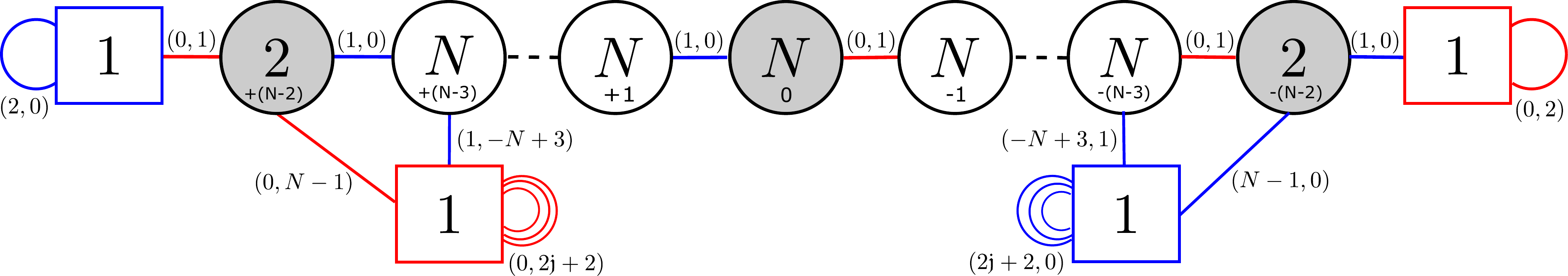}
	\caption{The quiver obtained by dualizing the end nodes of the quiver in figure \ref{fig:GMSUNl3}. Here $\fj=1, \cdots, N-3$.}
	\label{fig:GMSUNl4}
\end{figure}

If we now dualize the nodes at the left and the right ends of the quiver in figure \ref{fig:GMSUNl3}, we obtain the quiver of figure \ref{fig:GMSUNl4}. 

\begin{figure}[h]
	\centering
	\includegraphics[width=6.2in]{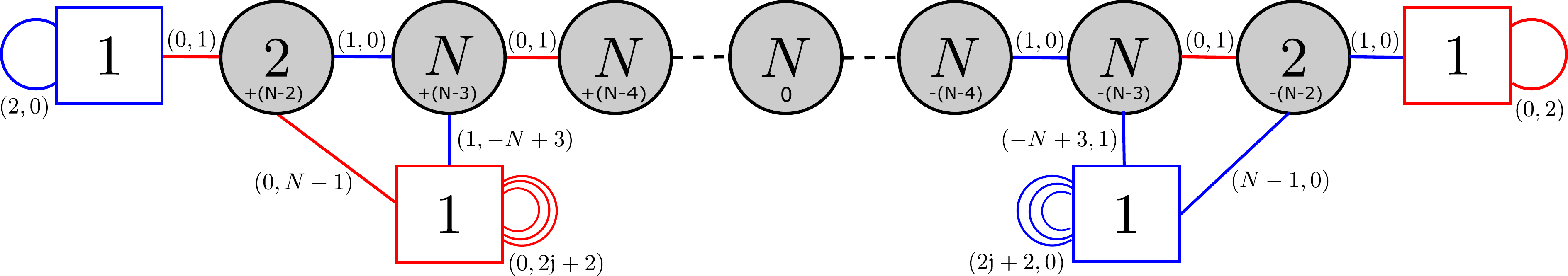}
	\caption{The quiver of figure \ref{fig:GMSUNl4} can be dualized to the one shown in this figure.}
	\label{fig:GMSUNl5}
\end{figure}

\begin{figure}[h]
	\centering
	\includegraphics[width=6.2in]{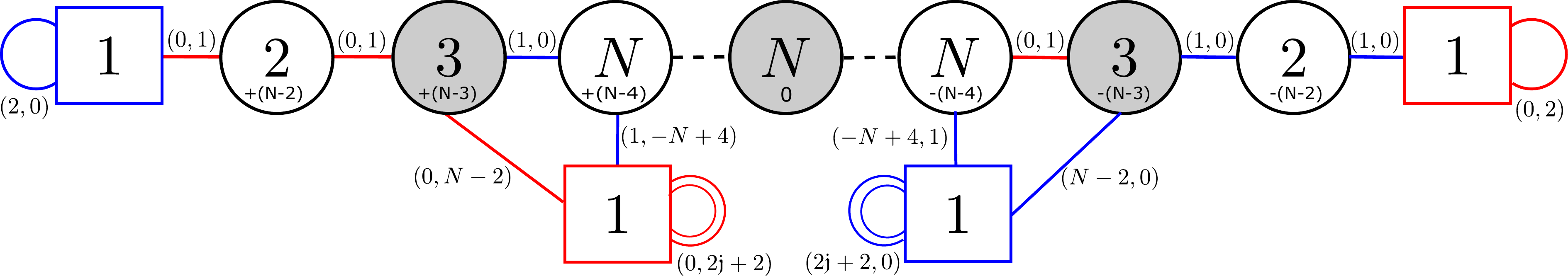}
	\caption{The quiver obtained by dualizing the penultimate nodes on either sides of the quiver in figure \ref{fig:GMSUNl5}. Here $\fj=1, \cdots, N-4$. }
	\label{fig:GMSUNl6}
\end{figure}

\begin{figure}[h]
	\centering
	\includegraphics[width=6.3in]{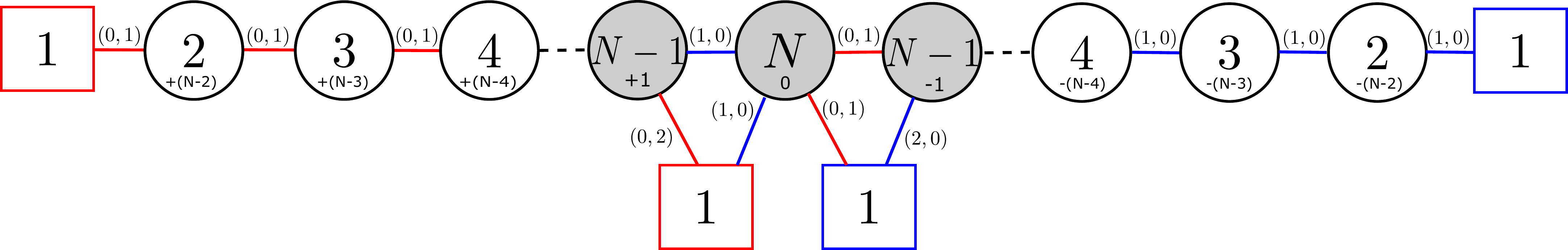}
	\caption{Repeated action of Seiberg duality on the quiver in figure \ref{fig:GMSUNl6} can mutate it into the quiver shown here. All the singlets become massive and integrated out.}
	\label{fig:GMSUNlultimate}
\end{figure}

We will now have to again go through the series of dualities mentioned in the previous paragraph, this time stopping when we dualize the $\pm(N-4)$-th nodes. This gives us the quiver of figure \ref{fig:GMSUNl5}. 
Dualizing the penultimate nodes on either sides of this quiver gives the quiver that can be represented by figure \ref{fig:GMSUNl6}. 
We can now repeat the series of dualities outlined earlier (starting by dualizing the $0$-th node, followed by dualizing the $(\pm1)$-st node and so on) multiple times such that we ultimately land on a linear quiver that corresponds to figure \ref{fig:GMSUNlultimate}. 
Dualizing the $0$-th node of this quiver then lands us on the duality frame of figure \ref{fig:evidenceofconfinementlowenergy} which is the result we sought. 

\subsection{Infinitely many $\CN=1$ duals for $SU(N)$ SQCD with $2N$ flavors}
\begin{figure}[h]
	\centering
	\begin{subfigure}[b]{6in}
	\centering
	\includegraphics[height=0.5in]{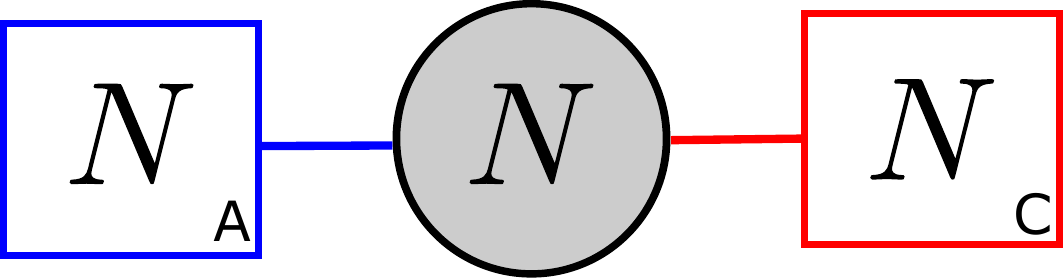}
	\caption{A quiver description dual to the $SU(N)$ SQCD with $2N$ flavors.}
	\label{fig:SQCDSUn}
	\end{subfigure}
	
	\begin{subfigure}[b]{6in}
	\centering
	\includegraphics[width=5.9in]{TNmDualofSQCD}
	\caption{A quiver description of $\CU_N^{(2)}$ obtained by gluing the two copies of $\widetilde{T}_{N}^{(2)}$. }
	\label{fig:TNmDualofSQCD}
    	\end{subfigure}
	
 	\caption{Some of the dual descriptions for the 4-punctured sphere theory with $(p, q)=(1, 1)$. Here we have maximal punctures of each color and minimal punctures of each color.}
	\label{fig:SQCDSUnDuals}
\end{figure}

Let us now consider the case of SQCD with $SU(N)$ gauge group and $2N$ flavors. From the class $\CS$ point of view, what we need to do is to start with 4-punctured (all maximal, 2 $+$ and 2$-$ color) sphere with $(p, q)=(1, 1)$ as in the section \ref{subsec:gaugedTn}, and then partially close the two maximal punctures of each color. This will result in replacing the $T_N$ block we glued to the end of the quivers by bifundamental hypermultiplets of $SU(N) \times SU(N)$. See the figure \ref{fig:SQCDSUnDuals}.

The matter content for the theory $\CU_N^{(m)}$ similar to the figure \ref{fig:TNmDualofSQCD} is given in the table \ref{table:matterUm}. 
\begin{table}[h]
\centering
\begin{tabular}{|c|c|c|c|c|c|c|c|}
	\hline
	& $SU(N)^{\pm}_i$ & $SU(N)_A $ & $U(1)_B$ & $SU(N)_C $ & $U(1)_D$ & $(J_+, J_-)$ \\
	\hline \hline
	$\phi^{+}_i$ $(1 \le i \le m)$ & $\textrm{adj}$ & & & & & $(0, 2)$  \\
	$q^+ , \tilde{q}^+$ $(i=m)$ & $\square, \bar{\square}$ &  & $1, -1$ & & & $(1, -N+1)$ \\
	$\mu^{+}_i$ $(0 \le i \le m)$ & adj & & & & & $(2, 0)$ \\
	$\tilde{\mu}^{+}_i$ $(1 \le i \le m-1)$ & adj & & & & & $(2, 0)$ \\
	$\mu_A$ & & adj & & & & $(2, 0)$ \\  
	$\mu^{+, (i)}_{\fj, \fj}$ $(2 \le i \le m+1 )$ & & & & & & $(2, -2\fj)$ \\
	$M^{+, (i)}_{\fj, -\fj}$ $(2 \le i \le m+1 )$ & & & & & & $(0, 2\fj + 2)$ \\
	\hline
	$\phi^{-}_i$ $(1 \le i \le m)$ & $\textrm{adj}$ & & & & & $(2, 0)$  \\
	$q^-, \tilde{q}^-$ $(i=m)$ & $\square, \bar{\square}$ & & &  & $1, -1$ & $(-N+1, 1)$ \\
	$\mu^{-}_i$ $(0 \le i \le m)$ & adj & & & & & $(0, 2)$ \\
	$\tilde{\mu}^{-}_i$ $(1 \le i \le m-1)$ & adj & & & & & $(0, 2)$ \\
	$\mu_C$ & & & & adj & & $(0, 2)$ \\  
	$\mu^{-, (i)}_{\fj, \fj}$ $(2 \le i \le m+1 )$  & & & & & & $(-2\fj, 2)$ \\
	$M^{-, (i)}_{\fj,-\fj}$ $(2 \le i \le m+1 )$ & & & & & & $(2\fj+2, 0)$ \\
	\hline
\end{tabular}
\caption{`Matter content' of the $\CU_N^{(m)}$ theory. Here $1 \le \fj \le N-1$.}
\label{table:matterUm}
\end{table}
The superpotential is given by 
\be
W = W'_+ + W'_- + \tr \mu^{+}_0 \mu^{-}_0 \ ,
\ee
where
\be
W'_\s = \sum_{k=1}^m \Tr \phi^\s_k (\mu^\s_k - \tilde{\mu}^\s_k ) + \sum_{i=2}^{m+1} \sum_{\fj=1}^{N-1} \mu^{\s, (i)}_{\fj, \fj} M^{\s, (i)}_{\fj, -\fj} \ ,
\ee
with
\be
\mu_m^\s = q^\s \tilde{q}^\s - \frac{1}{N} \tr (q^\s \tilde{q}^\s) \ , \quad \hat{\mu}^{\s, (m+1)}_{\fj,\fj} = \tr \tilde{q}^{\s} q^{\s}(\phi^{\s}_m)^{N-j-1} \ .
\ee

\begin{figure}[h]
	\centering
	\includegraphics[width=5.5in]{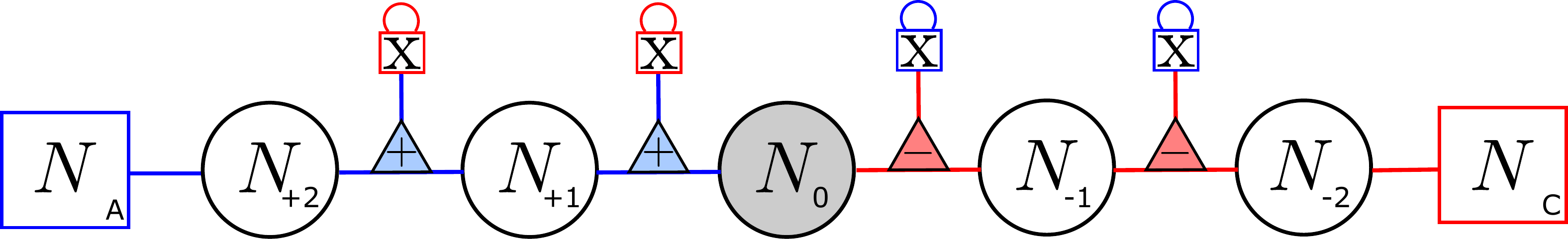}
	\caption{Another quiver description obtained by gluing two copies of $\widetilde{T}_N^{(2)}$. We call this as $\widehat{\CU}_N^{(2)}$. The theory will undergo cascading RG glow to the SQCD. }
	\label{fig:SQCDSUNDual}
\end{figure}

\paragraph{Anomaly coeffecients}
As an intermediate step, let us consider the Higgsed $T_N^{(m)}$ theory by Higgsing one of the punctures. Let us call it $\widetilde{T}_N^{(m)}$. This theory is given by the UV curve $\CC^{(m+1,-m)}_{0, 3}$ with $n_+=3$ where 2 of the punctures are maximal the other is minimal. The quiver diagram of the theory is the left half of figure \ref{fig:SQCDSUNDual} with central gauge group ungauged. When $m=0$, it becomes a theory of free $SU(N)_A \times SU(N)_G$ bifundamental hypermultiplets with $U(1)_B$ baryonic symmetry. The anomalies of this theory are given as:
\be
\begin{array}{c|cc}
  J_+ ,  J_+^3 && m(1-N) \\
  J_- , J_-^3  &&  m(N-1)-2N^2  \\
  J_+^2 J_- && \frac{1}{3}(4N^3-N-3)\\
  J_+ J_-^2 && -\frac{1}{3}(4N^3-N-3) \\
  J_- SU(N)_{A}^2, J_- SU(N)_{G}^2 && -N  \\
  J_+ SU(N)_{A}^2, J_+ SU(N)_{G}^2 && 0  \\
  J_+ U(1)_B^2 && 0\\
  J_- U(1)_B^2 && -2N^2 \\
  J_+^2 U(1)_{B}, J_-^2 U(1)_{B} && 0\\
\end{array}
\ee
Here $A$ and $G$ are the two maximal punctures while $B$ is the name we used for the minimal puncture. The anomalies of the $T_N^{(m)}$ theory  with all its colors inverted can be obtained by interchanging the roles of $J_+$ and $J_-$ in the above table.

We now compare the anomaly coefficients of our proposed dual theories. For $\CU_N^{(m)}$, we find:
\be
\begin{array}{c|cc}
  J_+ ,  J_+^3 , J_- , J_-^3  &&  -N^2-1  \\
  J_+^2 J_- , J_+ J_-^2&&  N^2-1 \\
  J_+ SU(N)_{A}^2, J_- SU(N)_{C}^2 && 0  \\
  J_- SU(N)_{A}^2, J_+ SU(N)_{C}^2 && -N  \\
  J_+ U(1)_B^2, J_- U(1)_D^2 && 0\\
  J_+ U(1)_D^2, J_- U(1)_B^2 && -2N^2\\
  J_+^2 U(1)_{B,D}, J_-^2 U(1)_{B,D} && 0\\
\end{array}
\ee
As before we find that these coefficients are independent of $m$ and match perfectly with those of $SU(N)$ SQCD with $2N$ flavors.

\paragraph{Cascading RG flows to SQCD}
As in the case of the section \ref{subsec:gaugedTn}, let us consider a dual description for the $\widetilde{T}_N$ theory itself to show that it flows to the same theory as the $SU(N)$ SQCD with $2N$ flavors.  
The `matter content' of the theory $\CU_N^{(m)}$ (figure \ref{fig:SQCDSUNDual}) is quite similar as in section \ref{subsec:gaugedTn}, but we get $SU(N)_A \times U(1)_B \times SU(N)_C \times U(1)_D \times U(1)_R \times U(1)_\CF$ global symmetry instead. It is described in the table \ref{table:matterUhm}. 
\begin{table}[h]
	\centering
    	\begin{tabular}{|c|c|c|c|c|c|c|c|}
    	\hline
        & $SU(N)^{\pm}_i$ & $SU(N)_A $ & $U(1)_B$ & $SU(N)_C $ & $U(1)_D$ & $(J_+, J_-)$ \\
        \hline \hline
        $\phi^{+}_i$ $(1 \le i \le m)$ & $\textrm{adj}$ & & & & & $(0, 2)$  \\
        $q^+, \tilde{q}^+$ $(i=m)$ & $\square, \bar{\square}$ & $\square, \bar {\square}$ & $1, -1$ & & & $(1, 0)$ \\
        $\mu^{+}_i$ $(0 \le i \le m)$ & adj & & & & & (2, 0) \\
        $\tilde{\mu}^{+}_i$ $(1 \le i \le m)$ & adj & & & & & (2, 0) \\
        $\mu^{+, (i)}_{\fj, \fj}$ $(1 \le i \le m)$ & & & & & & $(2, -2\fj)$ \\
        $M^{+, (i)}_{\fj, -\fj}$ $(1 \le i \le m)$ & & & & & & $(0, 2\fj + 2)$ \\
	\hline
	$\phi^{-}_i$ $(1 \le i \le m)$ & $\textrm{adj}$ & & & & & (2, 0)  \\
        $q^-, \tilde{q}^-$ $(i=m)$ & $\square, \bar{\square}$ & & & $\square, \bar{\square}$ & $1, -1$ & (0, 1) \\
        $\mu^{-}_i$ $(0 \le i \le m)$ & adj & & & & & (0, 2) \\
        $\tilde{\mu}^{-}_i$ $(1 \le i \le m)$ & adj & & & & & (0, 2) \\
        $\mu^{-, (i)}_{\fj, \fj}$ $(1 \le i \le m)$ & & & & & & $(-2\fj, 2)$ \\
        $M^{-, (i)}_\fj$ $(1 \le i \le m)$ & & & & & & $(2\fj+2, 0)$ \\
	\hline
	\end{tabular}
	\caption{The `matter content' of the $\widehat{\CU}_N^{(m)}$ theory. }
	\label{table:matterUhm}
\end{table}

The set of chiral operators in the $T_N$ theory contains (anti-)trifundamental operator $Q_{ijk}$ and $\tilde{Q}^{ijk}$. When an oppositely colored puncture of the $T_N$ block is closed, the operators $Q_{ijk}, \tilde{Q}^{ijk}$ split into $N$ bifundamental operators $Q_{ij(\ell)}, \tilde{Q}^{ij (\ell)}$ with $-\frac{N-1}{2} \le \ell \le \frac{N-1}{2}$, and the corresponding charges being $(J_+,J_-) = (N-1, -2\ell)$ or $(-2\ell, N-1)$ depending on the choice of color. These operators will be important to our analysis and we will label those coming from the $i$-th block in figure \ref{fig:SQCDSUNDual} as $Q^{\s,(i)}_\ell, \tilde{Q}^{\s,(i)}_\ell$ suppressing indices.

The superpotential for the theory is given as
\be \label{eq:hatNUmW}
 W = W'_+ + W'_- + \tr \mu^{+}_0 \mu^{-}_0 + \sum_{k=1}^{m} \tr \hat{\mu}^{+}_k \hat{\mu}^{-}_k \ ,
\ee
where
\be
 W'_\s = \sum_{k=1}^m \Tr \phi^\s_k (\mu^\s_k - \tilde{\mu}^\s_k ) + \sum_{i=1}^m \sum_{\fj=1}^{N-1} \mu^{\s, (i)}_{\fj, \fj} M^{\s, (i)}_{\fj, -\fj} \ ,
\ee
with
\be
 \mu_m^\s = q^\s \tilde{q}^\s - \frac{1}{N} \tr (q^\s \tilde{q}^\s) \ ,
\ee
and
\be
\hat{\mu}^{\s}_k =\left(\prod_{i=1}^{k} Q^{\s,(i)}_{\frac{N-1}{2}}\big)\phi^\s_{k}\big(\prod_{i=1}^{k} \widetilde{Q}^{\s,(i)}_{\frac{N-1}{2}}\right)  \ .
\ee
Here we formed the gauge invariant operators $\mu_m^\s$ so as to transform as the adjoint of $SU(N)_{\pm m}$ according to whether $\s = \pm$ while $\hat{\mu}^{\s}_k$ is constructed such that it transforms as the adjoint of $SU(N)_0$.

By applying a sequence of dualities, we have showed earlier that the central $SU(N)_0$-node confines.  From this, we conjecture that the $SU(N)_0$-node undergoes confinement with $N^2$ mesonic operators $\tilde{Q}^{\pm ,(1)}_{\frac{N-1}{2}} Q^{\mp,(1)}_{\frac{N-1}{2}}$ and quantum deformed moduli space given by 
\be
\det \big( \tilde{Q}^{\pm ,(1)}_{\frac{N-1}{2}} Q^{\mp,(1)}_{\frac{N-1}{2}}\big)- ``(\mu ^{+, (1)}  _{\fj, \fj =1}\mu ^{-, (1)}_{\fj, \fj =1})^{\frac{1}{2}N(N-1)} "= \Lambda _0^{b(N-1)} \ ,  \label{Ndeform} 
\ee 
where $\Lambda _0 ^b$ is the $SU(N)_0$ instanton factor, with the exponent $b$ determined by 
\be
b=3N-2k=3\epsilon _{UV}N , \qquad\hbox{where}\quad k=-3{\rm Tr} R _{UV}SU(N)_0^2=\frac{3}{2}(1-\epsilon _{UV})N .
\ee
The scaling dimensions of the two sides of \eqref{Ndeform} agree, upon using $\Delta =\frac{3}{2}R_{UV}$, where $R_{UV}$ is the superconformal R-charge before gauging $SU(N)_0$.   Gauging $SU(N_0)$ breaks the separate $U(1)_{\cal F_\pm}$ to $U(1)_{\cal F}=U(1)_{\cal F_+}-U(1)_{\cal F_-}$, with $U(1)_A=U(1)_{\cal F_+}+U(1)_{\cal F_-}$ anomalous.  The ${\rm Tr}U(1)_ASU(N)_0^2=N$ anomaly implies that $\Lambda _0^b$ carries charge $+2N$ under $U(1)_A$, which is consistent with the $U(1)_A$ charge of the product of operators on the LHS of \eqref{Ndeform}.
The operators on the LHS of \eqref{Ndeform} carry $U(1)_{R_{IR}}$ charge zero, as required for a quantum deformed chiral ring relation (and that is why other  $Q^{\pm,(i)}_{\ell }$, $\tilde Q^{\pm,(i)}_{\ell }$  do not appear in \eqref{Ndeform}).

The first and second term in the LHS of \eqref{Ndeform} are analogs of $\det \CM$ and $\CB\tilde{\CB}$ in SQCD with $N_f = N_c$. We put the second term in quotes because we have not fully determined the dependence on the $\mu_{\fj, \fj}^{\pm}$ beyond what is fixed by the symmetries. 
In any case, the $F$ terms of superpotential \eqref{eq:hatNUmW} sets the operators $\mu ^{\pm , (i)}  _{\fj, \fj}$ to zero, setting the terms in quotes to zero in \eqref{Ndeform}.  
On the deformed space  \eqref{Ndeform}, the $Q^{\pm,(1)}_{\frac{N-1}{2}}$ and $\tilde Q^{\pm,(1)}_{\frac{N-1}{2}}$ thus have non-zero expectation value. 
Then $\phi_1^+$ and $\phi_1^-$ will become massive via the last term of \eqref{eq:hatNUmW} with $k=1$. Moreover, the $SU(N)_{+1} \times SU(N)_{-1}$ gauge symmetry is broken down to the diagonal $SU(N)$, which will again undergo confinement. This is an iterative cascade of RG flows, reducing $m$ in each step, eventually flowing to $SU(N)$ SQCD with $2N$ flavors with a quartic superpotential in the IR.

\section{Superconformal index} \label{sec:index}

The superconformal index for a $\CN=1$ superconformal field theory is defined as
\be
 I (\fp, \fq, \xi; \vec{x}) = \Tr (-1)^F \fp^{j_1 + j_2 + \frac{R_0}{2}} \fq^{j_2 - j_1 + \frac{R_0}{2}} \xi^{\CF} \prod_i x_i^{F_i} \ .
\ee
where we introduced the fugacity $\xi$ for the $U(1)_\CF$ which is present for generic class $\CS$ theories. For the theory having a Lagrangian description in the UV, the index can be simply computed by multiplying the contributions from each matter multiplets in the UV and then by integrating over the gauge group. The contribution of each matter multiplets is calculated using the exact $R$-charge in the IR \cite{Romelsberger:2005eg}. In our case, the only possible non-anomalous $U(1)$ symmetry that can mix with $R$-symmetry in the IR is $U(1)_\CF$. Therefore we can obtain the index using the UV $R$-charge as long as we keep the fugacity $\xi$ turned on. Once we know the exact $R$-charge $R = R_0 + \e \CF$, we can simply redefine $\xi \to \xi (\fp \fq)^{\e/2}$ to obtain the true superconformal index.

\subsection{Topological field theory and superconformal index}

For an $\CN=1$ SCFT in class $\CS$, the superconformal index can be written in terms of a correlation function of the 2d (generalized) topological field theory living on the UV curve. This topological field theory is related to a deformation of 2d Yang-Mills theory \cite{Gadde:2009kb,Gadde:2011ik,Gadde:2011uv, Gaiotto:2012xa, Beem:2012yn,Rastelli:2014jja}. The index can be written as
\be \label{eq:tqftidx}
 I (\fp, \fq, \xi; \vec{a}_{i}) = \sum_\lambda (C_\lambda^+)^p (C_\lambda^-)^q \prod_{i=1}^{n} \psi^{\rho_i, \s_i}_\lambda (\vec{a}_{i}) \ ,
\ee
where $(p, q)$ are the degrees of the line bundles and $n$ is the number of punctures, which should satisfy the relation $p+q=2g-2+n$. Here we suppressed the $\fp, \fq, \xi$ dependence and the sum is over the representations $\lambda$ of $\Gamma$ labelling the six-dimensional $(2, 0)$ theory.

The basis function $\psi_\lambda^{\rho, \s} (\vec{a})$ corresponding to the puncture labelled by the embedding $\rho: SU(2) \to \G$ and color $\s$ can be written in the following form
\be
 \psi_{\lambda}^{\rho, \s} (\vec{a}) = K_{\rho}(\vec{a}; t_\s) P_{\lambda} ( \vec{a} t_\s^{\rho}) \ ,
\ee
where $t_\s = \xi^\s \sqrt{\fp \fq}$ and we suppressed the $\fp, \fq$ dependence. The $K$-factor does not depend on $\lambda$, but the form of the function depends on the type of puncture. $P_\lambda$ is a symmetric function of $\vec{a}$ which in certain limit reduces to the Macdonald polynomial. The argument $\vec{a} t^{\rho}_{\s}$ is determined by the embedding $\rho$ of $SU(2)$ into $\Gamma$ labelling the puncture (see \cite{Mekareeya:2012tn}). The structure constant can be written as $C^\s_\lambda = (\psi^{\varnothing, \s}_\lambda )^{-1}$ in terms of the basis function $\psi$'s.

Let us compute the index of the $T_N^{(m)}$ starting from the theory given by the UV curve $\CC_{0, m+3}^{(m+1, 0)}$ with $(n_+, n_-)=(3, m)$ where we know how to write the index from the TQFT:
\be
 I[\CC_{0, m+3}^{(m+1, 0)}] = \sum_{\lambda} (C_\lambda^+)^{m+1} \prod_{i=1}^{3} \psi_{\lambda}^{+} (\vec{a}_i) \prod_{j=1}^{m} \psi_{\lambda}^{-} (\vec{b}_i) \ . 
\ee
Now, we want to Higgs all the $-$ punctures. Complete Higgsing or closing of a puncture is implemented via replacing the wave function $\psi_\lambda^{\rho, \s} (\vec{b})$ corresponding to the puncture to close by $\psi_\lambda^{\varnothing, \s} (t_\s^\rho)$. From the relation $C^\s_\lambda = (\psi^{\varnothing, \s}_\lambda )^{-1}$, we see that the degree of the normal bundle corresponding to the color $\s$ reduces upon Higgsing. We get
\be \label{eq:idxTpq}
 I[T_{N}^{(m)}] (\fp, \fq, \xi; \vec{a}_{i}) = \sum_\lambda \frac{(C_\lambda^+)^{m+1}}{(C_\lambda^-)^{m}} \psi^{+}_\lambda (\vec{a}_{1}) \psi^{+}_\lambda (\vec{a}_{2}) \psi^{+}_\lambda (\vec{a}_{3}) \ ,
\ee
where we suppressed $\rho_i$ to denote full punctures. One can also flip all the colors $\pm$ in the components to get the same index with $\xi \to \xi^{-1}$. This is of the same form as the equation \eqref{eq:tqftidx}, from which we can plug in $(p, q) = (m+1, -m)$ with $3$ $+$ colored punctures. 

Once we have the equation \eqref{eq:idxTpq}, it is a piece of cake to show that the index is the same for the dual theories, independent of $m$. Gluing two copies of $T_{N}^{(m)}$ with opposite color by a cylinder to form the theory corresponding to the 4-punctured sphere with $(p, q)=(1, 1)$, the index can be written as
\be
I(\vec{a}, \vec{b}, \vec{c}, \vec{d}) &=& \sum_{\lambda, \mu} \frac{(C_\lambda^+)^{m+1}}{(C_{\lambda}^-)^{m}} \psi_\lambda^+(\vec{a})\psi_\lambda^+(\vec{b}) \left( \oint [d\vec{z}] I_{\textrm{vec}}(\vec{z}) \psi_\lambda^+(\vec{z}) \psi_{\mu}^-(\vec{z}) \right)
 \frac{(C_{\mu}^-)^{m+1}}{(C_{\mu}^+)^{m}} \psi_{\mu}^-(\vec{c})\psi_{\mu}^-(\vec{d}) \nn \\
 &=& \sum_{\lambda} C_\lambda^+ C_{\lambda}^- \psi_\lambda^+(\vec{a})\psi_\lambda^+(\vec{b})\psi_{\lambda}^-(\vec{c})\psi_{\lambda}^-(\vec{d}) \ . 
\ee
We here used the fact that wave functions are orthonormal:
\be
 \oint [d\vec{z}] I_{\textrm{vec}}(\vec{z}) \psi_\lambda^+(\vec{z}) \psi_{\mu}^-(\vec{z}) = \delta_{\lambda \mu} \ ,
\ee
where $I_{\textrm{vec}}(\vec{z})$ is the contribution to the index from a $\CN=1$ vector multiplet. 
Therefore for any choice of $m \in \IZ$ the gluing gives us the same index as that of the theory described by 2 full + punctures and 2 full - punctures and $(p, q)=(1, 1)$. It describes the two copies of $T_N$ theory glued by $\CN=1$ vector multiplet. The same argument goes through when we Higgs or partially close the full punctures of each color to minimal punctures to get the SQCD.

In the paper \cite{Beem:2012yn}, the superconformal index for the generic $(p, q)$ was proposed from the structure of the (generalized) topological field theory, initially without concrete SCFTs that realize the indices.  The SCFT that we discuss here gives such a concrete realization.

\subsection{Direct computation for the $SU(2)$ theories}
The proof of the previous section holds as long as the index of the $T_N$ theory can be written in terms of the basis wave function $\psi_\lambda (\vec{a})$. Here, we confirm the TQFT formula for $T_2^{(m)}$ theories \eqref{eq:idxTpq} by directly computing the index using the matter content of section \ref{sec:su2}. 

The index for a chiral multiplet with $(J_+, J_-)$ charge is given as
\be
 I^{(J_+, J_-)}_{\textrm{chi}}(\fp, \fq, \xi; \vec{z}) 
 = \prod_{\vec{v} \in \CR} \Gamma( (\fp\fq)^{\frac{R_0}{2}} \xi^{\CF} \vec{z}^{v}; \fp, \fq ) 
= \prod_{\vec{v} \in \CR} \Gamma( (\fp\fq)^{\frac{J_+ + J_-}{4}} \xi^{\frac{J_+ - J_-}{2}} \vec{z}^{\vec{v}}; \fp, \fq ) \ ,
\ee
where $\vec{v}$ are the weight vectors of the representation $\CR$ of the symmetry group the chiral multiplet is charged under. Here the notation $\vec{z}^{\vec{v}} $ is a short-hand for $\prod_i z_i^{v_i}$. Here, we used the elliptic gamma function which is defined as
\be
 \G (z; \fp, \fq) = \prod_{m, n =0}^{\infty} \frac{1 - z^{-1} \fp^{m+1} \fq^{n+1}}{1 - z \fp^m \fq^n} \ ,
\ee
to write the index in a concise form. We will suppress the $\fp, \fq$ dependence of $\Gamma(\vec{z}; \fp, \fq)$ whenever possible.

The vector multiplet contribution to the index is given by
\be
 I_{\textrm{vec}} (\fp, \fq; \vec{z}) 
 = \frac{1}{|\CW|} \prod_{\vec{\a} \in \Delta_G} \Gamma(\vec{z}^{\vec{\a}})^{-1}  \ ,
\ee
where $\CW$ is the Weyl group of $G$ and $\Delta_G$ is the set of root lattices of $G$. We also included the Haar measure for the gauge group $G$ to the vector multiplet index for convenience. For the $SU(N)$ gauge group, we get
\be
 I_{\textrm{vec}} (\fp, \fq; \vec{z}) = \frac{(\fp; \fp)^{N-1} (\fq; \fq)^{N-1} }{N!} \prod_{i \neq j} \frac{1}{\Gamma(z_i/z_j)} \ ,
\ee
where $i, j = 1, \cdots N$ and $\prod_i z_i = 1$. Here $(z; q)$ is the $q$-Pochhammer symbol which is defined to be
$(z; q) = \prod_{m=0}^{\infty} (1 - z q^m)$.

\paragraph{$T_2^{(m)}$ theory}
Let us compute the superconformal index of the $T_2^{(1)}$ theory discussed in section \ref{subsec:T2m1}.
We would like to compute the index in the UV using the description given as in figure \ref{fig:T22m1quiver} and show that it agrees with the TQFT formula.
The index on the electric side can be written as
\be
 I(\fp, \fq, \xi; a, b, c) &=& \oint \frac{dz}{2\pi i z} I_{\textrm{vec}}(z) I_{\textrm{chi}}^{(0, 2)}(z^{\pm 2, 0}) I_{\textrm{chi}}^{(1, 0)}(z^\pm a^\pm b^\pm) I_{\textrm{chi}}^{(1, -1)} (z^\pm c^\pm) I_{\textrm{chi}}^{(0, 4)} (1) \\
  &=& \kappa \oint \frac{dz}{2\pi i z} \frac{\G(z^{\pm 2, 0} (\fp \fq)^\half \xi^{-1})}{2 \G(z^{\pm 2})} \G(z^\pm a^\pm b^\pm (\fp \fq)^{1/4} \xi^{-\half}) \G(z^\pm c^\pm \xi) \nn
 \G( \fp \fq \xi^{-2})   \ ,
\ee
where $\kappa = (\fp; \fp)(\fq; \fq)$. We use a short-hand notation of $\pm$ to denote multiple products involving each sign. For example $f(a^\pm b^\pm) \equiv f(ab)f(a b^{-1})f(a^{-1} b) f(a^{-1} b^{-1}) $. Also, $f(z^{\pm 2, 0})$ means $f(z^2)f(z^{-2})f(z^0)$.

One tricky part here is choosing the correct contour for this integral. Usually, one picks the contour to be the unit circle and assumes $|\fp|, |\fq| < 1$ and $|\xi| = |a|=|b|=|c|=1$ so that we pick up the poles only inside the unit circle. This works as long as there is no chiral multiplet with $R_0$ or $R$ charge less than equal to zero. But if there is a chiral multiplet having $R_0 \le 0$, some of the poles may lie along the unit circle. In \cite{Agarwal:2014rua}, it was argued that one should take $|\xi^f (\fp \fq)^{r/2}| < 1 $ for the chiral multiplet with $R_0$-charge $r$ and $\CF$-charge $f$. Therefore, we need to include all the poles of the form $x \xi^f \fp^{\frac{r}{2} +m} \fq^{\frac{r}{2} + n}$ with $x$ being products of the fugacities corresponding to the gauge/flavor symmetries.

In our case, we have the poles of the form $z = (a^\pm b^\pm \xi^{1/2} (\fp \fq)^{1/4} \fp^m \fq^n)^\pm $  with $m, n \in \IZ_{\ge 0}$ from the chiral multiplets with $(J_+, J_-) = (1, 0)$ and poles of the form $z = (c^\pm \xi^{-1} \fp^m \fq^n)^\pm$ from the chirals with $(J_+, J_-) = (1, -1)$. Among the first set of poles, $z=a^\pm b^\pm \xi^{1/2} (\fp \fq)^{1/4} \fp^m \fq^n$ are the ones inside the unit circle and the other half of the poles are outside the contour. For the second set of poles, $z=c^\pm \xi^{-1} \fp^m \fq^n $ are the ones inside the contour. 

The index for the $T_2^{(m)}$ can be written as
\be \label{eq:T2mIdx}
I^{(m)}= \oint \prod_{i=1}^m \left( \frac{dz_i}{2\pi i z_i} I_{\textrm{vec}}(z_i) I_{\textrm{chi}}^{(0, 2)} (z^{\pm 2, 0}) I_{\textrm{chi}}^{(1, -1)} (z_{i-1}^\pm z_i^\pm ) I^{(0, 4)}_{\textrm{chi}}(1) \right) I_{\textrm{chi}}^{(1, 0)}(z_m^\pm a^\pm b^\pm) \ ,
\ee
where $z_0 = c$.
We confirmed that this indeed gives us the same index as the TQFT prediction of \eqref{eq:idxTpq} at the first few leading orders in $\fp$ and $\fq$ for $m=1, 2$.
If the dualities hold, we have the identity
\be
 \oint \frac{dz}{2\pi i z} I_{\textrm{vec}}(z) I^{(m)}(\xi) I^{(m)}(\xi^{-1})
 =  \oint \frac{dz}{2\pi i z}  I_{\textrm{vec}}(z) I_{\textrm{chi}}^{(1, 0)} (z^\pm a^\pm b^\pm) I_{\textrm{chi}}^{(0, 1)} (z^\pm c^\pm d^\pm) \ ,
\ee
where we glued two $T_2^{(m)}$ with opposite $\CF$ charges. We have verified this identity to hold for $m=1, 2$ at the leading orders in $\fp$ and $\fq$.

\paragraph{SQCD vs $\widehat{\CU}_N^{(m)}$ theory}
Let us compute the index in the dual frame $\widehat{\CU}_N^{(m)}$. In this frame, we should be able to see $SU(8)$ flavor symmetry since it cascades to the SQCD in the IR. In order to see this from the index, first we refine the index \ref{eq:T2mIdx} as
\be \label{eq:T2mfake}
\tilde{I}^{(m)} (\vec{a}) = \oint \prod_{i=1}^m \left( \frac{dz_i}{2\pi i z_i} I_{\textrm{vec}}(z_i) I_{\textrm{chi}}^{(0, 2)} (z^{\pm 2, 0}) I_{\textrm{chi}}^{(1, -1)} (z_{i-1}^\pm z_i^\pm ) I^{(0, 4)}_{\textrm{chi}}(1) \right)\prod_{n=1}^4 I_{\textrm{chi}}^{(1, 0)}(z_m^\pm a_n) , \quad
\ee
where $\prod_{i=1}^4 a_i = 1$. Here we introduced the fugacities for the $SU(4)$ flavor symmetry $a_{i=1, 2, 3}$. And then we find 
\be
 \oint \frac{dz}{2\pi i z} I_{\textrm{vec}}(z) \tilde{I}^{(m)}(\vec{a}, \xi) \tilde{I}^{(m)}(\vec{b}, \xi^{-1})
 =  \oint \frac{dz}{2\pi i z}  I_{\textrm{vec}}(z) \prod_{m=1}^4 I_{\textrm{chi}}^{(1, 0)} (z^\pm a_m) I_{\textrm{chi}}^{(0, 1)} (z^\pm b_m) , \quad
\ee
where we also refined the index for the SQCD. One can easily check the index preserves $SU(8)$ flavor symmetry by relabelling the fugacities. 

We should keep in mind that $\tilde{I}^{(m)}$ in \eqref{eq:T2mfake} is not a genuine index of the theory, since $T_2^{(m)}$  itself does not have the $SU(4)$ symmetry. There is a cubic coupling which breaks $SU(4) \to SU(2)^2$, and this coupling cannot be tuned to zero as we have discussed in section \ref{subsec:T2pq}. But after gluing two copies of $T_2^{(m)}$, we have exactly marginal deformations which includes the point with enhanced symmetry.

\section{Conclusion and outlook}

Guided by the construction of 4d QFTs from M5 branes wrapping Riemann surfaces, we constructed an infinite set of dual theories of $4d$ $\CN=1$ $SU(N)$ SQCD with $2N$ flavors. These theories are parametrized by an integer $m \in \IZ_{\ge 0}$ and involve $2m$ copies of the $T_N$ theory of \cite{Gaiotto:2009we}, $2N$ quarks/anti-quarks along with $2m(N-1)$ singlet chiral superfields as their building blocks.  As a check of the dualities we compared their central charges, anomaly coefficients and superconformal indices.  Along the way, we constructed a family of new $\CN=1$ SCFTs with $SU(N)^3$ flavor symmetries, which generalize the $\CN=2$ $T_N$ theory. 

The dual theories discussed here can be used to construct more duals, for example by applying them to the magnetic dual of  \cite{Seiberg:1994pq}. This will result in adding extra chiral multiplets transforming as adjoints of global symmetries $SU(N)_{A, C}$ and cubic superpotential terms. We can also consider the swapped dual of \cite{Gadde:2013fma}, and also Argyres-Seiberg type duals of \cite{Agarwal:2013uga,Agarwal:2014rua}. Moreover, as we have discussed in the section \ref{subsec:T23m2}, even the building block $T_N^{(m)}$ itself has many different dual descriptions,  so the number of duals grows rapidly with $m$.

One question is how to generalize our dualities to $N_f \neq 2N$. This may be possible e.g. by considering a mass deformation of the $T_N^{(m)}$ theory, as was done in the $T_N$ case \cite{Hayashi:2014hfa}. From the class $\CS$ perspective, this involves understanding dualities in the presence of irregular punctures.  Another direction would be a more detailed study of phase structure and chiral ring of the new theories. The spectral curve of the generalized Hitchin system associated to the $\CN=1$ theories \cite{Bonelli:2013pva,Xie:2013rsa,Yonekura:2013mya, Giacomelli:2014rna,Xie:2014yya} will be useful. 
It will be also interesting to generalize our construction of $T_N^{(m)}$ to $D$ and $E$ type theories and also with outer-automorphism twists using the $\CN=2$ results \cite{Tachikawa:2009rb, Tachikawa:2010vg, Chacaltana:2011ze,Chacaltana:2012ch,Chacaltana:2013oka,Chacaltana:2014jba, Chacaltana:2015bna}, as well as possible generalizations using the theories of \cite{Gaiotto:2015usa,Franco:2015jna}, which will provide analogous infinitely many duals for other gauge groups.

\acknowledgments
JS would like to thank Dan Xie for motivating him to understand the class $\CS$ theories with general twists and pointing out the possibility of an infinite number of duals. PA and JS thank Ibrahima Bah and Kazunobu Maruyoshi for the related collaborations and discussions. This work is supported by the DOE grant FG03-97ER40546.

\bibliographystyle{ytphys}
\bibliography{refs}

\end{document}